\documentclass[a4paper,11pt]{article}
\usepackage{jheppub} 
\def\CA{{\cal A}}

\def\CL{{\cal L}}

\def\CN{{\cal N}}
\def\CO{{\cal O}}

\def\CW{{\cal W}}
\def\CS{{\cal S}}

\DeclareMathOperator{\Tr}{Tr}

\newcommand{\vev}[1]{ \left\langle {#1} \right\rangle }

\def\diag{\mathop{\rm diag}\nolimits}

\def\SO{\mathop{\rm SO}}

\def\SU{\mathop{\rm SU}}
\def\U{\mathrm{U}}

\def\SL{\mathop{\rm SL}}
\def\tr{\mathop{\rm tr}}
\def\Im{\mathop{\rm Im}}
\def\Re{\mathop{\rm Re}}

\def\beq#1\eeq{\begin{align}#1\end{align}}

\title{Supersymmetric gauge theory, (2,0) theory and twisted 5d Super-Yang-Mills}
\author[]{Kazuya Yonekura}
\affiliation[]{School of Natural Sciences, Institute for Advanced Study, 1 Einstein Drive,\\ Princeton, NJ 08540 USA}
\abstract{Twisted compactification of the
6d $\CN=(2,0)$ theories on a punctured Riemann surface give a large class of 
4d $\CN=1$ and $\CN=2$ gauge theories, called class~${\cal S}$. We argue that nonperturbative dynamics of class~${\cal S}$
theories are described by 5d 
maximal Super-Yang-Mills (SYM) twisted on the Riemann surface.
In a sense, twisted 5d SYM might be regarded as a ``Lagrangian'' for class~${\cal S}$ theories on ${\mathbb R}^{1,2} \times S^1$.
First, we show that twisted 5d SYM gives generalized Hitchin's equations which was proposed recently.
Then, we discuss how to identify chiral operators with quantities in twisted 5d SYM. 
Mesons, or holomorphic moment maps, 
are identified with operators at punctures which are realized as 
3d superconformal theories $T_\rho[G]$ coupled to twisted 5d SYM.
``Baryons'' are identified qualitatively through a study of 4d $\CN=2$ Higgs branches.
We also derive a simple formula for dynamical superpotential vev which is
relevant for BPS domain wall tensions. 
With these tools, we examine many 
examples of 4d $\CN=1$ theories with several phases such as
confining, Higgs, and Coulomb phases, 
and show perfect agreements between field theories and twisted 5d SYM.
Spectral curve is an essential tool to solve generalized Hitchin's equations,
and our results clarify the physical information encoded in the curve.
}

\begin{document} 
\maketitle
\flushbottom
%%%%%%%%%%%%%%%%%%%%%%%%%%%%%%%%%%%%%%%%%%%%%%%%%%%%%%%%%%%

%%%%%%%%%%%%%%%%%%%%%%%%%%%%%%%%%%%%%%%%%%%%%%%%%%%%%%%%%%%
\section{Introduction and summary}

M5 brane is an important object in M-theory, but numerous studies have shown that it is also relevant
for four dimensional field theories. 
It was used to give solutions of Coulomb branches of a large class of $\CN=2$ field theories~\cite{Witten:1997sc} and 
also used to reveal strong coupling dynamics of $\CN=1$ theories~\cite{Hori:1997ab,Witten:1997ep}.
Furthermore, even new types of four dimensional field theories can be constructed from M5 branes~\cite{Maldacena:2000mw,Gaiotto:2009we}.
The important building block of those theories is the so-called $T_N$ theory~\cite{Gaiotto:2009we}.
It was discovered through the study of $\CN=2$ $S$-dualities generalizing the Argyres-Seiberg duality~\cite{Argyres:2007cn}
which involves the $E_6$ theory~\cite{Minahan:1996fg}. The $T_N$ theory for $N \geq 3$ does not have a Lagrangian description
at present. The existence of those theories enlarge the landscape of 4d quantum field theories.

The low energy world volume theory of $N$ coincident M5 branes is given by the mysterious six dimensional $\CN=(2,0)$ theories of the $A_{N-1}$ type.
More generally, by taking 6d $\CN=(2,0)$ theories of ADE type and compactifying it on a Riemann surface $C$ with punctures,
we get four dimensional field theories in the low energy limit. 
Theories of this class are known as class~$\CS$~\cite{Gaiotto:2009we,Gaiotto:2009hg}.
Depending on how to twist the theory on the Riemann surface,
we get $\CN=2$ or $\CN=1$ gauge theories.

Although direct field theory analysis of class~$\CS$ theories is often difficult, there is a way to study those theories in a unified way.
The key idea is to compactify the theories on a circle $S^1$~\cite{Kapustin:1998xn,Gaiotto:2009hg}.
The six dimensional theories are put on ${\mathbb R}^{1,2} \times S^1 \times C$. If we compactify on $C$ first,
we get four dimensional theories on ${\mathbb R}^{1,2} \times S^1$. However, if we change the order of compactification
and compactify the $S^1$ direction first, we get five dimensional maximal Super-Yang-Mills (SYM) on 
${\mathbb R}^{1,2} \times C$. In the case where the four dimensional theory has $\CN=2$ supersymmetry,
the condition for supersymmetric vacua gives Hitchin's equations~\cite{Hitchin:1986vp,Hitchin:1987mz}.\footnote{
See e.g., \cite{Kapustin:2006pk,Gukov:2006jk,Gukov:2008sn} for readable physics discussions on Hitchin's equations.}
The Coulomb moduli space of the four dimensional field theory on ${\mathbb R}^{1,2} \times S^1$ 
is given by the moduli space of solutions of Hitchin's equations.
In the case in which $\CN=1$ supersymmetry remains, we get generalized Hitchin's equations
which has been proposed in \cite{Xie:2013gma,Bonelli:2013pva},
as we explicitly derive in section~\ref{sec:2} from 5d SYM.

We argue that the 5d SYM twisted on the Riemann surface, as a whole, gives the description
of class $\CS$ theories. 
There is a proposal~\cite{Douglas:2010iu,Lambert:2010iw} 
(see also e.g., \cite{Bolognesi:2011nh,Tachikawa:2011ch,Kim:2011mv,Bern:2012di,Lambert:2012qy,Bak:2013bba})
that 5d SYM contains all the degrees of freedom in
the $\CN=(2,0)$ theories compactified on a circle.\footnote{ 
5d SYM on a curved five dimensional manifold is also considered in the context of localization calculations in e.g.,
\cite{Kallen:2012cs,Hosomichi:2012ek,Kawano:2012up,Kallen:2012va,Kim:2012ava,Terashima:2012ra,Kallen:2012zn,Imamura:2012xg,
Kim:2012tr,Fukuda:2012jr,Imamura:2012bm,Kim:2012qf,Minahan:2013jwa,Yagi:2013fda,Lee:2013ida,Cordova:2013bea,Cordova:2013cea,Kim:2013nva}
and relations to the proposal of \cite{Douglas:2010iu,Lambert:2010iw} are discussed.
Our work has a relation to \cite{Fukuda:2012jr} if we compactify ${\mathbb R}^{1,2}$ to $S^3$.
It would be interesting to investigate this direction further.} 
If this is the case, twisted 5d SYM on ${\mathbb R}^{1,2}  \times C$ should be somehow equivalent to 
the class~$\CS$ theories on ${\mathbb R}^{1,2} \times S^1$
in the limit in which the area $\CA$ of the Riemann surface goes to zero.
(See \cite{Gaiotto:2011xs} for finite area effects.)
The limit $\CA \to 0$ is rather singular; 
the effective coupling constant of 5d SYM, which is normalized to be dimensionless by using $\CA$, goes to infinity at the scale of compactification on $C$.
Then UV divergences of 5d SYM~\cite{Bern:2012di} may become a serious problem.
However, at least some protected quantities such as holomorphic quantities have trivial dependence on the gauge coupling and $\CA$,
and hence we expect these quantities can be computed by using twisted 5d SYM.
In this sense, we might regard twisted 5d SYM as a ``Lagrangian'' of class $\CS$ theories which often have no Lagrangian description
in terms of four dimensional fields.

As mentioned above, Hitchin's equations give solutions to the moduli space of Coulomb branches of $\CN=2$ theories.
We will see in section~\ref{sec:5} that twisted 5d SYM also reproduces moduli spaces of Higgs and mixed Higgs-Coulomb branches.
Twisted 5d SYM was used also in \cite{Benini:2010uu,Gaiotto:2011xs} for related purposes.
In the $\CN=1$ case, generalized Hitchin's equations and its spectral curve play the important role. Different approaches
were discussed in \cite{Xie:2013gma,Xie:2013rsa} and in \cite{Bonelli:2013pva}, 
and in this paper we always follow the approach of \cite{Xie:2013gma,Xie:2013rsa}.
The field theory dynamics of $\CN=1$ class~$\CS$ theories or similar theories are discussed in 
\cite{Maruyoshi:2009uk, Benini:2009mz, Tachikawa:2011ea, Bah:2011je, Bah:2011vv, Bah:2012dg, Beem:2012yn, Gadde:2013fma,
Bah:2013qya, Maruyoshi:2013hja, Bah:2013aha, Maruyoshi:2013ega}. 
As demonstrated explicitly in \cite{Xie:2013rsa}, moduli spaces of solutions of generalized Hitchin's equations
almost reproduce moduli spaces of those field theories. Based on twisted 5d SYM, we will make this description of moduli spaces more complete.
We also derive a formula for dynamically generated superpotential vacuum expectation value (vev) 
which is relevant for BPS tensions of domain walls~\cite{Dvali:1996xe}.
The formula simplifies the one obtained by Witten~\cite{Witten:1997ep} from M-theory.
Our construction does not directly rely on M-theory; we only need the 6d $\CN=(2,0)$ theories,
and hence any gauge group $G$ of ADE type is possible.

\paragraph{Summary of results.}
Let us summarize the main results obtained in this paper. First, the 5d space-time of the twisted 5d SYM is given by ${\mathbb R}^{1,2} \times C$
and the coordinates of ${\mathbb R}^{1,2}$ and $C$ are denoted as $x$ and $z$, respectively. The supersymmetry of the twisted 5d SYM
is realized as 3d $\CN=2$ supersymmetry by regarding $x$ as space-time coordinates and $z$ as ``internal'' coordinate.
The chiral fields of the twisted 5d SYM are given by $A_{\bar{z}}(x,z,\theta)$ and $\Phi_i(x,z,\theta)~(i=1,2)$, where
$\theta$ is the superspace coordinates of 3d $\CN=2$ supersymmetry. 
The field $A_{\bar{z}}$ takes values in the
canonical bundle $K=T^* C$ of the Riemann surface $C$, and $\Phi_i$ takes values in a rank two bundle $F$ over the Riemann surface
which is constrained as $\det F=K$. We will mainly focus on the case~\cite{Bah:2011vv} $F=L_1 \oplus L_2$ 
for two line bundles $L_1$ and $L_2$ with $L_1 \otimes L_2=K$. From the 3d point of view in which $z$ is regarded as an internal index, 
the gauge group ${\mathcal G}$ is the group of maps $C \to G$, where $G$ is a usual ADE type group.
The gauge transformation is given by $A_{\bar{z}} \to g^{-1}A_{\bar{z}} g+g^{-1} \partial_{\bar{z}}g$ and $\Phi_i \to g^{-1}\Phi_i g$ 
for $g \in {\mathcal G}$.
There is also the vector multiplet $V(x,z,\theta)$ of the above gauge group.

When there are no punctures, the Kahler potential and superpotential from the 3d point of view are given by
\beq
K&= -\frac{2}{g_5^2}\int \sqrt{g} |d^2z| \Tr \left(g^{z\bar{z}} A_{\bar{z}} A_{z} + h^{i{\bar j}}  \Phi_i \bar{\Phi}_{\bar{j}} \right),\\
W&= -\frac{\sqrt{2}i}{g_5^2} \int dz \wedge d{\bar z}\  \epsilon^{ij} \Tr \left( \Phi_i D_{\bar z} \Phi_j \right), 
\eeq
where $g^{z\bar{z}}$ and $h^{i \bar{j}}$ are the metrics of $K$ and $F$ respectively, $|d^2z|=i dz \wedge d \bar{z}$,
$g_5$ is the 5d gauge coupling, $\epsilon^{ij}$ is the totally antisymmetric tensor with $\epsilon^{12}=1$,
$\Tr$ is a negative-definite inner product on the space of Lie algebra,
and $D_{\bar z} \Phi_j =\partial_{\bar z} \Phi_j +[A_{\bar{z}}, \Phi_j] $.
There are also couplings between the chiral fields and the vector field $V$ determined by the above gauge transformation and the Kahler potential. 

The conditions for supersymmetric vacua are obtained from the above Kahler potential and superpotential as
\beq
& g^{z\bar{z}}F_{z\bar{z}} + h^{i\bar{j}}[\Phi_{i}, {\Phi}_{\bar{j}} ^\dagger] =0 ,~~~
 D_{\bar z} {\Phi}_i =0,  ~~~
\epsilon^{ij} [{\Phi_i}, {\Phi}_j] =0,  \\
&D_{\bar{z}} \sigma =0, ~~~
[\sigma,\Phi_i]=0 ,
\eeq
where the $\sigma$ is the real adjoint scalar field contained in the vector multiplet $V$.
The first line is generalized Hitchin's equations, while the second line is also required by the twisted 5d SYM.
When the gauge group $G$ is broken down to $H$ by a solution of generalized Hitchin's equations,
we can turn on $\sigma$ and dual photons in the Cartan subalgebra of $H$. In particular, if $H$ is trivial,
we get $\sigma=0$ and generalized Hitchin's equations capture the moduli space completely.
When $\sigma$ is nonzero, the moduli fields coming from $\sigma$ 
correspond to the moduli coming from the chiral operators~\cite{Gaiotto:2009gz} $Q_{i_1 i_2 i_3}$ of the $T_N$ theory and their relatives.

Punctures are interpreted~\cite{Chacaltana:2012zy} as 
$3d$ $\CN=4$ superconformal theories $T_\rho[G]$ introduced by Gaiotto-Witten~\cite{Gaiotto:2008ak}
coupled to the twisted 5d SYM. 
Let us focus on the case $\rho=0$ for simplicity. (See section~\ref{sec:3} for general $\rho$.)
The $T[G]$ theory has the Higgs branch global symmetry $G$ and Coulomb branch global symmetry $G^\vee$,
where for the ADE type groups, $G^\vee$ is the same as $G$. 
The corresponding holomorphic moment maps are denoted as 
$\mu_H^{(3d)}$ for the Higgs branch and $\mu_C^{(3d)}$ for the Coulomb branch. 
The superpotential coupling to the twisted 5d SYM is given by
\beq
W
 \supset -\sqrt{2}  \left( \sum_{a \in A} \Tr \left(\mu_{a,H}^{(3d)} \Phi_1(z_a) \right)-\sum_{b \in B} \Tr \left(\mu_{b,H}^{(3d)} \Phi_2(z_b) \right)
\right), 
\eeq
where $z_a~(a \in A)$ and $z_b~(b \in B)$ are positions of punctures on the Riemann surface, and
$\mu_{a,H}^{(3d)} $ and $\mu_{b,H}^{(3d)} $ are the Higgs branch moment maps of $T[G]$ living at the respective punctures.
Generalized Hitchin's equations with these source terms tell us that regular singularities of the Higgs fields $\Phi_{1,2}$ at the punctures are given as
\beq
\Phi_2 \to  \frac{c_1 \mu_{a,H}^{(3d)} }{z-z_a}~~~(z \to z_a),  ~~~~~~~
\Phi_1 \to  \frac{c_1 \mu_{b,H}^{(3d)} }{z-z_b}~~~(z \to z_b),
\eeq
where $c_1=-g_5^2/4\pi$ is a constant.
By adding a mass term $\tr (m \mu_C^{(3d)})$, the vev of $\mu_{H}^{(3d)} $ is given as $\mu_{H}^{(3d)} \propto m$.
Therefore, the above formulas reproduce the usual regular singularities. 
For irregular singularities, we just borrow the results from the literature, e.g.,~\cite{Gaiotto:2009hg,Nanopoulos:2010zb,Cecotti:2011rv,Bonelli:2011aa,
Xie:2012hs,Gaiotto:2012rg,Kanno:2013vi}.

Let us focus on the case $G=\SU(N)$ for concreteness. (See section~\ref{sec:generalgroup} for general $G$.)
Field theory information is extracted from the twisted 5d SYM in the following way.
\begin{enumerate}
\item Flavor symmetries of field theory are identified as the Coulomb branch $G^\vee=\SU(N)$ symmetries of copies of $T[\SU(N)]$ at the punctures,
and hence the moment maps of these symmetries are given by $\mu_{a,C}^{(3d)}$ and $\mu_{b,C}^{(3d)}$. 
The vevs of these operators can be extracted from solutions for the Higgs fields $\Phi_{1,2}$.
At the puncture $z \to z_a$, we have
\beq
\det (x-c_2 \Phi_1) \to \det (x -\mu_{a,C}^{(3d)}),~~~~\Phi_2 \to \frac{c_3m_a}{z-z_a}~~~~~(z \to z_a)
\eeq
where $c_2 =\sqrt{2}/4\pi$ and $c_3=g_5^2/16\pi^2$ are constants. At $z \to z_b$, similar equations hold with $\Phi_1 \leftrightarrow \Phi_2$.
The first equation means that the characteristic polynomial of the matrix $\mu_{a,C}^{(3d)}$ 
is equal to the characteristic polynomial of the non-singular Higgs field at $z_a$.
From these formulas and solutions of generalized Hitchin's equations, we can obtain chiral ring relations such as
deformed moduli constraints in Higgs phase.

\item The vev of superpotential for a given solution of the generalized Hitchin's equations is given as
\beq
W|_{\rm solution}& =c_4 \sum_{a \in A} \oint_{|z-z_b|=\epsilon} dz \Tr (\Phi_1 \Phi_2) %\nonumber \\
=-c_4\sum_{b \in B} \oint_{|z-z_a|=\epsilon} dz \Tr (\Phi_1 \Phi_2),
\eeq
where $c_4=-2i\sqrt{2}/g_5^2$ is a constant and $\epsilon$ is an infinitesimal number.
The two expressions above are equivalent by Cauchy's theorem. This superpotential vev gives BPS tensions of strongly coupled domain walls
in confining phase.

\item The spectral curve is defined by
\beq
0= \frac{1}{N!}(x_{i_1} -\Phi_{i_1})^{\alpha_1}_{~\beta_1} \cdots (x_{i_N} -\Phi_{i_N})^{\alpha_N}_{~\beta_N}
\epsilon_{\alpha_1\cdots \alpha_N} \epsilon^{\beta_1 \cdots \beta_N}.
\eeq
where $i_k=1,2$ and $(x_1,x_2)$ are the coordinates of the fiber of the rank two bundle $F (=L_1 \oplus L_2)$.
This equation defines a curve on the total space of $F$.
It contains informations of moduli space of field theory as in \cite{Xie:2013rsa}.\footnote{
The spectral curve here might look different from the one in \cite{Xie:2013rsa}, but they are essentially the same.}
This equation also gives the Seiberg-Witten curve~\cite{Seiberg:1994rs,Seiberg:1994aj} 
which determines the holomorphic gauge coupling of low energy massless $\U(1)$ fields in $\CN=1$ Coulomb phase~\cite{Intriligator:1994sm}. 

\end{enumerate}

For explicit examples, see section~\ref{sec:5} for $\CN=2$ theories and section~\ref{sec:6} for $\CN=1$ theories.
The above equations are written in the normalization which is natural in 5d SYM and 3d field theory.
There is more natural normalization from the point of view of 6d $(2,0)$ theory and 4d field theory which is discussed 
in sections~\ref{sec:2} and \ref{sec:3} and used in section~\ref{sec:6}.

The rest of this paper is organized as follows. In section~\ref{sec:2}, we perform the twisting explicitly and derive the twisted 5d SYM.
We also discuss generalized Hitchin's equations and spectral curve. In section~\ref{sec:3}, we introduce punctures.
Formulas for the vevs of mesons or moment maps, and also the superpotential, are derived there. In section~\ref{sec:4},
we review the compactification of effective action of 4d $\CN=1$ theories on $S^1$ similar to the discussion in \cite{Seiberg:1996nz}.
We argue some nonrenormalization theorems which claim that vevs of holomorphic quantities are independent of the radius of $S^1$.
In section~\ref{sec:5}, Higgs branches of $\CN=2$ generalized quiver gauge theories are studied. 
We will see the meaning of $\sigma$ in the vector multiplet of the twisted 5d SYM.
In section~\ref{sec:6}, a lot of $\CN=1$ examples are examined using our new tools.
Appendix~\ref{app:A} reviews $\CN=1$ class $\CS$ theories relevant for this paper. % and $\CN=1$ dualities.
Appendix~\ref{app:B} gives explicit computations of spectral curves.

%%%%%%%%%%%%%%%%%%%%%%%%%%%%%%%%%%%%%%%%%%%%%%%%%%%%%%%%%%%
\section{Twisted 5d Super-Yang-Mills and generalized Hitchin systems}\label{sec:2}

The class of 4d theories we study in this paper is obtained from 
the 6d $\CN=(2,0)$ theories compactified on a punctured Riemann surface $C=C_{g,n}$,
where $g$ is the genus and $n$ is the number of punctures. As discussed in the introduction, let us further compactify the theory 
on $S^1$ with radius $R$.
Now the $(2,0)$ theories are on the space ${\mathbb R}^{1,2} \times S^1 \times C$.
Suppose that our purpose is to study field theory quantities whose dependence on $R$ are well-controlled, 
such as holomorphic quantities as we discuss in detail in section~\ref{sec:4}. 
In this case, we may consider a very small $R$ limit such that $R$ is much smaller than the length scale of $C$.
Then we change the order of compactification; we first compactify the $(2,0)$ theories on $S^1$ and then on $C$.
The compactification of the $(2,0)$ theories on $S^1$ gives 5d maximal SYM. 
Therefore, we are left with the 5d SYM compactified on $C$, which can be described explicitly by Lagrangian.
The change of the order of compactification induces~\cite{Kapustin:1998xn} a mirror symmetry~\cite{Intriligator:1996ex} from the 3d point of view.
Thus, in the case of 8 superchages,
hypermultiplets of twisted 5d SYM are related to vector multiplets of 4d field theory on $S^1$, and vice versa. 

\subsection{SUSY transformations in twisted 5d SYM}
To preserve some of the supersymmetry, the compactification must be done with twisting~\cite{Witten:1988ze}.
We perform a twisted compactification of 5d SYM on ${\mathbb R}^{1,2} \times C$
which preserves at least 4 superchages (i.e. 4d $\CN=1$ or 3d $\CN=2$),
and give explicit supersymmetry transformations. The final result will be very simply summarized in terms of
the superpotential \eqref{eq:infsuper} and the Kahler potential \eqref{eq:infkahler},
so the reader who is only interested in the result can skip this subsection and go to subsection~\ref{sec:kahlersuper}.
The calculation here will be somewhat similar to that of the twisting in \cite{Kapustin:2006pk}.

\paragraph{5d SYM from 10d SYM.}

The 5d maximal SYM can be obtained from dimensional reduction of the 10d maximal SYM.
Our notation and conventions are as follows. 
We use the metric signature $-+ \cdots +$.
We use indices $I, J, \cdots$ for 10 dimensions, i.e. $I=0,1,\cdots,9$,
and use $\mu, \nu, \cdots$ for $0,1,2,3$. 
For later convenience, we also define
\beq
z =\frac{ x^4+ix^5}{\sqrt{2}},~~~u^1= \frac{x^6+ix^7}{\sqrt{2}},~~~
u^2=\frac{x^8+ i x^9}{\sqrt{2}}. \label{eq:complexcoordinate}
\eeq
In these complex coordinates, upper and lower indices on tensors are related as, e.g., $A^z=A_{\bar{z}}$ etc.

The ten dimensional gamma matrices are denoted as $\Gamma_I$. We also define $\Gamma_{I_1 \cdots I_n}=\Gamma_{[I_1} \cdots \Gamma_{I_n]}$,
where $[,~]$ represents anti-symmetrization.
The charge conjugation matrix in 10d is denoted as $C$, which satisfies 
\beq
(\Gamma_I)^T=-C\Gamma_I C^\dagger,~~~C^T=-C,
\eeq
where the superscript $T$ represents transpose. 
The Majorana-Weyl conditions for a spinor $\epsilon$ are given as
\beq
(\Gamma_{01\cdots 9}) \epsilon &=\epsilon, \\
\epsilon^\dagger \Gamma_0 &=\epsilon^T C. \label{eq:realitycondition}
\eeq
If we use a basis in which all the gamma matrices are real, we might take $C=\Gamma_0$ and then $\epsilon$ is  real.
But we will not assume this specific basis.

Lie group generators are taken to be anti-Hermitian. For example, the field strength tensor is $F=dA+A \wedge A$ or
$F_{IJ}=\partial_I A_J- \partial_J A_I+[A_I, A_J]$. The simbol $\Tr$ will denote a negative definite inner product on Lie algebras.
For $\SU(N)$, it is just the trace in the fundamental representation. For a simply laced group $G$, if we take an $\SU(2)$ subalgebra,
then it coincides with the trace in the fundamental representation of $\SU(2)$.

The action of the 10d SYM is given by
\beq
I_{10}=\frac{1}{e^2_{10}} \int d^{10} x \Tr \left( \frac{1}{2}F_{IJ}F^{IJ} - i \lambda^T C \Gamma^I D_I \lambda \right)
\eeq
where $\lambda$ is the Majorana-Weyl gaugino, and $e_{10}$ is the ten dimensional gauge coupling.
This action is invariant under the following supersymmetry transformation;
\beq
\delta A_I = i\epsilon^T C \Gamma_I \lambda ,~~~~~~~
\delta \lambda = \frac{1}{2} \Gamma^{IJ} F_{IJ} \epsilon.
\eeq

The 5d SYM can be obtained by simply taking all the fields to be independent of five coordinates.
We take $x^0,x^1,x^2, x^4,x^5$ as the space-time directions of 5d SYM, and then all the fields are independent of the remaining directions
$x^3, x^6,x^7,x^8,x^9$. The motivation for this unusual convention is as follows. When we relate 5d SYM on ${\mathbb R}^{1,2} \times  C$
to the 6d $\CN=(2,0)$ theories on ${\mathbb R}^{1,2} \times S^1 \times C$, we want to take the $S^1$ direction as $x^3$.
Then, in the large radius limit of $S^1$, the $x^0, x^1, x^2, x^3$ become the coordinates of four dimensional flat Minkowski space 
${\mathbb R}^{1,3}$.
We also note that the coordinate $z=(x^4+ix^5)/\sqrt{2}$ defined in \eqref{eq:complexcoordinate} will be the holomorphic coordinate of 
the Riemann surface $C$.

\paragraph{Twisting.}
The 5d SYM has the Lorentz and R-symmetry groups $\SO(1,4) \times \SO(5)$ which is the subgroup of the 10d Lorentz group $\SO(1,9)$.
We consider the subgroup $\SO(1,2) \times \U(1)_z \times \U(2)_u \subset\SO(1,4) \times \SO(5) $,
where $\U(1)_z$ and $\U(2)_u=\U(1)_u \times \SU(2)_u$ are the rotation groups of $z$ and $(u^1, u^2)$ 
defined in \eqref{eq:complexcoordinate} respectively,
and $\SO(1,2)$ is the Lorentz group of $x^0, x^1, x^2$. The 16 dimensional Majorana-Weyl spin representation of $\SO(1,9)$ is decomposed as
\beq
{\bf 16} \to \sum_{\pm} \left[ ({\bf 2}, {\bf 1})^{\pm\frac{1}{2}, +1} \oplus ({\bf 2}, {\bf 2})^{\pm\frac{1}{2}, 0} \oplus ({\bf 2}, {\bf 1})^{\pm\frac{1}{2}, -1}      \right]
\label{eq:decomposition16}
\eeq
where, for example, $({\bf 2}, {\bf 1})^{\pm\frac{1}{2}, +1}$ means that it transforms as ${\bf 2}$ under $\SO(2,1)$, 
as ${\bf 1}$ under $\SU(2)_u \subset \U(2)_u$, and has charges $\pm \frac{1}{2}$ and $+1$ under $\U(1)_z$ and $\U(1)_u \subset \U(2)_u$, respectively.

By compactifying the 5d SYM on $C$, the $\U(1)_z$ becomes the 
rotation group of the tangent bundle of $C$. To preserve some of the supersymmetries, we introduce background gauge field
$\omega_u$ coupled to the R-symmetry $\U(2)_u$. This connection $\omega_u$ defines a rank two vector bundle which we denote as $F$.
The connections are supposed to satisfy the relation
\beq
\omega+\tr \omega_u=0, \label{eq:vanishingconnection}
\eeq
where $\omega$ is the connection of the tangent bundle  $TC$, and $\tr \omega_u$ is the connection on the determinant bundle $\det F$
of the rank two bundle $F$. This condition means that $\det F=K$, where $K=T^*C$ is the canonical (cotangent) bundle of $C$.
Then, the spin bundle in the representations
\beq
{\bf 2}_\ell \equiv    ({\bf 2}, {\bf 1})^{+\frac{1}{2}, +1} ,~~~~ {\bf 2}_r \equiv ({\bf 2}, {\bf 1})^{-\frac{1}{2}, -1} 
\eeq
becomes a trivial bundle and we can preserve the supersymmetries corresponding to these representations.

We denote the supersymmetry transformation parameters in the representations ${\bf 2}_\ell$ and ${\bf 2}_r$ as 
$\epsilon_\ell$ and $\epsilon_r$, respectively.
They satisfy, e.g., 
\beq
& \Gamma_{z\bar{z}}\epsilon_\ell= \Gamma_{u^1 \bar{u}^1}\epsilon_\ell= \Gamma_{u^2\bar{u}^2}\epsilon_\ell=\epsilon_\ell , 
~~~~~~~~ \Gamma_z \epsilon_\ell=\Gamma_u \epsilon_\ell=0     , \label{eq:useful1} \\
& \Gamma_{z\bar{z}}\epsilon_r= \Gamma_{u^1\bar{u}^1}\epsilon_r= \Gamma_{u^2\bar{u}^2}\epsilon_r=-\epsilon_r   , 
~~~~~\Gamma_{\bar{z}} \epsilon_r=\Gamma_{\bar{u}} \epsilon_r=0  .    \label{eq:useful2} 
\eeq

\paragraph{SUSY transformation.}

For convenience, we define the ``holomorphic three form'' $\Omega$. Let $z^i$ be complex coordinates, 
\beq
z^1=u^1,~~~~z^2=u^2,~~~~z^3=z.
\eeq
Then, we define
\beq
\Omega=\frac{1}{3!} \Omega_{ijk} dz^i \wedge dz^j \wedge dx^k= du^1 \wedge du^2 \wedge dz.
\eeq
In components, $\Omega_{123}=1$ etc. Note that if indices are raised, $\Omega^{\bar{1}\bar{2}\bar{3}}=1$ etc.
The $\bar{\Omega}$ is defined as the complex conjugate of $\Omega$.
The following calculation may be easier to understand if we consider the subgroup $\SO(1,3) \times \SO(6) \subset \SO(1,9)$
which is not a true symmetry group of the 5d SYM but is technically convenient.

The gaugino $\lambda $ transforms under the representation \eqref{eq:decomposition16}.
It is convenient to parametrize $\lambda$ as 
\beq
\lambda &=  -\lambda_{\ell}-\lambda_{r} + \frac{1}{4} \Omega^{\bar{i}\bar{j}\bar{k}}  \Gamma_{\bar{i}\bar{j}} \psi_{\ell, \bar{k}}  
  +\frac{1}{4} \bar{\Omega}^{{i}{j}{k}}  \Gamma_{{i}{j}} \psi_{r, {k}}. \label{eq:gauginodecompose}
\eeq
Here, $\lambda_\ell$ and $\psi_{\ell, \bar{i}}$ transform in the representation ${\bf 2}_\ell$, and $\lambda_r$ and $\psi_{r,i}$ transform in
the representation ${\bf 2}_r$.
The coefficients are chosen to agree with the conventions of Wess and Bagger~\cite{Wess:1992cp} later.
The reality condition \eqref{eq:realitycondition} gives 
\beq
\lambda_\ell^\dagger \Gamma_0 &=\lambda_r^T C,~~~~\psi_{\ell, \bar{i}}^\dagger \Gamma_0=\psi_{r,i}^T C, \label{eq:realitycondition2}
\eeq
and equations with $\ell \leftrightarrow r$. In this sense, ${\bf 2}_\ell$ and ${\bf 2}_r$ are complex conjugates of each other.

Let us study the supersymmetry transformation under $\epsilon_\ell$, which we denote as $\delta_\ell$.
The variation of the gaugino $\lambda$ is given as
\beq
\delta_\ell \lambda 
&=\left( \frac{1}{2}F_{\mu\nu} \Gamma^{\mu\nu} +F_{i \bar{i}} \Gamma^{i\bar{i}} +\frac{1}{2}F_{ij} \Gamma^{ij}+F_{\mu i}\Gamma^{\mu i}  \right)\epsilon_\ell 
\nonumber \\
&=\left( \frac{1}{2}F_{\mu\nu} \Gamma^{\mu\nu} -F_{i \bar{i}} +\frac{1}{2}F_{ij} \Gamma_{\bar{i}\bar{j}}
-\frac{1}{8}\bar{\Omega}^{ijk}F_{\mu i}\Gamma_{jk} \Gamma^\mu \Gamma_{\bar{1}\bar{2}\bar{3}}  \right)\epsilon_\ell ,
\eeq
where we have used \eqref{eq:useful1}.
Comparing this with \eqref{eq:gauginodecompose}, we get
\beq
\delta_\ell \lambda_\ell &=\left( -\frac{1}{2}F_{\mu\nu} \Gamma^{\mu\nu} +F_{i \bar{i}}   \right) \epsilon_\ell, \\
\delta_\ell \lambda_r &=0, \\
\delta_\ell \psi_{\ell, \bar{i}} &=\bar{\Omega}_{\bar{i}\bar{j}\bar{k}} F_{jk} \epsilon_\ell, \\
\delta_\ell \psi_{r, i} &= -\frac{1}{2} F_{\mu i}  \Gamma^\mu \Gamma_{\bar{1}\bar{2}\bar{3}}  \epsilon_\ell.
\eeq
On the other hand, the variation of the gauge field is given as
\beq
\delta_\ell A_I  
&= i \epsilon_\ell^TC \Gamma_I \left( -\lambda_{r} +\frac{1}{4} \Omega^{\bar{i}\bar{j}\bar{k}}  \Gamma_{\bar{i}\bar{j}} \psi_{\ell, \bar{k}}  \right),
\eeq
where we have again used \eqref{eq:useful1}, and we have also used the fact that $\lambda_\ell$ and $\epsilon_\ell$ transform as $({\bf 2}, {\bf 1})$
under $\SO(3,1) \times \SU(3) \subset \SO(9,1)$ which rotate $(x^0,x^1,x^2, x^3)$ and $(z^1,z^2,z^3)$,
and hence $\epsilon_\ell^TC \Gamma_\mu  \lambda_\ell=\epsilon_\ell^TC \Gamma_{\bar i}  \lambda_\ell
=\epsilon_\ell^TC \Gamma_{i}  \lambda_\ell=0$.
From these facts, we obtain
\beq
\delta_\ell A_\mu &= -  i \epsilon_\ell^TC \Gamma_\mu \lambda_r, \\
\delta_\ell A_{\bar{i}} &= \frac{i}{2} \epsilon_\ell^TC \Gamma_{\bar{1}\bar{2}\bar{3}} \psi_{\ell, \bar{i}}, \\
\delta_\ell A_{{i}}  &=0.
\eeq

The supersymmetry transformations above  can be rewritten in the form of the transformations of
3d ${\cal N}=2$ supersymmetry, which is obtained from 4d ${\cal N}=1$ supersymmetry by dimensional reduction on the $x^3$ direction.
For this purpose, we need some definitions. First, notice that on the space of ${\bf 2}_\ell \oplus {\bf 2}_r$, we have
\beq
(\Gamma_{123}+\Gamma_{\bar{1}\bar{2}\bar{3}})^2 \epsilon_{\ell, r} =-8\epsilon_{\ell,r}.
\eeq
We represent this equation by writing
\beq
(\Gamma_{123}+\Gamma_{\bar{1}\bar{2}\bar{3}})^2  \cong -8.
\eeq
With this in mind, we define
\beq
\gamma_\mu =\frac{1}{2\sqrt{2}} (\Gamma_{123}+\Gamma_{\bar{1}\bar{2}\bar{3}}) \Gamma_\mu, ~~~~~~
C_4=\frac{i}{2\sqrt{2}} C (\Gamma_{123}+\Gamma_{\bar{1}\bar{2}\bar{3}}) .
\eeq
They are interpreted as the four dimensional gamma matrices and the charge conjugation matrix, respectively.
For example, they satisfy
\beq
&\{ \gamma_\mu, \gamma_\nu \} \cong 2g_{\mu\nu},~~~~~\gamma_{\mu\nu} \cong \Gamma_{\mu\nu}, \\
&(\gamma_I)^T \cong -C_4\gamma_I C_4^\dagger,~~~C^T=-C.
\eeq
The reality condition \eqref{eq:realitycondition2} is now given as
\beq
\lambda_\ell^\dagger (-i\gamma_0) &=\lambda_r^T C_4,~~~~\psi_{\ell, \bar{i}}^\dagger (-i\gamma_0)=\psi_{r,i}^T C_4. \label{eq:realitycondition3}
\eeq

Next, we introduce auxiliary fields $D$ and $F_{\bar{i}}$. 
We set
\beq
D&=-iF_{i\bar{i}},  \label{eq:Dterm}\\
F_{\bar{i}}&=\frac{1}{\sqrt{2}}\bar{\Omega}_{\bar{i}\bar{j}\bar{k}} F_{jk}. \label{eq:Fterm}
\eeq

Now we can rewrite the supersymmetry transformations.
We can interpret $(A_\mu, \lambda_\ell, \lambda_r, D)$ as a vector multiplet. Their transformation law is
\beq
\delta_\ell A_\mu &=  \epsilon_\ell^TC_4 \gamma_\mu \lambda_r, \\
\delta_\ell \lambda_\ell &=\left( -\frac{1}{2}F_{\mu\nu} \gamma^{\mu\nu} + iD \right) \epsilon_\ell, \\
\delta_\ell \lambda_r &=0. 
\eeq
The multiplets $(A_{\bar{i}}, \psi_{\ell, {\bar i}}, F_{\bar{i}})$ can be interpreted as chiral multiplets.
Their transformation law is 
\beq
\delta_\ell A_{\bar{i}} &= \sqrt{2} \epsilon_\ell^TC_4 \psi_{\ell, \bar{i}}, \\
\delta_\ell A_{{i}}  &=0, \\
\delta_\ell \psi_{\ell, \bar{i}} &=\sqrt{2} F_{\bar i} \epsilon_\ell, \\
\delta_\ell  \psi_{r, i}  &= \sqrt{2} D_\mu A_{ i}  \gamma^\mu  \epsilon_\ell ,
\eeq
where we have written $F_{\mu i}=D_\mu A_{i}$.
These are precisely the transformation laws of 3d ${\cal N}=2$ (or dimensionally reduced 4d ${\cal N}=1$) supersymmetry.\footnote{
See page 50 of Wess and Bagger~\cite{Wess:1992cp}. To make comparison, we identify $\lambda_\ell \to \lambda_\alpha$,
$\lambda_r \to \bar{\lambda}^{\dot{\alpha}}$, $\lambda_\ell^TC_4 \to \lambda^\alpha$, $\lambda_r^TC_4 \to \bar{\lambda}_{\dot{\alpha}}$.
The gamma matrices should be given as
\beq
\gamma_\mu=i\left(
\begin{array}{cc}
0 & \sigma_\mu \\
\bar{\sigma}_\mu & 0
\end{array}
\right).
\eeq
We also need to change the orientation so that $\ell \leftrightarrow r$.}

The above calculation has been done very similarly to a compactification of the 10d SYM on a Calabi-Yau threefold.
But in our case, $A_{\bar{u}^1}$ and $A_{\bar{u}^2}$ are not gauge fields, but scalar fields in the adjoint representation.
Due to the background field $\omega_u$ coupled to the R-symmetry $\U(2)_u$,
these adjoint fields take values in the rank 2 bundle $F$,
on which $\omega_u$ is a connection.
Let us rename the chiral multiplets $A_{\bar{u}^1}$ and $A_{\bar{u}^2}$ as
\beq
&\Phi_1=A_{\bar{u}^1},~~~\psi_1=\psi_{\ell, \bar{u}^1},~~~F_1=F_{\bar{u}^1},  \nonumber \\
&\Phi_2=A_{\bar{u}^2},~~~\psi_2=\psi_{\ell,\bar{u}^2},~~~F_2=F_{\bar{u}^2}.
\eeq
The anti-chiral multiplets are denoted as $\bar{\Phi}_{1,2},~\bar{\psi}_{1,2},~\bar{F}_{1,2}$.
Because we are taking gauge group generators $T_a$ to be anti-hermitian, 
${\bar \Phi}$ is defined as ${\bar \Phi}=T_a (\Phi^a)^*=-\Phi^\dagger$.
The chiral multiplet which includes $A_{\bar{z}}$ and $\psi_{\ell,\bar{z}}$ will be denoted by its lowest component $A_{\bar{z}}$, and the vector multiplet
which includes $ \lambda_\ell, \lambda_r$ and $A_\mu$ will be denoted by $V$.

The above analysis is valid for an orthogonal basis of the bundles $K$ and $F$. However, it is more convenient to use a holomorphic basis.
Let $g_{z\bar{z}}$ be the metric of $C$, and  $h^{i\bar{j}}~(i, \bar{j}=1,2)$
be the metric of $F$.
Then, \eqref{eq:Dterm} and \eqref{eq:Fterm} are given more precisely as
\beq
D&=-i \left( g^{z\bar{z}}F_{z\bar{z}} -h^{i {\bar j}}[\Phi_i, \bar{\Phi}_{\bar{j}}]   \right),  \label{eq:Dtermagain}\\
F_{\bar{z}}&=\frac{1}{\sqrt{2}} [\bar{\Phi}_{\bar{i}}, \bar{\Phi}_{\bar{j}}] \epsilon^{\bar{i}\bar{j}} , \label{eq:Ftermagain1} \\
g_{z\bar{z}}h^{i\bar{j}}F_{i} &=-\sqrt{2}  \epsilon^{\bar{j}\bar{k}} D_z \bar{\Phi}_{\bar{k}}, \label{eq:Ftermagain2} ,
\eeq
where $\epsilon^{ij}$ is the totally anti-symmetric tensor with $\epsilon^{12}=1$. The metrics are introduced so that the equations transform covariantly.

\subsection{Kahler and superpotential}\label{sec:kahlersuper}
The result of the previous subsection is summarized in the following way. From the three dimensional point of view,
we have chiral multiplets $\Phi_1(x,z,\theta),\Phi_2(x,z,\theta)$ and $A_{\bar{z}}(x,z,\theta)$ and a vector multiplet $V(x,z,\theta)$,
where $\theta$ is the superspace coordinate.
The $x$ is the space-time coordinates, while $z$ is regarded as an ``internal coordinate'' from the point of view of the 3d $\CN=2$ supersymmetry.
The fields $\Phi_i~(i=1,2)$ take values in the bundle $F \otimes {\rm ad}(E)$, where ${\rm ad}(E)$ is the vector bundle associated to the gauge group
in the adjoint representation, and $F$ is a rank two holomorphic vector bundle such that its determinant bundle is the same as the canonical bundle,
$\det F=K=T^* C$. This $F$ was introduced in the twisting.

The $D$-term and the $F$-terms of the previous subsection are reproduced by the Kahler potential and superpotential given by
\beq
K_{3d}&= -\frac{2}{g_5^2}\int \sqrt{g} |d^2z| \Tr \left(g^{z\bar{z}} A_{\bar{z}} A_{z} + h^{i{\bar j}}  \Phi_i \bar{\Phi}_{\bar{j}} \right), \label{eq:infkahler} \\
W_{3d}&=- \frac{2}{g_5^2} \int |d^2z| \frac{1}{\sqrt{2}} \epsilon^{ij} \Tr \left( \Phi_i D_{\bar z} \Phi_j \right), \label{eq:infsuper}
\eeq
where $|d^2z| = idz \wedge d\bar{z}$, $g_{z\bar{z}}$ is the metric of $C$, $h^{i\bar{j}}$ is the metric of the bundle $F$, 
$D_{\bar{z}}\Phi_i=\partial_{\bar{z}}\Phi_i+[A_{\bar{z}},\Phi_i]$, 
$\epsilon^{ij}$ is the totally antisymmetric tensor with $\epsilon^{12}=1$, $\sqrt{g}=g_{z\bar{z}}$, and $g_5$ is the 5d gauge coupling.
From them, the equations for the $F$-terms \eqref{eq:Ftermagain1} and \eqref{eq:Ftermagain2} are derived straightforwardly.
Note that the Kahler potential explicitly depends on the metrics, but the superpotential is independent of them, thanks to the
condition $\det F=K$.

The $D$-term \eqref{eq:Dtermagain} can be obtained in the following way. (See \cite{Wess:1992cp} for an explanation of how to couple
general Kahler potentials to general gauge groups.)
From the three dimensional point of view, we can interpret the theory as a 3d gauge theory whose
gauge group ${\cal G}$ is the group of maps $C \to G$. This is an infinite dimensional gauge group.
The fields $A_{\bar z}$ and $\Phi_i$ take values in 
an infinite dimensional space ${\cal W}$.
The transformation of them under ${\cal G}$ is given as
\beq
\delta_\alpha A_{\bar z}=-D_{\bar{z}} \alpha,~~~\delta_\alpha \Phi_{i}=[\alpha, \Phi_{i}], \label{eq:gaugetransf}
\eeq
where $\alpha=\alpha(z)$ is an infinitesimal gauge transformation parameter.
The space ${\cal W}$ is a flat Kahler manifold with the Kahler potential \eqref{eq:infkahler}, and ${\cal G}$ is a symmetry group of this Kahler manifold.
Let us compute the moment map under the transformation \eqref{eq:gaugetransf}.
First, the Kahler form $\omega$ (excluding $g_5$) is given as
\beq
\omega= -\int \sqrt{g} |d^2z| \Tr \left(ig^{z\bar{z}}  \delta A_{\bar{z}} \wedge \delta A_{{z}}+ i h^{i\bar{j}}  \delta \Phi_i \wedge \delta \bar{\Phi}_{\bar{j}}  
\right), \label{eq:kahlerform}
\eeq
where $\delta$ means the exterior derivative on the space ${\cal W}$.
Let $V(\alpha)$ be the vector field which generates \eqref{eq:gaugetransf} on the space ${\cal W}$.
The moment map $\mu(\alpha)$ is defined as $\delta \mu(\alpha) =\iota_{V(\alpha)}\omega$,
where $\iota$ means to contract the vector $V(\alpha)$ with the two form $\omega$, giving a one form. Explicitly, it is given as
\beq
\iota_{V(\alpha)}\omega= -\int \sqrt{g} |d^2z| \Tr \left(-ig^{z\bar{z}} ((D_{\bar{z}} \alpha)  \delta A_{{z}}-(D_{{z}} \alpha) \delta A_{\bar{z}} )
+ i h^{i\bar{j}} ( [\alpha, \Phi_{i}] \delta \bar{\Phi}_{\bar{j}} -[\alpha, \bar{\Phi}_{\bar{i}}] \delta {\Phi}_{{j}}  )  \right). 
\eeq
From $\delta \mu(\alpha) =\iota_{V(\alpha)}\omega$, we get
\beq
\mu(\alpha) 
&=i \int \sqrt{g} |d^2z| \Tr \alpha \left(  g^{z\bar{z}}F_{z\bar{z}} -h^{i\bar{j}}[\Phi_i, \bar{\Phi}_j]    \right) \nonumber \\
&\equiv   -\int \sqrt{g} |d^2z| \Tr \alpha \mu.
\eeq
If the gauge multiplet $A_\mu,\lambda,\bar{\lambda},D$ (where $ \lambda=\lambda_{\ell}, \bar{\lambda}=\lambda_r$) has the kinetic term,\footnote{
We take vector multiplets $V^a$ to be real, and define $V=T_a V^a$ with anti-hermitian generators $T_a$.
Then $W_\alpha$ is defined as $W_\alpha =\frac{i}{8} {\bar D}^2 (e^{2i V} D e^{-2i V})$, which is slightly different from the Wess and Bagger convention. }
\beq
-\int d^2\theta \int \sqrt{g} |d^2z| \frac{1}{2g_5^2}\Tr (W^\alpha W_\alpha) \equiv \frac{1}{4g_5^2}\int d^2\theta [W^\alpha W_\alpha],
\eeq
then the equation of motion of the auxiliary field $D$ gives $D=\mu$.
The equation \eqref{eq:Dtermagain} is precisely this equation of motion of $D$.

In total, the 3d Lagrangian is given by
\beq
{\cal L}_{3d}= \int d^2\theta d^2\bar{\theta} K_{3d} +\int d^2 \theta W_{3d}+\int d^2\theta \frac{1}{4g_5^2}[W^\alpha W_\alpha] +{\rm h.c.} 
\eeq
The Kahler potential $K$ should be coupled appropriately to the vector multiplet as in \cite{Wess:1992cp}.

Before closing this subsection, let us comment on the dependence on the radius $R$ of the compactified direction $x^3$.
As explained in \cite{Gaiotto:2009hg}, we redefine $\Phi_i$ as
\beq
\Phi_i^{\rm (6d)} =R^{-1}\Phi_i,\label{eq:Phinorm}
\eeq
so that $\Phi_i^{\rm (6d)} $ has mass dimension two.
In this normalization, $\Phi_i^{\rm (6d)} $ can have direct interpretation as the scalars of the 6d ${\cal N}=(2,0)$ theories.
Furthermore, $g_5$ is given as
\beq
\frac{1}{g^2_5}= \frac{1}{8\pi^2 R}.
\eeq
Taking into account the fact that the action is multiplied by $2 \pi R$ by dimensional reduction,
we define the four dimensional superpotential as
\beq
W_{ 4d} &\equiv \frac{1}{2 \pi R}W_{3d} \nonumber \\
&=\frac{1}{\sqrt{2} (2\pi)^3 i} \int dz \wedge d \bar{z}\ \epsilon^{ij} \Tr \left( \Phi_i^{\rm (6d)}  D_{\bar z} \Phi_j^{\rm (6d)}  \right). \label{eq:4dsuperpot}
\eeq
Notice that this formula is independent of $R$ and the metrics $g_{z\bar{z}}$ and $h^{i\bar{j}}$. 
This superpotential is well-defined thanks to $\det F=K=T^* C$.

\subsection{Generalized Hitchin's equations and spectral curve}\label{sec:generalizedH}
Let us consider a field configuration such that the 3d Lorentz invariance is preserved. 
Then the conditions for supersymmetry to be preserved in the above Lagrangian are given as
\beq
0&= g^{z\bar{z}}F_{z\bar{z}} - h^{i\bar{j}}[\Phi_{i}, \bar{\Phi}_{\bar{j}} ]    ,  \label{eq:HitchinD}\\
0&= D_{\bar z} {\Phi}_i ,\label{eq:HitchinF1}  \\
0&=\epsilon^{ij} [{\Phi_i}, {\Phi}_j], \label{eq:HitchinF2}  \\
0&=D_{\bar{z}} \sigma, \label{eq:Hitchinhid1} \\
0&=[\sigma,\Phi_i] , \label{eq:Hitchinhid2} 
\eeq
where $\sigma=A_3$ is the real adjoint scalar in the 3d $\CN=2$ vector multiplet.
Solutions of these equations describe the vacuum moduli space of the low energy three dimensional theory.
We should also divide the space of solutions by the gauge group ${\cal G}$.

\subsubsection{Spectral curves}\label{sec:spectral}

Let us suppose $\sigma=0$ for a while.
The equations \eqref{eq:HitchinD}, \eqref{eq:HitchinF1}, and \eqref{eq:HitchinF2} are precisely 
the generalized Hitchin's equations~\cite{Xie:2013gma,Bonelli:2013pva}.
As we have seen in the previous subsection, the equation \eqref{eq:HitchinD} comes from the $D$-term condition of the gauge group ${\cal G}$.
As is usually done in physics literature, instead of imposing \eqref{eq:HitchinD} and dividing by ${\cal G}$, we may divide the space ${\cal W}$
by the complexification ${\cal G}_{\mathbb C}$ of ${\cal G}$.\footnote{There should be some stability condition as in the original Hitchin systems.
We neglect the stability in this paper, assuming that it is not essential in generic situations we are going to study in this paper.} 
In any case, we get a Kahler quotient ${\cal W}//{\cal G}$.
On this quotient space, we need to impose \eqref{eq:HitchinF1} and \eqref{eq:HitchinF2} which are holomorphic.
Note that \eqref{eq:HitchinF1} and \eqref{eq:HitchinF2} are covariant under ${\cal G}_{\mathbb C}$, so they are consistent with
the Kahler quotient. In this way, we get a Kahler manifold for the moduli space, as expected from the 3d ${\cal N}=2$ supersymmetry.

In the following discussion of this subsection, we take the gauge group $G=\SU(N)$.
There is a very useful way to study the moduli space of solutions of (generalized) Hitchin's equations, known as spectral curve.

\paragraph{Spectral curve for $\CN=2$ theory.}
First we briefly review the case of the original Hitchin systems, which describe Coulomb branches of $\CN=2$ theories
as discussed in ~\cite{Donagi:1995cf}. 
In this section, we only consider the case in which there is no puncture.

In the original Hitchin systems, we take $F=\CO \oplus K$, where $\CO$ represents the trivial bundle.
We set $\Phi_1=0$, and there is only one adjoint Higgs field $\Phi=\Phi_2$ which is a section of $K \otimes {\rm ad}(E)$.
Then we write down an equation
\beq
0=\det (x-\Phi(z))=x^N+\sum_{k=2}^N \phi_k(z) x^{N-k}, \label{eq:originalcurve}
\eeq
where $x$ is the coordinate of the fiber of the canonical bundle $K$. We omit the unit matrix; more precisely
the equation is $\det (x \cdot {\bf 1}_N-\Phi(z))=0$ for the $N \times N$ unit matrix ${\bf 1}_N$. 
The equation \eqref{eq:originalcurve} defines a curve in the total space $(z,x)$
of the canonical bundle $K$. This curve is called the spectral curve.
Notice that the $\phi_k$ is a holomorphic section of the line bundle $K^{k}$, $\partial_{\bar{z}} \phi_k=0$, because of the equation $D_{\bar{z}}\Phi=0$.

The moduli space of solutions of Hitchin's equations, denoted as $M_H$, has the following structure.
We define the base $B_H$ as 
\beq
B_H=\bigoplus_{k=2}^N H^0(C, K^{k}),
\eeq
where $H^0(C, K^{k})$ is the space of holomorphic sections of $K^k$.
Then, let us define a map $\pi:M_H \to B_H$ as
\beq
\pi : (A_{\bar{z}}, \Phi) \mapsto \{ \phi_k \}_{2 \leq k \leq N},
\eeq
where $(A_{\bar{z}}, \Phi)$ is a solution of the Hitchin's equations, and $\phi_k$ are defined in \eqref{eq:originalcurve}.
Then, it is known that (i) the map $\pi$ is surjective, $\pi(M_H)=B_H$, and (ii) at a generic point $p \in B_H$, the fiber
$\pi^{-1}(p)$ is a complex torus given by (a subspace of) the Jacobian variety\footnote{
We will describe the Jacobian variaty very explicitly in subsection~\ref{sec:N2twisted} in a simple case.} 
of the spectral curve \eqref{eq:originalcurve}.
See \cite{Donagi:1995cf} for explanations. 

The physical meaning of the above structure of $M_H$ is the following. The sections $\phi_k$ parametrize Coulomb branches of $\CN=2$
theories. Thus $B_H$ is the Coulomb moduli space. When 4d space is compactitfied on $S^1$, massless $\U(1)$ gauge fields are dual
to complex scalars whose target space is a complex torus, as we will review in section~\ref{sec:4}.
The complex torus given by the fiber $\pi^{-1}(p)$ is precisely the moduli space of these dual complex scalars.
The spectral curve \eqref{eq:originalcurve} is precisely the Seiberg-Witten curve describing low energy 
holomorphic gauge coupling matrix of massless $\U(1)$ vector fields.
We will briefly review M5 brane interpretation below.

\paragraph{Spectral curve for $\CN=1$ theory.}
Now let us consider $\CN=1$ theories described by the generalized Hitchin's equations.
In this case, the Higgs fields $\Phi_i$ take values in the rank two bundle $F$. Then, for example, an equation like $\det (x-\Phi_1)=0$
is not covariant under the $\U(2)$ which acts on $F$. 
A generalization of the spectral curve which is covariant under $\U(2)$ is given by
\beq
0=P_{i_1 \cdots i_N} \equiv \frac{1}{N!}(x_{i_1} -\Phi_{i_1})^{\alpha_1}_{~\beta_1} \cdots (x_{i_N} -\Phi_{i_N})^{\alpha_N}_{~\beta_N}
\epsilon_{\alpha_1\cdots \alpha_N} \epsilon^{\beta_1 \cdots \beta_N}, \label{eq:U(2)generalcurve}
\eeq
where $(x_1, x_2)$ are the coordinates of the fiber of the bundle $F$ and $i_1, \cdots, i_N$ take values 1 or 2.
For example, the above definition gives $P_{11\cdots 1}=\det (x_1 -\Phi_1)$ and $P_{22\cdots 2}=\det (x_2 -\Phi_2)$.

There are $N+1$ equations in \eqref{eq:U(2)generalcurve} since $i_1, \cdots, i_N$ are totally symmetric.
Because there are only 3 variables $z, x_1,x_2$, the equations \eqref{eq:U(2)generalcurve} look overdetermined
and not defining a curve in the total space of $F$. However, \eqref{eq:U(2)generalcurve} really defines a curve thanks to the commuting condition
\eqref{eq:HitchinF2}. 
Due to this equation, we can diagonalize $\Phi_1$ and $\Phi_2$ simultaneously by complex gauge transformations if 
their eigenvalues are generic. We get
\beq
\Phi_i \to \diag(\lambda_{i,1}(z),\cdots, \lambda_{i,N}(z)).
\eeq
Now the curve $\Sigma=\{(z,x_1,x_2) \in F :P_{i_1 \cdots i_N}=0 \}$ is equivalently given as
\beq
\Sigma &=\{(z,x_1,x_2) \in F :P_{i_1 \cdots i_N}(z,x_1,x_2)=0 \} \nonumber \\
&=\{(z,x_1,x_2) \in F : (x_1, x_2)=(\lambda_{1,k}(z), \lambda_{2,k}(z)),~k=1,\cdots,N \} ,\label{eq:generalcurve2}
\eeq
as one can check explicitly from the definition \eqref{eq:U(2)generalcurve}.
Therefore, $\Sigma$ defines is an $N$-covering of the base Riemann surface $C$. The branching points of the curve are
the points where some of the eigenvalues degenerate, i.e., typically, some of the $N$ sheets meet smoothly at those degenerate points to form a 
single connected curve $\Sigma$.\footnote{In some cases, the curve $\Sigma$ is not connected but has several connected components,
as in the Higgs branches of 4d $\CN=2$ theories discussed in section~\ref{sec:N2twisted}.}

The curve \eqref{eq:generalcurve2} has a clear meaning in terms of M5 branes.
Suppose $N$ M5 branes are wrapping a cycle in a Calabi-Yau threefold.
We assume that the Calabi-Yau locally looks like the total space of the bundle $F$, and the holomorphic cycle is given by the zero section of $F$,
which is the Riemann surface $C$.
Then, the eigenvalues of $\Phi_1$ and $\Phi_2$ are interpreted as the positions of these M5 branes in the transverse directions to the world volume of 
the M5 branes. As in \cite{Witten:1997sc}, the $N$ M5 branes become a single M5 brane wrapping a holomorphic curve in the Calabi-Yau threefold.
The curve \eqref{eq:generalcurve2} can be interpreted as this curve of the M5 brane.

From the above interpretation in terms of the M5 branes, we can clearly see the following physical identifications. 
First, the space of normalizable deformations of the curve $\Sigma$ corresponds to the moduli space of low energy 4d gauge theory.
Second, if the genus of the curve $\Sigma$, $g_{\Sigma}$, is nonzero, then this curve is precisely the Seiberg-Witten curve which
gives the low energy holomorphic coupling matrix of massless $\U(1)$ gauge fields.
This is because the two-form gauge field in the world volume of a single M5 brane gives massless $\U(1)$ gauge fields
in the low energy 4d field theories~\cite{Klemm:1996bj,Witten:1997sc}.

Therefore the curve \eqref{eq:generalcurve2} is very important 
for extracting the physical information from generalized Hitchin systems.
A method of determining the curve is explained in detail in \cite{Xie:2013rsa}.
In appendix~\ref{app:B}, we will explain another method which directly uses \eqref{eq:U(2)generalcurve}.

Generalized Hitchin systems have not yet been studied mathematically. Although we have used the M5 brane interpretation above,
it would be very interesting to deduce the above properties purely mathematically. 
We just summarize some properties which are expected from the M5 brane intuition.
We do not try to give a proof, or we do not even attempt to make our claims mathematically precise.  

Let us define $\phi^{(k)}_{i_1 \cdots i_k}$ as
\beq
\phi^{(k)}_{i_1 \cdots i_k}=\frac{(-1)^k}{k!(N-k)!}(\Phi_{i_1})^{\alpha_1}_{~\beta_1} \cdots (\Phi_{i_k})^{\alpha_k}_{~\beta_k}
\epsilon_{\alpha_1\cdots \alpha_k \alpha_{k+1} \cdots \alpha_N} \epsilon^{\beta_1 \cdots \beta_k \alpha_{k+1} \cdots \alpha_N}. \label{eq:defgeneralcoeff}
\eeq
The $\phi^{(k)}$ is a section of ${\rm Sym}^k F$, i.e., the symmetric part of $F \otimes \cdots \otimes F$ where $F$ appears $k$ times.
Then the curve \eqref{eq:U(2)generalcurve} is given by
\beq
0=(x^N)_{i_1 \cdots i_N}+\sum_{k=2}^N \phi^{(k)}_{(i_1 \cdots i_k} (x^{N-k})_{i_{k+1} \cdots i_N)}, \label{eq:expandedgeneralcurve}
\eeq
where $(x^{k})_{i_1 \cdots i_k}=x_{i_1}\cdots x_{i_k}$, and $(,)$ represents symmetrization.

Let $M_{GH}$ be the moduli space of solutions of generalized Hitchin's equations. We also define a map 
$\pi:M_{GH} \rightarrow  \bigoplus_{k=2}^N H^0(C, {\rm Sym}^k F )$ such that
\beq
&\pi :  (A_{\bar{z}},\Phi_i) \mapsto \{ \phi^{(k)}_{i_1 \cdots i_k} \}_{2 \leq k \leq N} ,
\eeq
where $(A_{\bar{z}},\Phi_i)$ is a solution of the generalized Hitchin's equations. 
We define the base $B_{GH}$ as 
\beq
B_{GH}=\pi(M_{GH}) \subset  \bigoplus_{k=2}^N H^0(C, {\rm Sym}^k F ).
\eeq 
We expect (but do not prove) the following properties;
\begin{enumerate}
\item The base $B_{GH}$ is given by the subspace of $\bigoplus_{k=2}^N H^0(C, {\rm Sym}^k F )$ such that 
the equations \eqref{eq:expandedgeneralcurve} define a consistent curve in the total space of $F$. In other words,
for each set $\{ \phi^{(k)}_{i_1 \cdots i_k} \}_{2 \leq k \leq N}$ for which the curve \eqref{eq:expandedgeneralcurve} makes sense
as an $N$ covering of the base Riemann surface $C$,
there exists a solution of the generalized Hitchin's equations satisfying \eqref{eq:defgeneralcoeff}.
\item For a generic point of $B_{GH}$, $p \in B_{GH}$, the fiber $\pi^{-1}(p)$ is a complex torus given by (a subspace of)
the Jacobian variety of the curve \eqref{eq:expandedgeneralcurve}.
\end{enumerate}

The second claim above may be shown in a similar way to the proof sketched in \cite{Donagi:1995cf}.
The crucial difference from the $\CN=2$ case is that $B_{GH}$ is not the same as the linear space $ \bigoplus_{k=2}^N H^0(C, {\rm Sym}^k F )$.
$B_{GH}$ is a nonlinear space in general. This is because \eqref{eq:expandedgeneralcurve} is an overdetermined set of equations
(i.e., $N+1$ equations for three variables $(z,x_1,x_2)$), and hence there should be nontrivial relations among parameters 
inside $\phi^{(k)}_{i_1 \cdots i_k}$. This property makes it possible to reproduce very rich dynamics of $\CN=1$ theories.

In the above discussion, we have assumed that there are no punctures. When we include punctures,
there are new degrees of freedom living at the punctures, and the above statements need modification.
Inclusion of punctures will be discussed in section~\ref{sec:3}.

\paragraph{Twisted Higgs bundle.}
There is a particularly simple branch of the full moduli space $M_{GH}$.
Suppose that the bundle $F$ has a holomorphic sub-bundle $L$, $L \subset F$, with rank one.
For example, if $F$ is a direct sum of two line bundles $F=L_1 \oplus L_2$, we can take $L=L_1$ or $L=L_2$.
Then, we have a branch in which $\Phi$ is taken as a section of the bundle $L \otimes {\rm ad}(E)$.
In this case, the commuting condition \eqref{eq:HitchinF2} is trivially satisfied since we are considering a rank one subspace of $F$.
This case is studied mathematically and it is called twisted Higgs bundle; see e.g., \cite{markman1994spectral}.

We can write down the spectral curve as
\beq
0=\det (x -\Phi)=x^N+\sum_{k=2}^N \tilde{\phi}_k(z) x^{N-k},\label{eq:twistedcurve}
\eeq
where $\tilde{\phi}_k$ is a holomorphic section of $L^k$.
The moduli space of a twisted Higgs bundle, $M_{TH}$, has the following structure. We define the base $B_{TH}$ as
$B_{TH}=\bigoplus_{k=2}^N H^0(C,L^k)$ and the map $\pi : M_{TH} \rightarrow B_{TH}$ as
\beq
\pi: (A_{\bar{z}},\Phi) \mapsto \{ \tilde{\phi}_k \}_{2 \leq k \leq N}.
\eeq
Then, (i) the map $\pi$ is surjective, $\pi(M_{TH})=B_{TH}$, and (ii) a generic fiber $\pi^{-1}(p)$ is given by a complex torus which is
(a subspace of) the Jacobian variety of the curve \eqref{eq:twistedcurve}. In particular, there are no constraints on $\tilde{\phi}_k$.
This structure is similar to the $\CN=2$ case, just by replacing $K^k $ with $ L^k$.

\subsubsection{Turning on $\sigma$}\label{sec:sigma}

Up to now, we have assumed that the real adjoint scalar $\sigma$ in the vector multiplet is zero.
Here we discuss what happens when they are nonzero.

First, note that any gauge invariant polynomial of $\sigma$, such as $\Tr \sigma^k~(k=2,\cdots,N)$ is a section of the trivial bundle.
Thus, from \eqref{eq:Hitchinhid1}, we conclude that they are constants on the Riemann surface $C$.
Up to gauge transformations, we can diagonalize it, and then $\sigma$ has constant eigenvalues.
Therefore, we can consider $\sigma$ to be a constant matrix on $C$.
The constant vev of $\sigma$ breaks the gauge group $G$ to a subgroup $H'$ whose elements commute with $\sigma$.
The equations $D_{\bar{z}} \sigma=0$ and $[\sigma, \Phi_i]=0$ impose that $A_{\bar{z}}$ and $\Phi_i$
take values in this subgroup $H'$. They satisfy generalized Hitchin's equations with the gauge group $H'$.

Another, equivalent way of describing the situation is the following. Let us consider a subset of solutions of generalized Hitchin's equations such that 
the pair $(A_{\bar{z}},\Phi_i)$ breaks the gauge group $G$ to a subgroup $H$. (Do not confuse this $H$ with the above $H'$.)
Then we can turn on a nonzero constant vev for $\sigma$ in the Cartan subalgebra of $H$.
This vev generically breaks $H$ to $\U(1)^r$, where $r$ is the rank of $H$.
In 3d, we can dualize $\U(1)^r$ vector fields to $r$ dual photons which are real scalars.
The $r$ real scalars coming from $\sigma$ are combined with the $r$ dual photons to give $r$ complex scalars.
In this way, we get a new branch of the moduli space of the twisted 5d SYM, which cannot be captured by generalized Hitchin systems alone.
We will get more insight on this type of branches in sections~\ref{sec:5} and \ref{sec:6}.

%%%%%%%%%%%%%%%%%%%%%%%%%%%%%%%%%%%%%%%%%%%%%%%%%%%%%%%%%%%
\section{Punctures, operators and superpotential}\label{sec:3}

In this section, we study punctures in the context of the twisted 5d SYM.
Our main purposes are; (i) to get identification of chiral operators, especially holomorphic moment maps or mesons,
and (ii) to derive a formula for dynamically generated superpotential vev. 

\subsection{$T_\rho[\SU(N)]$ theories}\label{sec:TSUNreview}
Here we focus on the case $G=\SU(N)$ for concreteness. More general gauge groups will be discussed in subsection~\ref{sec:generalgroup}.
Although $\CN=1$ supersymmetry allows a large class of punctures~\cite{Xie:2013gma},
we will focus on the locally half-BPS punctures of \cite{Gaiotto:2009we}.

We take the following point of view \cite{Chacaltana:2012zy}. 
At a puncture $z=z_p$, there is a three dimensional $\CN=4$ superconformal theory $T_\rho[\SU(N)]$ introduced by 
Gaiotto and Witten~\cite{Gaiotto:2008ak}. This 3d theory has codimension two in 5d space-time and is located at $z_p$.
At this point, the 3d theory is coupled to twisted 5d SYM in a way which preserves half of the supersymmetry locally. 
Let us review basic things about $T_\rho[\SU(N)]$ theories.
 
\begin{figure}
\begin{center}
\includegraphics[scale=0.3]{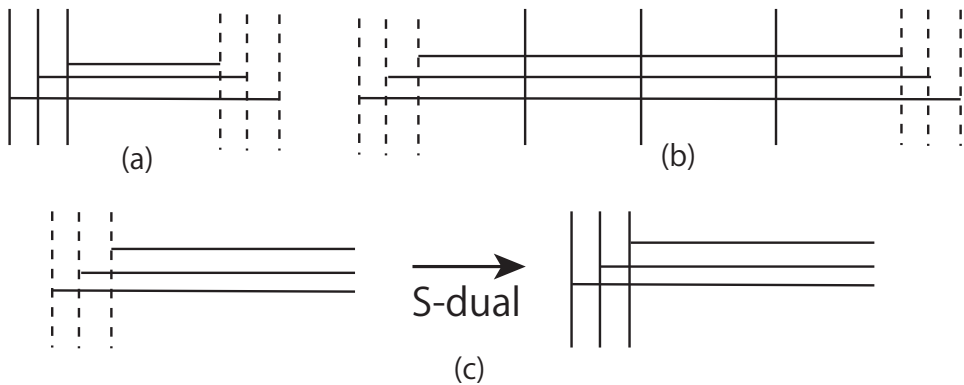}
\caption{(a):The brane realization of the $T[\SU(N)]$ theory in type IIB string theory for $N=3$. Horizontal lines are D3 branes, vertical solid lines are
NS5 branes, and vertical dashed lines are D5 branes. The NS5 branes and the D5 branes are extended to different directions in ten dimensions,
but we do not explicitly show that in the figure.
The $\U(k)$ vector multiplets in \eqref{eq:TSUNquiver} are coming from D3 branes suspended between adjacent NS5 branes.
(b):A 4d quiver superconformal gauge theory in type IIA brane construction. Horizontal lines are D4 branes, vertical solid lines are
NS5 branes, and vertical dashed lines are D6 branes.
There are two maximal punctures realized at the two sets of $N$ D6 branes at the ends of D4 branes, and several other simple punctures.
(c):D4 branes ending on D6 branes give a maximal puncture of figure~(b). 
By $S^1$ compactification of the $x^3$ direction and taking a $T$-dual, we get D3 branes ending on D5 branes in type IIB string theory.
The $S$-dual of it gives the $T[\SU(N)]$ theory at the end of the D3 branes, which is realized by D3 branes suspended between NS5 branes.
}\label{fig:TSUN}
\end{center}
\end{figure}

First we consider the $T[\SU(N)]$ theory, i.e., $\rho=0$,
which gives a maximal puncture when it is coupled to twisted 5d SYM.
The $T[\SU(N)]$ theory is the low energy limit of a 3d $\CN=4$ quiver gauge theory.
The quiver is given as
\beq
\U(1)-\U(2)-\cdots-\U(N-1)-\SU(N)_H|_{\rm flavor}. \label{eq:TSUNquiver}
\eeq
where $\SU(N)_H$ is a flavor symmetry and other $\U(k)~(k=1,\cdots,N-1)$ are gauge symmetries.
Between each adjacent groups $\U(k)$ and $\U(k+1)$ (or $\SU(N)_H$ for $k+1=N)$, there are bifundamental multiplets $A_k$
and $B_k$ in the representations $ {\bf (k+1)} \times \overline{\bf k} $ and ${\bf k} \times \overline{\bf (k+1)}$ of $\U(k+1) \times \U(k)$, respectively.
In type IIB string theory, this theory is realized by the brane construction as in figure~\ref{fig:TSUN}-(a). 

The fact that a maximal puncture of \cite{Gaiotto:2009we} is realized by a copy of the $T[\SU(N)]$ theory can be seen
along the lines of \cite{Benini:2009gi,Chacaltana:2012zy,Gaiotto:2011xs,Benini:2010uu}.
Here we sketch the reasoning.
Let us consider a 4d linear quiver superconformal theory constructed as in figure~\ref{fig:TSUN}-(b) in type IIA string theory~\cite{Witten:1997sc}.
A maximal puncture is realized as $N$ D4 branes ending on $N$ D6 branes.
In this paper, we are compactifying the $x^3$ direction. In this case we can take $T$-dual in this direction, and 
get a configuration in which D3 branes are ending on D5 branes. Taking $S$-dual as in figure~\ref{fig:TSUN}-(c),
we get the $T[\SU(N)]$ theory at the end of the bunch of D3 branes. 
Therefore, we may interpret that the $T[\SU(N)]$ is the $S$-dual of the maximal puncture.
What we are actually doing in this paper, in the type IIA context, is to consider the linear quiver configuration, uplift it to M-theory,
and compactify it on the $x^3$ direction and going again to type IIA string theory 
by regarding the $S^1$ of $x^3$ as the M-theory circle~\cite{Kapustin:1998xn}. Our twisted 5d SYM is realized by D4 branes in type IIA string theory.
The $S$-dual of type IIB string theory used above is naturally realized by
the exchange of M-theory circle from the $x^{10}$ direction to the $x^3$ direction. 
The $S$-dual induces mirror symmetry from the 3d point of view~\cite{Hanany:1996ie}.

Let us see some properties of the $T[\SU(N)]$ theory.
From the brane construction of figure~\ref{fig:TSUN}-(a) and using the interpretation of mirror symmetry as $S$-duality of type IIB 
string theory~\cite{Hanany:1996ie},
one can see that it is self-dual under mirror symmetry; its Higgs branch and Coulomb branch have completely the same structure,
and their difference is that they are acted by different $R$-symmetries, which we denote as $\SO(3)_X$ and $\SO(3)_Y$.
The hyperkahler holomorphic moment map of the Higgs branch $\SU(N)_H$ symmetry is given by
\beq
\mu_H^{\rm (3d)}=A_{N-1}B_{N-1}-\frac{1}{N} \tr (A_{N-1}B_{N-1}), \label{eq:higgsbranchmoment}
\eeq
where $A_k$ and $B_k$ are considered as $(k+1) \times k$ and $k \times (k+1)$ matrices, respectively.
By mirror symmetry, the Coulomb branch also has an $\SU(N)_C$ flavor symmetry, which is realized quantum mechanically.
We denote its homomorphic moment map as $\mu_C^{\rm (3d)}$.

Let us next consider more general $T_\rho[\SU(N)]$ theories.
Those theories are classified by an embedding of $\SU(2)$ in $\SU(N)$,
\beq
\rho: \SU(2) \rightarrow \SU(N).
\eeq
Under this embedding $\rho$, the fundamental representation of $\SU(N)$ is decomposed into irreducible representations of $\SU(2)$
as ${\bf N} \to {\bf n_1}+{\bf n_2}+\cdots+{\bf n_\ell}$, where ${\bf n_i}$ is the $n_i$ dimensional spin $(n_i-1)/2$ representation of $\SU(2)$.
Without loss of generality, we can assume $n_1 \geq n_2 \geq \cdots \geq n_\ell$. When this theory is coupled to the twisted 5d SYM,
it gives a puncture corresponding to the partition $(n_1,\cdots,n_\ell)$ discussed in \cite{Gaiotto:2009we}.

The $T_\rho[\SU(N)]$ theory has a quiver description and brane construction. However, for our purpose,
it is enough to note the following fact \cite{Gaiotto:2008ak,Benini:2009gi,Chacaltana:2012zy}. Let us take a
copy of the $T[\SU(N)]$ theory and give a nilpotent vev to the Coulomb branch operator 
$\mu_C^{\rm (3d)}$ as
\beq
\vev{\mu_C^{\rm (3d)}}  \propto \rho(\sigma^+), \label{eq:nilpotentvev}
\eeq
where $\sigma^+=\sigma^1+i\sigma^2$ is the raising operator of $\SU(2)$. 
Then, the low energy limit of this theory is the $T_\rho[\SU(N)]$ theory with some decoupled Nambu-Goldstone multiplets associated with
the spontaneous symmetry breaking of the Coulomb branch $\SU(N)_C$ symmetry by $\langle \mu_C^{\rm (3d)} \rangle$.\footnote{
The reason is as follows. In the setup of \cite{Gaiotto:2008sa}, 4d ${\CN=4}$ SYM is put on a half space $x^3 \geq 0$.
The hyperkahler moment map $\vec{\mu}_C^{(3d)}$ which is a triplet of an $\SO(3)_X$ $R$-symmetry is
given by $\vec{\mu}_C^{(3d)} \propto \vec{X}(0)$ in an appropriate $S$-dual frame, where $\vec{X}$ is the adjoint scalars in the triplet of $\SO(3)_X$.
By giving a vev $\vec{X}(0) \propto \rho(\vec{\sigma})$, the vev becomes effectively infinity in the IR limit and it gives a Nahm pole of $\vec{X}$ at $x^3=0$.
This Nahm pole is the necessary ingredient of the $T_\rho[\SU(N)]$ theory~\cite{Gaiotto:2008ak}.
}

The adjoint representation of $\SU(N)$ is decomposed into $\SU(2)$ representations as
\beq
{\rm adj} & \to \bigoplus_{a \in A} {\bf  (2j_a+1)}
\label{eq:adjdecomposition}
\eeq
where ${\bf  (2j_a+1)}$ is the spin $j_a$ representation, and $\{ j_a \}_{a \in A}$ represents the set of all spins appearing in the decomposition.
Then $\mu_C$ is decomposed as
\beq
\mu_C^{\rm (3d)}=\rho(\sigma^+)+\sum_{a \in A} \sum_{m=-j_a}^{j_a} T_{a, m}\ \mu_C^{a,m},
\eeq
where $T_{a,m}~(-j_a \leq m \leq j_a)$ are generators of $\SU(N)$ corresponding to the decomposition \eqref{eq:adjdecomposition}.
They satisfy e.g., $[\rho(\sigma^3/2),T_{a,m}]=mT_{a,m}$ and $[\rho(\sigma^+),T_{a,m}] \propto T_{a,m+1}$.
We have set the proportionality factor in \eqref{eq:nilpotentvev} to be unity for simplicity.
Because of the vev $\rho(\sigma^+)$, one can see that the fields $\mu_C^{a,m}$ for $m>-j_a$ are just Nambu-Goldstone multiplets.
By a complex $\SU(N)_C$ transformation $\mu_C \to U\mu_CU^{-1}$ which eliminates the Nambu-Goldstone multiplets, we get
\beq
\mu_C^{\rm (3d)}=\rho(\sigma^+)+\sum_{a \in A} T_{a, -j_a}\ \mu_C^{a,-j_a}.\label{eq:slodowy}
\eeq
A subset of Lie algebra of this form is called the Slodowy slice ${\cal S}_{\rho(\sigma^+)}$ 
transverse to a nilpotent orbit $\CO_{\rho(\sigma^+)}$. See \cite{Gaiotto:2008sa} for more details on this slice.

\subsection{Coupling $T_\rho[\SU(N)]$ to twisted 5d SYM}\label{sec:coupleTSUN}
In this subsection, for the moment, 
we restrict ourselves to the 4d $\CN=2$ (or 3d $\CN=4$) case so that we can see our formalism in the context of perhaps more
familiar $\CN=2$ Gaiotto type theories~\cite{Gaiotto:2009we}. 
Let us very briefly review this case (see e.g., ~\cite{Kapustin:2006pk}).
The bundle $F$ is taken as 
$\CO \oplus K$, where $\CO$ is the trivial bundle and $K$ is the canonical bundle.
From the point of view of the 3d $\CN=4$ supersymmetry, the pair $(V, \Phi_1)$ is a vector multiplet and the pair $(A_{\bar{z}},\Phi_2)$ is a
hypermultiplet. (The role of $\Phi_1$ and $\Phi_2$ can of course be exchanged by a simple relabeling of 1 and 2. The only important point is that
one of the $\Phi_i$ in the same multiplet as $V$ is a section of $\CO$, while the other one in the same multiplet as $A_{\bar{z}}$ is a section of $K$.)

The space of $(A_{\bar{z}},\Phi_2)$, which we denote as $\CW'$, 
spans an infinite dimensional flat hyperkahler manifold (see e.g., \cite{Hitchin:1986ea} for a review of hyperkahler manifold). The metric is
\beq
g_{\rm hk}=-\frac{1}{g_5^2} \int |d^2z| \Tr \left( \delta A_{\bar{z}} \otimes \delta A_{z}+\delta A_{z} \otimes \delta A_{\bar{z}} +
\delta \Phi_2 \otimes \delta \bar{\Phi}_2+ \delta \bar{\Phi}_2 \otimes \delta \Phi_2\right),
\eeq
where $\delta$ means exterior derivative on the space $\CW'$ as in section~\ref{sec:2}, and we have used the fact that $g^{z\bar{z}}\sqrt{g}=1$
on the Riemann surface.
There are three complex structures $I,J$ and $K$, which satisfy $I^2=J^2=K^2=-1$ and $JK=I,~KI=J,~IJ=K$.
They act on the tangent space of $\CW'$, and their actions are defined as 
\beq
 I^T(\delta A_{\bar{z}}, \delta \Phi_2, \delta {A}_{z}, \delta\bar{\Phi}_2)&=(i\delta A_{\bar{z}}, i\delta \Phi_2, -i\delta {A}_{z}, -i\delta\bar{\Phi}_2), \\
 J^T(\delta A_{\bar{z}}, \delta \Phi_2, \delta {A}_{z}, \delta\bar{\Phi}_2)&=(-\delta\bar{\Phi}_2, \delta {A}_{z},-\delta \Phi_2,\delta A_{\bar{z}}), \\
 K^T (\delta A_{\bar{z}}, \delta \Phi_2, \delta {A}_{z}, \delta\bar{\Phi}_2)&=(-i\delta\bar{\Phi}_2, i\delta {A}_{z}, i\delta \Phi_2,-i\delta A_{\bar{z}} ),
\eeq
where $I^T,J^T$ and $K^T$ are the transpose of $I,J$ and $K$ respectively. 
The $I$ is the complex structure which is present in the less supersymmetric 3d $\CN=2$ (or 4d $\CN=1$) case.

The Kahler forms associated to the complex structures $I,J$ and $K$ are given as $\omega_I=I^{T} \otimes 1(g_{\rm hk})$,
i.e., one of the indices of $I^T$ is contracted with one of the indices of $g_{\rm hk}$, and similarly for $J$ and $K$.
The $\omega_I$ is given by \eqref{eq:kahlerform} with the replacement 
$h^{i\bar{j}}\delta \Phi_i \wedge \delta \bar{\Phi}_{\bar{j}} \to g^{z\bar{z}} \delta \Phi_2 \wedge \delta \bar{\Phi}_2$. 
The Kahler forms associated to $J$ and $K$ are
combined as
\beq
\omega_J+i\omega_K=-\int |d^2z| 2\Tr [\delta A_{\bar{z}} \wedge \delta \Phi_2],
\eeq
where we have omitted the gauge coupling $g_5^{-2}$ for the moment.
The holomorphic moment map $\mu^{\rm (5d)}_{\mathbb C}(\alpha)$ under the gauge transformation \eqref{eq:gaugetransf} is defined by
$\delta \mu^{\rm (5d)}_{\mathbb C}(\alpha)=\iota_{V(\alpha)} (\omega_2+i\omega_3)$ as in the case of the real moment map in section~\ref{sec:2}, 
and it is given as
\beq
\mu^{\rm (5d)}_{\mathbb C}(\alpha) &=-\int |d^2z| 2\Tr[\alpha(D_{\bar{z}} \Phi_2 ) ] \nonumber \\
& \equiv -\int  \sqrt{g}|d^2z| \Tr[\alpha \mu^{\rm (5d)}_{\mathbb C} ].
\eeq

Now, let us put a copy of the $T[\SU(N)]$ theory at $z=z_p$ in the 5d space-time, and couple the Higgs branch $\SU(N)_H$ symmetry to the twisted 5d SYM.
The contribution to the holomorphic moment map from this sector is given as $\sqrt{g^{-1}} \delta^{2}(z-z_p) \mu_H^{(3d)}$,
where $\mu_H^{(3d)}$ is explicitly given by \eqref{eq:higgsbranchmoment}.
The total of the holomorphic moment maps 
$\mu_{\mathbb C}=g_5^{-2}\mu^{\rm (5d)}_{\mathbb C}+\sqrt{g^{-1}} \delta^{2}(z-z_p) \mu_H^{(3d)}$ is coupled to
the adjoint chiral multiplet $\Phi_1$ in the vector multiplet $(V,\Phi_1)$ as
\beq
W_{3d} &=-\int \sqrt{g} |d^2 z|     \sqrt{2} \Tr(\Phi_1 \mu_{\mathbb C}) \nonumber \\
&=-\sqrt{2} \left( \Tr \left(\mu_H^{(3d)} \Phi_1(z_p) \right)+ \int |d^2z| \frac{2}{g_5^2}\Tr(\Phi_1 D_{\bar{z}} \Phi_2  ) \right). \label{eq:coupledsuper}
\eeq
The second term is exactly the same as \eqref{eq:infsuper} which was already derived previously.
The first term gives a coupling between $T[\SU(N)]$ and the twisted 5d SYM.

Now suppose that $\Phi_1(z_p)$ has a generic diagonal vev. Then the first term of \eqref{eq:coupledsuper} gives a 
mass term for the $T[\SU(N)]$ theory. Let us study the effect of this mass term on the Coulomb branch of the $T[\SU(N)]$.
For this purpose, we use mirror symmetry. As mentioned earlier, the mirror of $T[\SU(N)]$ is also $T[\SU(N)]$, but the Coulomb branch is mapped
to the Higgs branch, and masses are mapped
to FI-parameters. By holomorphy, the holomorphic masses are mapped to holomorphic FI terms in the superpotential.
Let us denote the FI parameter of $\U(1) \subset \U(k)$ in \eqref{eq:TSUNquiver} as $\xi_k$. The superpotential
contains terms $\sqrt{2}\xi_k \tr \phi_k$, where $\phi_k$ is the adjoint scalar of $\U(k)$. Then, the $F$-term conditions of $\phi_k$ give
\beq
&B'_kA'_k-A'_{k-1}B'_{k-1}=\xi_k~~~(2 \leq k \leq N-1), \nonumber \\
&B'_1 A'_1=\xi_1.
\eeq
where we denote bifundamentals in the mirror side as $A'_k$ and $B'_k$.
We define $M'=A'_{N-1}B'_{N-1}$.
From the equations above, we get
\beq
M'(M'-\xi_{N-1}) &=(A'_{N-1}B'_{N-1}A'_{N-1}B'_{N-1}-\xi_{N-1}A'_{N-1}B'_{N-1}) \nonumber \\
&=A'_{N-1}A'_{N-2}B'_{N-2}B'_{N-1}. \nonumber 
\eeq
Similarly, noting that for $k \geq 2$
\beq
&B'_k B'_{k+1}\cdots B'_{N-1} (A'_{N-1}B'_{N-1}-\sum_{i=k}^{N-1} \xi_i) 
=B'_k B'_{k+1}\cdots B'_{N-2}(A'_{N-2}B'_{N-2}-\sum_{i=k}^{N-2} \xi_i) B'_{N-1} \nonumber \\
&=\cdots 
=A'_{k-1}B'_{k-1}B'_{k+1} \cdots B'_{N-1} \nonumber 
\eeq
and also $B'_1(A'_1B'_1-\xi_1)=0$, we get~\cite{Gaiotto:2008ak}
\beq
M'(M'-\xi_{N-1})(M'-\xi_{N-1}-\xi_{N-2})\cdots (M'-\sum_{k=1}^{N-1} \xi_k)=0.
\eeq
This equation gives the characteristic polynomial of $M'$. In particular, the eigenvalues of $M'$ are given as 
$\sum_{i=k}^{N-1} \xi_i$.

The moment map $\mu^{(3d)}_C$ of the original theory is the moment map of the Higgs branch of the mirror theory, $\mu^{(3d)}_C=M'-\frac{1}{N} \tr M'$. 
The characteristic polynomial of $\mu^{(3d)}_C$ is given as
\beq
P(x) &\equiv \det (x-\mu^{(3d)}_C)=\prod_{k=1}^N (x-\lambda_k), \\
\lambda_k &=\sum_{i=k}^{N-1} \xi_i-\frac{1}{N} \sum_{i=1}^{N-1} i \xi_i,
\eeq
for arbitrary variable $x$.
Now recall that the FI parameters $\xi_k$ of this mirror theory 
come from the mass parameters of the original theory, which are given by the eigenvalues of $\Phi_1(z_p)$.
By the symmetry under the Weyl group of $\SU(N)$, or more directly from the brane picture of figure~\ref{fig:TSUN}-(a),
the $\lambda_k$ must be proportional to the eigenvalues of $\Phi_1(z_p)$.

The factor of proportionality in mirror symmetry between masses and FI parameters
can be determined by comparing a quark mass in one theory and a vortex mass in the mirror theory~\cite{Aharony:1997bx}.
As the most simple example, let us consider 3d $\CN=4$ $\U(1)$
gauge theory with one hypermultiplet $q,\tilde{q}$. The superpotential is $W=\sqrt{2}\phi (\tilde{q}q-\xi)$. One can explicitly calculate
(see e.g., \cite{Tong:2005un}) that
the BPS bound for a vortex mass is given by $4\pi \xi$. The mirror of this theory consists of a single hypermultiplet $p, \tilde{p}$
with a mass $m\tilde{p} p$. Therefore we get $m=4 \pi \xi$. By inspection of a brane construction of this simple theory and more
general theories, one obtains $\sqrt{2}\Phi_1(z_p)=4\pi \diag(\lambda_1, \cdots, \lambda_N)$ up to a phase which we neglect.
Thus we get 
\beq
\det (x-\mu^{(3d)}_C)=\det \left(x-\frac{\sqrt{2}}{4\pi} \Phi_1(z_p) \right). \label{eq:idmomentmap}
\eeq

In the class ${\cal S}$ theories which we are studying in this paper, there is a flavor $\SU(N)$ symmetry associated to each maximal puncture.
In the context of the twisted 5d SYM, this flavor symmetry associated to the puncture $z=z_p$ comes from
the Coulomb branch $\SU(N)_C$ symmetry of the $T[\SU(N)]$ theory at the puncture.
The operator $\mu^{(3d)}_C$ is the holomorphic moment map associated to this flavor symmetry.
Therefore \eqref{eq:idmomentmap} gives us the way to identify the vev of the moment map in field theory with a quantity in twisted 5d SYM.
In the following, we will often write 
\beq
\mu^{(3d)}_C \approx \frac{\sqrt{2}}{4\pi}\Phi_1(z_p). \label{eq:conjugate}
\eeq
The meaning of this equation is that the characteristic polynomial of both sides agree with each other.
If the eigenvalues are generic, i.e., if all the eigenvalues are distinct, \eqref{eq:conjugate} means that they  are conjugate matrices.

For more general $T_\rho[\SU(N)]$ theories, we simply use the renormalization group interpretation of the previous subsection~\ref{sec:TSUNreview}.
In this case we just need to restrict $\mu^{(3d)}_C$ to the form \eqref{eq:slodowy}.
Assuming that the eigenvalues of $\Phi_1(z_p)$ are generic, we define the orbit of $\lambda \equiv \sqrt{2}\Phi_1(z_p)/4\pi$ as the set
\beq
\CO_\lambda=\left\{ g_{\mathbb C} \lambda g_{\mathbb C}^{-1}; g_{\mathbb C} \in \SU(N)_{\mathbb C} \right\},
\eeq
where $\SU(N)_{\mathbb C}=\SL(N)$ is the complexification of $\SU(N)$.
Then $\mu^{(3d)}_C $ is in the intersection of $\CO_\lambda$ and the Slodowy slice ${\cal S}_{\rho(\sigma^+)}$,
 ${\cal S}_{\rho(\sigma^+)} \cap \CO_\lambda$.
See e.g., \cite{Gukov:2006jk,Gukov:2008sn,Gaiotto:2008ak,Gaiotto:2009hg,Chacaltana:2012zy} for 
explanations on non-generic cases where some of the eigenvalues degenerate.

Before closing this subsection, let us consider the dependence on the radius $R$ of $S^1$.
As in section~\ref{sec:2}, the three dimensional Kahler and superpotential are related to the four dimensional ones as
\beq
(2\pi R) K_{4d}=K_{3d},~~~~~~~(2\pi R) W_{4d}=W_{3d},
\eeq
where $K_{\rm 3d}$ contains both the bulk contribution of the twisted 5d SYM and the contribution from the Coulomb branch of the $T[\SU(N)]$
theory. Moment maps are proportional to Kahler metrics, so the vevs of 4d moment maps are related to 3d moment maps as
\beq
\mu^{(4d)}=\frac{1}{2 \pi R } \mu^{(3d)}_C \approx \frac{\sqrt{2}}{8\pi^2R} \Phi_1(z_p) = \frac{\sqrt{2}}{ 8\pi^2}  \Phi_1^{(6d)}(z_p),\label{eq:precisecorres}
\eeq
where $\Phi_1^{(6d)}(z_p)$ was introduced in section~\ref{sec:kahlersuper}. In this way, $\mu^{(4d)}$ is independent of $R$.

\subsection{Dynamical superpotential}\label{sec:supervev}

In the above discussion, we have concentrated on the 4d $\CN=2$ or 3d $\CN=4$ case with one puncture.
However, the class of theories discussed in appendix~\ref{app:A} is locally the same as this 3d $\CN=4$ case,
so we can easily generalize the result. We only consider the case $F=L_1 \oplus L_2$ as in the appendix.
Then $\Phi_1$ and $\Phi_2$ are sections of $L_1 \otimes {\rm ad}(E)$ and $L_2 \otimes {\rm ad}(E)$, respectively.

Suppose that $\Phi_1$ is coupled to copies of $T_\rho[\SU(N)]$ theories at the punctures $z_a ~(a \in A)$,
and $\Phi_2$ is coupled to them at $z_b~(b \in B)$, where $A$ and $B$ are sets labeling the punctures. 
The choice of whether to couple $T_\rho[\SU(N)]$ to $\Phi_1$ or $\Phi_2$ corresponds to what $\CN=1$ theory we consider; 
it is the choice of $\pm$ of punctures discussed in appendix~\ref{app:A}.
The superpotential is now given by
\beq
W_{3d} 
=-\sqrt{2} \left( \sum_{a \in A} \Tr \left(\mu_{a,H}^{(3d)} \Phi_1(z_a) \right)-\sum_{b \in B} \Tr \left(\mu_{b,H}^{(3d)} \Phi_2(z_b) \right)
+ \int |d^2z| \frac{2}{g_5^2}\Tr(\Phi_1 D_{\bar{z}} \Phi_2  ) \right), \label{eq:generalcoupledsuper}
\eeq
where the minus sign in the second term comes from the fact that the third bulk term is antisymmetric in $\Phi_1 \leftrightarrow \Phi_2$.
The normalizations of $\Phi_{1}$ and $\Phi_2$, and hence
the above couplings, depend on the basis of $L_1$ and $L_2$ at the punctures. 
We can, e.g., canonically normalize the Kahler potential of them at the positions
of the punctures. We will make some comments later on normalization of fields.

The $F$-term equations are given by
\beq
0&=\sum_{a \in A} \mu_{a,H}^{(3d)} \delta^2(z-z_a)+\frac{2}{g_5^2} D_{\bar{z}} \Phi_2, \label{eq:Ftermwpuncture1} \\
0&=\sum_{b \in B} \mu_{b,H}^{(3d)} \delta^2(z-z_b)+\frac{2}{g_5^2} D_{\bar{z}} \Phi_1. \label{eq:Ftermwpuncture2}
\eeq
Due to the delta function sources, the Higgs fields $\Phi_1$ and $\Phi_2$ develop singularities as
\beq
\Phi_2 \to -\frac{ g_5^2}{4 \pi} \frac{\mu_{a,H}^{(3d)} }{z-z_a}~~~(z \to z_a),  \label{eq:regsing1}\\
\Phi_1 \to -\frac{ g_5^2}{4 \pi } \frac{\mu_{b,H}^{(3d)} }{z-z_b}~~~(z \to z_b). \label{eq:regsing2}
\eeq
Here we assumed that the singular part of $A_{\bar{z}}$ (if any) commutes with the singular parts of $\Phi_1$ and $\Phi_2$.
These equations will be the expected singular behaviors at regular punctures as we will see later.

Using \eqref{eq:Ftermwpuncture1} and \eqref{eq:regsing2}, the superpotential \eqref{eq:generalcoupledsuper} evaluated at 
a solution of the $F$-term equations are given by
\beq
W_{3d}|_{\rm solution} &=\sqrt{2}  \sum_{b \in B} \Tr \left(\mu_{b,H}^{(3d)} \Phi_2(z_b) \right) \nonumber \\
&=\frac{2\sqrt{2} i}{g_5^2}\sum_{b \in B} \oint_{|z-z_b|=\epsilon} dz \Tr (\Phi_1 \Phi_2),
\eeq
where $\epsilon$ is an infinitesimal number.
This is the formula for dynamically generated superpotential vev. In field theory side, it is generated typically by
gaugino condensation in theories with a mass gap such as 4d pure $\CN=1$ SYM.
When there are distinct vacua, the difference of the superpotential vevs between two of the vacua
is a physical observable since it gives tensions of BPS domain walls~\cite{Dvali:1996xe}.

Let us check the $R$ dependence. The 4d superpotential is given by
\beq
W_{4d}|_{\rm solution}&=\frac{1}{2 \pi R} W_{3d}|_{\rm solution} \nonumber \\
&=-\sum_{b \in B} \oint_{|z-z_b|=\epsilon} \frac{dz}{2 \pi i }  \frac{\sqrt{2}}{4 \pi^2 } \Tr (\Phi_1^{(6d)} \Phi_2^{(6d)} ), \label{eq:normalizedsuper}
\eeq
where we have used $g_5^2=8\pi^2 R$.
This is independent of $R$. By using $ dz \wedge d\bar{z}  \partial_{\bar{z}} \Tr (\Phi_1^{(6d)} \Phi_2^{(6d)} )=0$ and integration by parts,
we can also rewrite the superpotential as
\beq
W_{4d}|_{\rm solution}
&=\sum_{a \in A} \oint_{|z-z_a|=\epsilon} \frac{dz}{2 \pi i}  \frac{\sqrt{2}}{4 \pi^2} \Tr (\Phi_1^{(6d)} \Phi_2^{(6d)} ). \label{eq:cauchyedsuper}
\eeq

In the above discussions, we have only considered regular singularities which can be realized by $T_\rho[\SU(N)]$
theories. However, we expect that the formula \eqref{eq:normalizedsuper} is valid even if there are irregular singularities.
The reason is that irregular singularities (which have type IIA brane realization) can be achieved by renormalization group flows from regular ones
by taking some of the masses to infinity.

\subsection{Regular singularity}\label{sec:regsing}

At a puncture $z_p$, we can add a mass term associated to the flavor symmetry at that puncture.
Suppose that the puncture is maximal and we add a mass term
\beq
W_{4d} \supset \tr (m \mu^{(4d)})  ~\leftrightarrow ~W_{3d} \supset \tr (m \mu^{(3d)}_C).
\eeq
Then, the mirror of \eqref{eq:conjugate} tells us that we get
\beq
\mu^{(3d)}_H \approx \frac{1}{4\pi}m.
\eeq
Combined with \eqref{eq:regsing1} and $g_5^2=8\pi^2 R$, we get
\beq
\Phi_2^{(6d)}=R^{-1}\Phi_2 \to \frac{1}{2} \frac{m}{z-z_p}.
\eeq
This is the standard behavior of regular singularity. 
The factor $1/2$ may look unusual, but it is necessary so that $m_{a,b}$ can be directly interpreted as quark masses in 4d $\CN=2$ theories.\footnote{
By carefully examining BPS tensions of M2 branes of M-theory, strings of $(2,0)$ theory and particles of 4d theories,
the Seiberg-Witten differential is determined as $\lambda_{\rm SW}= (x /\pi ) dz$~\cite{Gaiotto:2009hg}, 
where $\det (x-\Phi^{(6d)}(z))=0$ as in section~\ref{sec:spectral}.
Then, by computing $\oint \lambda_{\rm SW}$ for a massive quark with a large mass, one can see that singularities of $x$ must be 
$x \sim  m/(2z)$.    }

Using the relation \eqref{eq:precisecorres}, we get the superpotential vev as
\beq
W_{4d}|_{\rm solution}&=
\sum_{a \in A} \oint_{|z-z_a|=\epsilon} \frac{dz}{2 \pi i}  2 \Tr (\Phi_1^{(6d)} \mu_a^{(4d)} ) \nonumber \\
& =  \sum_{a \in A} \tr(m_a \mu_a^{(4d)}), \label{eq:rewrittensuper}
\eeq
where $m_{a,b}$ are mass matrices at the punctures $z_a$ and $z_b$. 
This formula is valid only for the vev.

We have been careful about the normalization of the fields in the above discussion, but their meaning is limited in 
4d $\CN=1$ theories. In the case of 4d $\CN=2$ theories, a holomorphic moment map $\mu^{(4d)}$ is in a current multiplet of
a global symmetry, and hence it is not renormalized. Correspondingly, a mass in 4d $\CN=2$ theories are not renormalized.
However, there is no such nonrenormalization for $\mu^{(4d)}$ and masses in 4d $\CN=1$ theories.

Let us see it more explicitly. Consider flavors of quarks $q$ and $\tilde{q}$ coupled to 4d $\CN=2$ or $\CN=1$ gauge multiplets.
In the $\CN=2$ case, it is known that the Kahler potential of the quarks is not renormalized~\cite{Argyres:1996eh}. 
Therefore, we can normalize 
the quarks so that they have canonical kinetic terms in the Kahler potential. The normalization of $\mu^{(4d)}=\tilde{q}q$ is also fixed.
However, in the $\CN=1$ case, the Kahler potential receives perturbative corrections, and in particular there are wave function renormalizations,
$K=Z q^\dagger q+\tilde{Z} \tilde{q}^\dagger \tilde{q}$.
These corrections are not holomorphic.
As long as we only consider holomorphic quantities, there is no way to fix the normalization of $q$ and $\tilde{q}$.
Canonically normalizing the quarks requires non-holomorphic wave function renormalizations of the quarks which depend on
renormalization scales. 

Although the normalizations of $m$ and $\mu^{(4d)}$ are not uniquely fixed, their product $m\mu^{(4d)}$ is renormalization group invariant
due to the usual nonrenormalization theorem of superpotential.
In general, exact results in supersymmetric field theories are invariant under wave function renormalization of fields~\cite{ArkaniHamed:1997ut}.

With the above facts in mind, we renormalize $\Phi_{1,2}^{(6d)}$ as
\beq
\tilde{\Phi}_1=\frac{\sqrt{2}}{ 8\pi^2}  \Phi_1^{(6d)},~~~~~\tilde{\Phi}_2=2\Phi_{2}^{(6d)},
\eeq
and we define their limits at $z \to z_{a,b}$ as
\beq
&\tilde{\Phi}_1 \to  \mu_a^{(4d)},~~~\tilde{\Phi}_2 \to \frac{m_a}{z-z_a}~~~(z \to z_a), \label{eq:nomalizedop1}\\
&\tilde{\Phi}_2 \to \mu_b^{(4d)},~~~\tilde{\Phi}_1 \to \frac{m_b}{z-z_b}~~~(z \to z_a). \label{eq:nomalizedop2}
\eeq 
The normalization of $m_a$ and $\mu_a^{(4d)}$ are the same as before, but the normalization of $m_b$ and $\mu_b^{(4d)}$
are changed in such a way that their products $m_b \mu_b^{(4d)}$ are unchanged. 
The superpotential \eqref{eq:normalizedsuper} or \eqref{eq:cauchyedsuper} is now given as
\beq
W_{4d}|_{\rm solution}=\sum_{a \in A} \oint_{|z-z_a|=\epsilon} \frac{dz}{2 \pi i }  \Tr (\tilde{\Phi}_1   \tilde{\Phi}_2 ). \label{eq:nomalizedsup}
\eeq
Note that the coefficients of \eqref{eq:nomalizedop1}, \eqref{eq:nomalizedop2} and \eqref{eq:nomalizedsup} are very simple now.

The reason that we are so serious about the numerical coefficients of the above equations
(aside from phases) is that we can actually check them.
In SQCD, numerical coefficients of exact results are really determined precisely~\cite{Finnell:1995dr}, 
at least in a certain class of renormalization schemes. See \cite{Intriligator:1995au} for those exact results including the coefficients.
Thus, by comparing the results of twisted 5d SYM and field theories, we can obtain important consistency checks including these coefficients.

\subsection{General gauge groups and $T_\rho[G]$ theories}\label{sec:generalgroup}

In this subsection we discuss the case of general gauge groups $G=A,D,E$ which appear in 
the 6d $\CN=(2,0)$ theories.\footnote{There is a further generalization including outer-automorphism twist~\cite{Tachikawa:2010vg}.
We do not study this direction in this paper.}
Our discussion will be very brief and the reader should consult \cite{Gaiotto:2008sa,Gaiotto:2008ak} for essential ingredients.

At each puncture, there is a copy of the 3d theory $T_\rho[G]$. First let us recall the definition of the simplest theory $T[G]$, i.e., $\rho=0$.
Consider 4d $\CN=4$ SYM with the theta angle taken to be zero. 
We divide the theory into two parts, $x^3<0$ and $x^3>0$. Then the theory can be regarded
as two 4d $\CN=4$ SYM defined on the half spaces $x^3<0$ and $x^3>0$ which are connected by a boundary condition at $x^3=0$ smoothly. 
Now we take $S$-dual of the theory in $x^3>0$.
In this region we get $\CN=4$ SYM with the dual gauge group $G^\vee$. For a simply laced gauge group $G=A,D,E$, 
the dual group $G^\vee$ is the same as $G$ (at the level of Lie algebra),
but we continue to write it as $G^\vee$ to distinguish between the original $\CN=4$ SYM and the dual one.
We have the $\CN=4$ SYM with gauge group $G$ for $x^3<0$ and gauge group $G^\vee$ for $x^3>0$, and there must be some
boundary condition at $x^3=0$. The boundary condition is such that there is a copy of the $T[G]$ theory at $x^3=0$ 
which is coupled to both of the $\CN=4$ SYM.
The Higgs branch of the $T[G]$ theory has a flavor $G$ symmetry, and the Coulomb branch of it has the flavor $G^\vee$ symmetry,
and each of these flavor symmetries are gauged by the corresponding bulk $\CN=4$ SYM.

For our purpose, we take a copy of the $T[G]$ theory and couple its Higgs branch moment map $ \mu_H^{(3d)}$ to twisted 5d SYM as
$ \sqrt{2} \Tr \left(\mu_H^{(3d)} \Phi_1(z_p) \right)$. Then we want to know the Coulomb branch moduli space of the $T[G]$ theory when 
$\Phi_1(z_p) $ is nonzero. 

This problem can be studied as follows. Let us consider the above setup of two $\CN=4$ SYM coupled to the $T[G]$ theory.
We ungauge the gauge group $G^\vee$ by introducing the Dirichlet boundary condition of gauge fields at, say $x^3=L>0$.
Furthermore, in the region $x^3<0$, we turn on the vev of the adjoint chiral field $\Phi_{\CN=4}$ in the vector multiplet of $\CN=4$ SYM on the half space.
We take the vev as $\vev{\Phi_{\CN=4}}=\Phi_1(z_p) $. Because $\Phi_{\CN=4}$ is coupled to the $T[G]$ theory as 
$\sqrt{2} \Tr \left(\mu_H^{(3d)} \Phi_{\CN=4} \right)$, we can realize the same situation in this 4d setup as in our twisted 5d SYM setup.
Actually, this is more than just an analogy. If we take the Riemann surface $C$ of the twisted 5d SYM to be a cigar geometry and put
the $T[G]$ theory at the tip of the cigar, then, by dimensional reduction on the $S^1$ direction of the cigar, we get 4d $\CN=4$ SYM
on a half space with the $T[G]$ theory at the boundary~\cite{Benini:2010uu,Chacaltana:2012zy}.

The Coulomb branch of the $T[G]$ theory in this 4d setup is determined as follows. We take $S$-dual of the $\CN=4$ SYM on $x^3<0$.
Then, recalling the way  the $T[G]$ theory was introduced above, one can see that we get a smooth $\CN=4$ SYM with the gauge group $G^\vee$
 in the entire region $x^3<L$. For a simply laced gauge group $G=A,D,E$, the vev of $\Phi_{\CN=4}$ is mapped to
 the vev of the dual adjoint field $ \Phi_{\CN=4}^\vee$ as (see e.g., \cite{Kapustin:2006pk})
 \beq
\vev{ \Phi_{\CN=4}^\vee}=\frac{(e^\vee)^2 }{ 4\pi }\vev{ \Phi_{\CN=4}} ,\label{eq:dualadjoint}
 \eeq
 where $e^\vee$ is the gauge coupling of the $\CN=4$ gauge group $G^\vee$. It is related to the gauge coupling $e$ of $G$
 as $(e^\vee)^2/4\pi=4\pi/ e^2$ since we have taken the theta angle to be zero. The equation \eqref{eq:dualadjoint}
 should be interpreted as the statement that their eigenvalues match, because the eigenvalues have the physical meaning
 as BPS masses of $W$-bosons and monopoles in $\CN=4$ SYM.
 The $S$-dual exchanges the masses of $W$-bosons and monopoles.

The moduli space of the system is described by Nahm's equations on the space $x^3<L$. 
We have imposed the Dirichlet boundary condition at $x^3=L$ and
we also impose the boundary condition $ \Phi_{\CN=4}^\vee \to \vev{ \Phi_{\CN=4}^\vee}$ at $x^3 \to -\infty$.
The result is that the moduli space, as a complex manifold, is the orbit of $ \vev{\Phi_{\CN=4}^\vee}$.
In particular, $ \Phi_{\CN=4}^\vee$ at $x=L$ is conjugate to $\vev{ \Phi_{\CN=4}^\vee} $. The holomorphic moment map is given as
\beq
\mu^{(3d)}_C=\frac{\sqrt{2}}{(e^\vee)^2} \Phi_{\CN=4}^\vee(x^3=L) \approx \frac{\sqrt{2}}{(e^\vee)^2} \vev{\Phi_{\CN=4}^\vee}
=\frac{\sqrt{2}}{4 \pi} \Phi_1(z_p) ,
\eeq
where $\approx $ means that both sides are conjugate by complexified $G_{\mathbb C}$, assuming that the eigenvalues are generic.
The first equality, including the coefficient $\sqrt{2}/(e^\vee)^2$, can be determined by careful calculation of the hyperkahler moment map.
The second equality is a consequence of the Nahm's equations. In the third equality we have used \eqref{eq:dualadjoint} and 
$\vev{\Phi_{\CN=4}}=\Phi_1(z_p)$. This result generalizes \eqref{eq:conjugate} which was derived for $\SU(N)$ to arbitrary 
simply laced gauge groups. It is pleasant that we get the same result from two different lines of arguments
in subsection~\ref{sec:coupleTSUN} and in this subsection.

For more general $T_{\rho^\vee}[G]$ theories, we simply note that the above discussions are almost unchanged
other than the fact that we need to include a Nahm pole $\rho^\vee/(y-L)$ at $y \to L$.
Then the moduli space is the intersection of the Slodowy slice ${\cal S}_{\rho^\vee}$ and the orbit $\CO_\lambda$ of $\lambda=(\sqrt{2}/4\pi) \Phi_1(z_p)$.

The discussions on superpotential vev in subsection~\ref{sec:supervev} and regular singularities in subsection \ref{sec:regsing}
are the same. Lagrangian descriptions of the $T_\rho[G]$ theories are not known for general $G$,
but for many purposes the equations \eqref{eq:nomalizedop1}, \eqref{eq:nomalizedop2} and \eqref{eq:nomalizedsup} are enough
and we do not need more explicit information about $T_\rho[G]$.

%%%%%%%%%%%%%%%%%%%%%%%%%%%%%%%%%%%%%%%%%%%%%%%%%%%%%%%%%%%
\section{Nonrenormalization theorems}\label{sec:4}

One of the main claims in this paper is that the holomorphic dynamics of 4d field theories of class ${\cal S}$ are described by classical solutions of
twisted 5d SYM. We are considering 6d $\CN=(2,0)$ theories on a Riemann surface $C$, and we further compactify the theory on $S^1$.
Taking the radius $R$ to be smaller than the length scale of $C$, denoted as $L$,
we get twisted 5d SYM on ${\mathbb R}^{1,2} \times C$. On the other hand,
taking the limit that the length scale $L$ to be much smaller than $R$, we get a 4d theory on ${\mathbb R}^{1,2} \times S^1$.
The limit that the 5d SYM is reliable ($R/L \to 0$) is different from the limit that we obtain purely 4d field theory ($R/L \to \infty$).
Therefore, it is not evident whether the 5d SYM can describe the dynamics of 4d field theory.

There are at least two points which need to be justified. First, we have to show that field theory quantities we are interested in
have rather trivial (or well controlled) dependence on $R$. 
If a quantity
receives complicated quantum corrections which depend on $R$ and cannot be controlled, 
we have no justification of the above argument at all to study that quantity.
Second, we have to justify that quantum corrections in twisted 5d SYM can be neglected, and we can treat it classically. 
In this section, we investigate these two points.
We will only assume 4d $\CN=1$ supersymmetry and discuss holomorphic quantities. 
If the theory has $\CN=2$ supersymmetry, we can also control the Kahler potential, but we do not discuss that in this paper.

\subsection{Nonrenormalization in field theory}\label{sec:fieldnonren}

Let us consider a 4d $\CN=1$ field theory. In the low energy limit, we assume that the theory is described by
neutral massless moduli fields $u^i$ and massless $\U(1)$ vector fields $V^I$. 
Massless charged particles may appear at some points of the moduli space, but we focus our attention on generic points of the moduli space where
there are no such massless charged particles. The low energy theory can be empty. For example, there is a mass gap
in $\CN=1$ pure SYM and there are no massless particles at low energies below the confinement scale. Our discussion below includes such cases.

The low energy effective Lagrangian is given by 
\beq
\CL_{\rm eff}^{(4d)}=\int d^2\theta d^2\bar{\theta}^2 K_{\rm eff}^{(4d)}(u,u^\dagger)+\int d^2\theta W_{\rm eff}^{(4d)}(u)+\int d\theta^2 \frac{\tau_{IJ}(u)}{8\pi i} W^{\alpha I} W^J_\alpha+{\rm h.c.}, \label{eq:4deffective}
\eeq
where $\tau_{IJ}(u)=(4 \pi i/e^2 +\theta/2 \pi)_{IJ}$ is the holomorphic gauge coupling matrix of the massless gauge fields $V^I$.
The theory is very weakly coupled (almost free) in the IR.
For example, if the theory is $\CN=1$ pure SYM, there are no massless fields and we only have
a constant superpotential $W_{\rm eff}^{(4d)}=N \Lambda^3$ generated by gaugino condensation.

Now we compactify the theory on $S^1$ and perform dimensional reduction at the classical level. 
We take the radius $R$ to be very large so that the effective theory \eqref{eq:4deffective}
is valid at the scale of compactification. 
The kinetic term for $\U(1)$ gauge fields in 4d are
\beq
 \int d^4 x \CL_{\rm eff}^{(4d)} \supset  \int  \left(   \frac{\tau_{IJ}}{4\pi i}      F^I_{+}  \wedge * F^J_{+}    -
 \frac{\bar{\tau}_{IJ}}{4\pi i} F^I_{-} \wedge * F^J_{-}      \right)
\eeq
where $F_{\pm}=(F \pm *F)/2$, the $*$ means the Hodge star, and we are using Euclidean signature $(++++)$.
By dimensional reduction to 3d, we get 
\beq
&2\pi R \int d^3 x \CL_{\rm eff}^{(4d)} %\nonumber \\
\supset  2 \pi R \int  \left(     e^{-2}_{IJ} \left(  F'^I  \wedge * F'^J  +  R^{-2} d a^I \wedge * da^J   \right)
 -\frac{i\theta_{IJ}}{4\pi^2} R^{-1} F'^I \wedge d a^J  \right),
\eeq
where $F'^I$ is the gauge field strength in three dimensions, and $a^I=R A^I_3$ is the gauge field in the $S^1$ direction.
This $a^I$ is the $\U(1)$ Wilson loop in the $S^1$ direction $2 \pi a^I=\oint d x^3 A_3^I$ and it has a period 
$a^I \cong a^I+1$.

In ${\mathbb R}^{1,2} \times S^1$, we can dualize the vector multiplets to chiral multiplets. Let us see it for the bosonic fields. 
We consider $F'_I$ as a fundamental variable in the path integral, and change the action as
\beq
 \int & \left(     2 \pi Re^{-2}_{IJ} \left(  F'^I  \wedge *  F'^J  +  R^{-2} d a^I \wedge * d a^J   \right)
 -\frac{i\theta_{IJ}}{2\pi}  F'^I \wedge d a^J  + ib_I dF'^I \right).
\eeq
The $b_I$ is a Lagrange multiplier scalar field to impose the Bianchi identity $dF'^I=0$.
If there exist some monopole-like objects with a magnetic density $j^I$ such that $dF'^I=2\pi j^I$ and $\int j^I \in {\mathbb Z}$,
we may change the $b_I$ term in the above action as $i b_I (dF'^I -2\pi j^I)$. 
Then we can see that $b_I$ has a periodicity $b_I \cong b_I+1$ due to $\int j^I \in {\mathbb Z}$. (The argument here is
heuristic and not rigorous. See \cite{Seiberg:1996nz} for more rigorous treatment.)

Integrating over $F'^I$, we get
\beq
&\frac{1}{2R} \int  \left(    \left( \frac{4 \pi}{ e^2} \right)_{IJ}  d a^I \wedge * d a^J  +    
\left( \frac{ e^2}{4 \pi} \right)^{IJ} \left(db_I+ \frac{\theta_{IK}}{2 \pi} da^K  \right) \wedge * \left(db_J+ \frac{\theta_{JL}}{2 \pi} da^L  \right) 
   \right) \nonumber \\
&= \int \left(\frac{ e^2}{8\pi R} \right)^{IJ} d \varphi_I \wedge * \varphi_J^\dagger +\cdots, \label{eq:dualphotonkin}
\eeq
where ellipsis denotes terms involving derivatives of $\tau_{IJ}(u)$, and we have defined complex scalar fields $\varphi_I$ as
\beq
\varphi_I= b_I+ \tau_{IJ} a^J.
\eeq
Because of the periodicity of $a^I$ and $b_I$, the scalars $\varphi_I$ have the periodicity
\beq
\varphi_I \cong \varphi_I+m_I +\tau_{IJ} n^J, \label{eq:torusscalar}
\eeq
where $m_I$ and $n^I$ are integers. This means that the scalars $\varphi_I$ live on a complex torus (or more precisely an Abelian variety) with the complex 
structure $\tau_{IJ}$.

Therefore we get an effective 3d theory described by $u^i$ and $\varphi_I$ with the Kahler potential and superpotential given as
\beq
K^{(3d)}_{\rm eff} &=2 \pi R K_{\rm eff}^{(4d)}(u,u^\dagger)+\frac{1}{R} ((\Im \tau)^{-1})^{IJ} \Im \varphi_I \Im \varphi_J,\\
W^{(3d)}_{\rm eff} &=2 \pi R W_{\rm eff}^{(4d)}(u). \label{eq:3dWfrom4d}
\eeq
One can check that the above Kahler potential for $\varphi_I$ reproduces the kinetic term \eqref{eq:dualphotonkin} and 
it is invariant under \eqref{eq:torusscalar} up to irrelevant holomorphic+anti-holomorphic terms.
Therefore, the total moduli space $M$ is spanned by $u^i$ and $\varphi_I$.
It has a fiber structure $\pi: M \to B$ where $B$ is the moduli space of $u^i$, $B=\{u^i : \partial W_{\rm eff}^{(3d)}/ \partial u^i =0 \}$,
and the fiber $\pi^{-1}(u)$ is the torus spanned by $\varphi_I$. 
This is the same structure discussed in subsection~\ref{sec:generalizedH}, and we identify $M=M_{GH}$ and $B=B_{GH}$ if 
$\sigma$ is zero and there are no punctures. The fiber structure $\pi : M \to B$ is generally true due to the above field theory analysis.

\paragraph{Nonrenormalization.}
In the above discussion, we have compactified the effective theory at the classical level.
Now we argue that the superpotential $W_{\rm eff}^{(3d)}$ and other holomorphic quantities are not renormalized by the compactification.

First, let us show that the superpotential cannot depend on $\varphi_I$. 
The superpotential must be a holomorphic function of $\varphi_I$ and $u^i$.
The important point is that there is no strong coupling gauge dynamics below the scale of compactification, since
by our assumption, the theory consists of neutral moduli fields and massless $\U(1)$ fields at the energy scale of compactification
and there are no charged fields at low energies on generic points of the moduli space of $u^i$.\footnote{
If there is a cubic term in the superpotential, $W_{\rm eff}  \sim u^3$, the 3d theory gets strongly coupled at low energies.
However, we believe that this kind of strong coupling does not affect the following discussions.
}
Then a singularity cannot appear in the superpotential as a function of $\varphi_I$ for a fixed generic $u^i$,
and the superpotential must be holomorphic.
However, there is no holomorphic function on a complex torus (or compact complex manifolds in general) other than constants.
We conclude that $W_{\rm eff}^{(3d)}$ does not depend on $\varphi_I$. The crucial point in the above discussion
is that the gauge group of 4d theory is broken and/or confined already above the compactification scale and there are only 
$\U(1)$'s and neutral moduli fields.
If the unbroken gauge group was non-Abelian at the compactification scale and was broken down 
to $\U(1)$'s at or below the compactification scale, the story would be completely different.
See \cite{Aharony:2013dha,Aharony:2013kma} for careful discussions on such a case.

Next, we argue that there is no renormalization at all to the superpotential $W_{\rm eff}^{(3d)}(u)=2\pi RW_{\rm eff}^{(4d)}(u)$.
For this purpose, we use holomorphy and symmetry argument~\cite{Seiberg:1993vc}. 
In the UV, our 4d theory consists of matters and gauge multiplets. By ``matters'', we mean free chiral multiplets and
also some isolated superconformal theories such as the $T_N$ theory. Gauge multiplets are coupled
to global symmetries of the matters. If we turn off all gauge and superpotential interactions, 
the matter sector is $\CN=1$ superconformal and has chiral primary operators $O_a$ whose dimensions $\Delta_a$ and $R$-charges $R_a$ are related 
as $\Delta_a=\frac{3}{2}  R_a$~\cite{Flato:1983te,Dobrev:1985qv,Minwalla:1997ka}. 

Now let us turn on a UV superpotential as
\beq
W^{(4d)}_{\rm UV} =\sum_{a} \xi_a O_a.
\eeq
We also turn on gauge interactions. Each gauge group has a holomorphic dynamical scale (or one instanton factor) 
$\Lambda^b=\mu^b \exp(-8\pi^2/g^2+i\theta)$,
where $b$ is the coefficient of one-loop beta function and $\mu$ is a renormalization scale.
Here, ``one-loop'' beta function means the beta function of a gauge coupling
in the limit that all the gauge and superpotential couplings are going to zero. This definition is well-defined even for theories
without Lagrangian descriptions such as the $T_N$ theory. Contributions from non-Lagrangian sector can be parametrized by 
two-point current correlators.

We assign mass dimensions and $R$-charges to the parameters so that the interactions preserve the scaling and $R$ symmetries
discussed above.
From the superpotential, it is easy to see that $\xi_a$ has dimension $\Delta_{\xi_a}=3-\Delta_a$ and $R$-charge $R_{\xi_a}=2-R_a$,
and hence it satisfies $\Delta_{\xi_a}=\frac{3}{2}R_{\xi_a}$. The mass dimension of $\Lambda^b$ is evidently $b$.
The $R$-charge of $\Lambda^b$ is determined by anomaly. In general, $R$-symmetry becomes anomalous
when matters are coupled to gauge fields, and this anomaly can be cancelled by a shift of the theta angle $\theta$.
This shift determines the $R$-charge of $\Lambda^b=\mu^b \exp(-8\pi^2/g^2+i\theta)$.
It is known~\cite{Anselmi:1997am} that this $R$-charge is  given by $\frac{2}{3}b$.
Therefore, we conclude that all the holomorphic operators and parameters have the relation $\Delta=\frac{3}{2} R$.

Now, notice that the radius $R$ of the circle $S^1$ has mass dimension $-1$ and $R$-charge $0$.
Thus this parameter has a ``wrong'' relation between mass dimension and $R$-charge.
Note also that the $R$ is the only quantity which has wrong mass dimension and $R$-charge and 
also could possibly appear in the superpotential. (For example, wave function renormalizations also have wrong scaling dimensions 
due to quantum corrections in general, but
they can be extended to real vector superfields and cannot appear in holomorphic quantities.)
However, since $W^{(3d)}_{\rm eff}/(2 \pi R)$ has mass dimension $3$ and $R$-charge $2$, the radius 
$R$ cannot appear in this quantity to preserve the spurious symmetries.
In the decompactifying limit $R \to \infty$, we should recover the 4d effective superpotential $W^{(3d)}_{\rm eff}/(2 \pi R) \to W^{(4d)}_{\rm eff}$.
Since it is independent of $R$, we get the exact relation $W^{(3d)}_{\rm eff}/(2 \pi R) = W^{(4d)}_{\rm eff}$.
Therefore we have established the nonrenormalization of the superpotential in the compactification.
By the same reasoning, the vevs of holomorphic operators and $\tau_{IJ}$ do not depend on $R$ and they are not renormalized.

We stress again that $W^{(3d)}_{\rm eff}/(2 \pi R) $ is equal to the IR effective superpotential of the 4d theory, and not UV tree level superpotential.
For example, in the case of $\CN=1$ pure SYM, there is no UV superpotential. 
But the IR superpotential is generated by gaugino condensation as
$W^{(4d)}_{\rm eff}=N\Lambda^3$. Therefore we get $W^{(3d)}_{\rm eff} = 2 \pi R N\Lambda^3$.

In the above discussion we have assumed that the radius $R$ is very large.
But the result should be valid for all the values of $R$ under the assumption that there is no phase transition as we change $R$.

\subsection{(Non)renormalization in twisted 5d SYM}

Here we argue that we can use classical equations of twisted 5d SYM.
As discussed in section~\ref{sec:2}, the structure of supersymmetry in twisted 5d SYM is the same as that of 3d $\CN=2$ supersymmetry.
In 3d $\CN=2$ case, the gauge coupling $g_5$ can be extended to real vector multiplets after some redefinition of chiral fields.
This is because the field strength $\Tr W^\alpha W_\alpha$ can be written as $-2\Tr W^\alpha W_\alpha=\bar{D}^2( \Tr \Sigma^2)$
for the gauge invariant operator $\Tr \Sigma^2$, where $\Sigma = \frac{i}{4} {\bar D}^\alpha (e^{2i V} D_\alpha e^{-2i V})$ is gauge covariant 
in 3d $\CN=2$ supersymmetry.
Then the kinetic term for the gauge field can be written as an integral $\int d^2\theta d^2 \bar{\theta}^2$, and hence $g_5^{-2}$ can be extended into 
a background real vector field.
Therefore, holomorphic quantities do not receive quantum corrections of $g_5$.
The twisted 5d SYM has another parameter $L$ which is the length scale of the Riemann surface $C$.
Because of the relation $\Delta=\frac{3}{2}R$ of the previous subsection, the only quantity which could potentially appear 
is the combination $R/L \propto g_5^2/L$. Since there is no correction due to $g_5$, $L$ cannot also appear.
Thus we can take a limit $L \to \infty$ and $g_5^2 \propto R \to 0$ to compute field theory holomorphic quantities.
In this limit, the twisted 5d SYM can be treated classically.
However, note that we have to treat the $T_\rho[G]$ theories at punctures quantum mechanically.
For example, the Coulomb branch $\SU(N)_C$ symmetry of the $T[\SU(N)]$ theory appears only quantum mechanically
at the low energy fixed point of \eqref{eq:TSUNquiver}.

A possible loophole of the above argument is the following. 
Below the compactification scale of $C$, we get a 3d theory. If the gauge group $G$ is broken to $\U(1)$'s and there are no massless charged fields,
the 3d theory remains weakly coupled in the IR. However, if some non-Abelian groups remain and/or there are massless charged fields,
the theory becomes strongly coupled in the far IR by renormalization group flows in the low energy 3d theory.

\paragraph{3d mirrors.}
Actually, it is interesting to study more explicitly the case that the gauge group is unbroken at the length scale of $C$. 
Suppose that the vevs of the fields $\Phi_1, \Phi_2, A_{\bar{z}}$ and $V$ are negligible at the compactification
scale of $C$, and the gauge group $G$ is unbroken at this scale. Then the correct physical procedure is to do 
Kaluza-Klein reduction of fields,
\beq
 A_{\bar{z}}(x,\theta,z) &= \sum_n A_{\bar{z}}^{(n)}(x,\theta) \psi_{\bar{z}}^{(n)}(z), \\
\Phi_{i}(x,\theta,z) &= \sum_n \Phi_{i}^{(n)}(x,\theta) \psi_{i}^{(n)}(z), \\
V(x,\theta,z) &= \sum_n V^{(n)}(x,\theta) \psi_{V}^{(n)}(z), 
\eeq
where $\psi(z)$'s are wave functions on $C$, and $\theta$ is the superspace coordinate. 

For simplicity, we only consider the case $F=L_1 \otimes L_2$.
We denote holomorphic sections of $L_i$ and $K$ as $s^{(n)}_i(z)$ and $s^{(n)}_K(z)$.
Then, by taking only zero modes in the above Kaluza-Klein decomposition, we get
\beq
 A_{\bar{z}}(x,\theta,z) &\to \sum_{n=1}^{g} A_{\bar{z}}^{(n)}(x,\theta) (s_{K}^{(n)})^*(z), \\
\Phi_{i}(x,\theta,z) &\to \sum_{n=1}^{g_i} \Phi_{i}^{(n)}(x,\theta) s_{i}^{(n)}(z), \\
V(x,\theta,z) & \to V(x,\theta) , 
\eeq
where $g_i=\dim H^0(C,L_i)$ and $g=\dim H^0(C,K)$.
Therefore, we get a 3d theory composed of the gauge multiplet $V$ of the gauge group $G$, 
adjoint chiral multiplets $A_{\bar{z}}^{(n)}$ and $ \Phi_{i}^{(n)}$,
and $T_\rho[G]$ theories.
Here we have assumed that the gauge symmetry of $A_{\bar{z}}$ on $C$ is fixed in an appropriate way
so that $A_{\bar{z}}$ can be treated just as matter field.

For example, let us consider the 3d $\CN=4$ case in which $L_1$ is the trivial bundle and $L_2=K$.
Then, the theory is an 3d $\CN=4$ theory with 
$g$ hypermultiplets $(A_{\bar{z}}^{(n)},\Phi_{2}^{(n)})~(n=1,\cdots,g)$ and the vector multiplet $(V,\Phi_1)$ 
coupled to $T_\rho[G]$ theories. Note that the couplings between $T_\rho[G]$ and $(V,\Phi_1)$ at a puncture $z=z_p$ become
just standard couplings of matters and vector multiplets of 3d $\CN=4$ theory after the dimensional reduction.

The above 3d theory has been obtained in the chain of dimensional reduction
\beq
&{\rm (2,0)~theory~on~}{\mathbb R}^{1,2} \times S^1 \times C \nonumber \\
\to& {\rm ~5d~SYM~on~}{\mathbb R}^{1,2}  \times C \nonumber \\
\to & {\rm ~3d~theory~on~}{\mathbb R}^{1,2}. \nonumber 
\eeq
On the other hand, we can also consider a 3d theory obtained as
\beq
&{\rm (2,0)~theory~on~}{\mathbb R}^{1,2} \times S^1 \times C \nonumber \\
\to& {\rm ~4d~theory~on~}{\mathbb R}^{1,2}  \times S^1 \nonumber \\
\to & {\rm ~3d~theory~on~}{\mathbb R}^{1,2}. \nonumber 
\eeq
Assuming that the above two processes lead to the same IR fixed point, we get 3d mirror symmetry between the above two theories.
Actually, the theory with $(V,\Phi_1)$,  $(A_{\bar{z}}^{(n)},\Phi_{2}^{(n)})~(n=1,\cdots,g)$ and
$T_\rho[G]$ was really obtained as the 3d mirror of the low energy limit of 4d theory on $S^1$ \cite{Benini:2010uu}
by using a different (but related) method. Our method may also give a large class of 3d $\CN=2$ mirrors.
Obviously it would be interesting to investigate it further, which we leave for future work.

\paragraph{Comment on (2,0) theory and 5d SYM.}
Before closing this section, let us comment on the relation between the $\CN=(2,0)$ theories and 5d SYM.
Throughout this paper we are assuming the existence of the $\CN=(2,0)$ theories and discussing its implications on 4d field theories.
However, we are only using 5d SYM on $C$ by forgetting about the compactified $S^1$ direction.  
Then, very naively, the moduli space of solutions of twisted 5d SYM might seem to correspond to 
the moduli space of a genuine 3d theory and not 4d theory on $S^1$, since we are forgetting the existence of $S^1$.
However, it is not the case. There is a crucial difference between a genuine 3d theory and a 4d theory compactified on $S^1$.
In a genuine 3d theory, a scalar field in a vector multiplet is just a scalar and there is no periodicity.
However, in a 4d theory on $S^1$, this scalar comes from the component of gauge field in the $S^1$ direction
and it has a periodicity due to gauge symmetry. Moduli spaces of Hitchin systems, which are derived from twisted 5d SYM, reproduce
this periodicity. This is because a generic fiber $\pi^{-1}(p)$ of the moduli spaces of Hitchin systems discussed in subsection~\ref{sec:spectral}
is a complex torus, and the fact that it is a torus is closely related to the periodicity \eqref{eq:torusscalar} which comes from gauge symmetries.
Therefore, in a sense, the 5d SYM ``remember'' the existence of the $S^1$ direction of the $\CN=(2,0)$ theories.
This is consistent with the proposal~\cite{Douglas:2010iu,Lambert:2010iw} that 
all the degrees of freedom of the $\CN=(2,0)$ theories on $S^1$ are contained in 5d SYM.

%%%%%%%%%%%%%%%%%%%%%%%%%%%%%%%%%%%%%%%%%%%%%%%%%%%%%%%%%%%
\section{ Higgs branch of ${  \CN=2}$ theories}\label{sec:5}

In this section, we apply twisted 5d SYM to Higgs branches of 4d $\CN=2$ theories.
We do not aim to determine the complete structure of Higgs branches. Rather than that,
our main purpose is to understand the role of the adjoint scalar $\sigma$ in the vector multiplet $V$ of twisted 5d SYM.
As we will see explicitly in section~\ref{sec:6}, twisted 5d SYM becomes most powerful when $\sigma$ is forced to be zero
by the vevs of other fields, but $\CN=2$ Higgs branches allow nonzero vevs $\sigma$ and we can get some insight about them
by studying $\CN=2$ Higgs branches. 
However, we will reproduce the moduli spaces on generic points of the Higgs branches.
We will also see that our formalism can be used to derive chiral ring relations involving holomorphic moment maps
for arbitrary gauge groups $G=A,D,E$.

\subsection{Field theory}
In this subsection we only consider the case $G=\SU(N)$.
Theories we are going to study are the generalized quiver gauge theories introduced in \cite{Gaiotto:2009we}
which can be constructed by copies of the $T_N$ theory. (This $T_N$ theory is different from the $T[\SU(N)]$ theory discussed in section~\ref{sec:3}.)
The results in this subsection are obtained in \cite{Maruyoshi:2013hja} 
(see also \cite{Hanany:2010qu,Tachikawa:2011ea}) and we review them for completeness. 

Let us first recall a few properties of the $T_N$ theory. It has flavor $\SU(N)_A \times \SU(N)_B \times \SU(N)_C$ 
symmetries. There are Higgs branch chiral operators $\mu_A$, $\mu_B$ and $\mu_C$ in the adjoint representations of the flavor groups
$\SU(N)_A$, $\SU(N)_B$ and $\SU(N)_C$ respectively. They are the holomorphic moment maps of the respective flavor groups.
The $T_N$ theory also has chiral operators $Q^{i_A i_B i_C}$ and $Q_{i_A i_B i_C}$
which are trifundamental and anti-trifundamental representations of 
$\SU(N)_A \times \SU(N)_B \times \SU(N)_C$ respectively~\cite{Gaiotto:2009gz}.\footnote{ 
There are more general operators $Q_{(k)}~(k=2,\cdots,N-1)$ in the $T_N$ theory~\cite{Maruyoshi:2013hja}, but we will not discuss them.
We believe that their existence does not affect the conclusions in this subsection.}
Here $i_A$, $i_B$ and $i_C$ are flavor indices. 

In the case $N=2$, the $T_2$ theory is just eight free chiral multiplets $Q^{i_A i_B i_C}$
in the trifundamental representation of $\SU(2)_A \times \SU(2)_B \times \SU(2)_C$.
In this case, $Q_{i_A i_B i_C} \propto \epsilon_{i_A j_A}\epsilon_{i_B j_B}\epsilon_{i_C j_C}Q^{j_A j_B j_C}$,
$(\mu_A)^{i_A}_{~j_A} \propto Q^{ i_A i_B i_C}Q_{j_A i_B i_C}$ and so on.

\paragraph{Chiral ring relations.}
There are many chiral ring relations of the operators~\cite{Gaiotto:2008nz,Benini:2009mz,Maruyoshi:2013hja}. 
First, let us define the characteristic polynomials of matrices $\mu_{A,B,C}$ as
\beq
P_X(x)=\det (x -\mu_X) 
~~~(X=A,B,C).
\eeq
%where $v_{X,0}=1$ and $v_{X,1}=0$. 
Then the chiral ring relations we will use are given as~\cite{Maruyoshi:2013hja}
\beq
P_A(x)=P_B(x)=P_C(x) \equiv P(x),  \label{eq:CR1} %~\left( {\rm or}~\tr \mu_1^k=\tr \mu_2^k=\tr \mu_3^k \right),
\eeq
\beq
(\mu_A)^{i_A}_{~j_A} Q^{j_A i_B i_C}=(\mu_B)^{i_B}_{~j_B} Q^{i_A j_B i_C}=(\mu_C)^{i_C}_{~j_C} Q^{i_A i_B j_C}, \nonumber  \\
(\mu_A)^{j_A}_{~i_A} Q_{j_A i_B i_C}=(\mu_B)^{j_B}_{~i_B} Q_{i_A j_B i_C}=(\mu_C)^{j_C}_{~i_C} Q_{i_A i_B j_C}, \label{eq:CR2}
\eeq
\beq
\left(Q^{i_A i_B i_C}Q_{j_A j_B i_C} \right)=
\left[ \left( \frac{P(x)-P(y)}{x-y} \right)(x=\mu_A \otimes {\bf 1}, ~y={\bf 1} \otimes \mu_B) \right]^{i_A i_B}_{j_A j_B} ,\label{eq:CR3}
%\sum_{k=0}^{N} v_{k} \sum_{m=0}^{N-k-1 } (\mu_A^{N-k-1-m})^{i_A}_{~j_A}  (\mu_B^m)^{i_B}_{~j_B}, 
\eeq
\beq
&\frac{1}{N!}Q^{i_{A,1}i_{B,1} i_{C,1}}\cdots 
Q^{i_{A,N}i_{B,N} i_{C,N}} \epsilon_{i_{B,1} \cdots  i_{B,N}}\epsilon_{i_{C,1} \cdots  i_{C,N}} 
= (\mu_A^0)^{(i_{A,1}}_{~j_{A,1}} \cdots  (\mu_A^{N-1})^{i_{A,N})}_{~j_{A,N}}
\epsilon^{j_{A,1} \cdots  j_{A,N}} ,\label{eq:CR4}
\eeq
%\beq
%&\frac{1}{(N-1)!}Q^{i_{A,1}i_{B,1} i_{C,1}}\cdots 
%Q^{i_{A,N-1}i_{B,N-1} i_{C,N-1}} \epsilon_{i_{B,1} \cdots i_{B,N-1} i_B}\epsilon_{i_{C,1} \cdots i_{C,N-1} i_C}  \nonumber \\
%=&Q_{i_A i_B i_C} (\mu_A^0)^{(i_{A,1}}_{~j_{A,1}}\mu_A^1)^{i_{A,2}}_{~j_{A,2}} \cdots (\mu_A^{N-2})^{i_{A,N-1})}_{~j_{A,N-1}} 
%\epsilon^{j_{A,1}j_{A,2} \cdots j_{A,N-1} i_A} 
%\eeq
%\beq
%&\frac{(-1)^{\frac{1}{2} N(N-1)}}{(N-1)!}Q_{i_{A,1}i_{B,1} i_{C,1}}\cdots 
%Q_{i_{A,N-1}i_{B,N-1} i_{C,N-1}} \epsilon^{i_{B,1} \cdots i_{B,N-1} i_B}\epsilon^{i_{C,1} \cdots i_{C,N-1} i_C} \nonumber \\
%=&Q^{i_A i_B i_C} (\mu_A^0)^{(j_{A,1}}_{~i_{A,1}}\mu_A^1)^{(j_{A,2}}_{~i_{A,2}} \cdots (\mu_A^{N-2})^{j_{A,N-1})}_{~i_{A,N-1}} 
%\epsilon_{j_{A,1}j_{A,2} \cdots j_{A,N-1} i_A}
%\eeq
where $\mu^k_X$ is the $k$-th power of the matrix $\mu_X$.
The meaning of \eqref{eq:CR3} is that we first compute a polynomial of $x$ and $y$ given as $(P(x)-P(y))/(x-y)$ and then substitute 
matrices $x=\mu_A \otimes {\bf 1}$ and $y={\bf 1} \otimes \mu_B $ and evaluate components $(i_A i_B j_A j_B)$.
The last equation \eqref{eq:CR4} is not explicitly written in \cite{Maruyoshi:2013hja}, 
but can be derived from more fundamental chiral ring relations written there.

\paragraph{Higgs branch of the $T_N$ theory.}
Let us use the above chiral ring relations to study the Higgs branch of the $T_N$ theory.
First, \eqref{eq:CR1} tells us that the eigenvalues of $\mu_A$, $\mu_B$ and $\mu_C$ are the same.
Assuming that the eigenvalues are generic, we have
\beq
U_X \mu_X U_X^{-1}=\diag(\lambda_1, \cdots, \lambda_N) \equiv \lambda~~(X=A,B,C),\label{eq:momenteigen}
\eeq
where $U_X \in \SL(N)_X$ and $\sum_{k=1}^N \lambda_k=0$. 

We define 
\beq
&\tilde{Q}^{i_A i_B i_C}=(U_A)^{i_A}_{~j_A}(U_B)^{i_B}_{~j_B}(U_C)^{i_C}_{~j_C}{Q}^{j_A j_B j_C}, \\~~~
&\tilde{Q}_{i_A i_B i_C}=(U_A^{-1})^{j_A}_{~i_A}(U_B^{-1})^{j_B}_{~i_B}(U_C^{-1})^{j_C}_{~i_C}{Q}_{j_A j_B j_C}.
\eeq
Using \eqref{eq:CR2}, we can see that the only nonzero components of $\tilde{Q}^{i_A i_B i_C}$ and $\tilde{Q}_{i_A i_B i_C}$ are given by
$i_A=i_B=i_C$,
\beq
\tilde{Q}^{kkk}=q^k, ~~~~~\tilde{Q}_{kkk}=q_k,\label{eq:Qeigen}
\eeq
and other components are zero.

Using \eqref{eq:CR3}, we get
\beq
q^k q_k=\prod_{\ell \neq k}(\lambda_k-\lambda_\ell),  \label{eq:qqrelation}
\eeq
where there is no sum or product over $k$.
Thus, all the $q_k$ are fixed in terms of $q^k$ and $\lambda_\ell$. Furthermore, \eqref{eq:CR4} gives us
\beq
\prod_{k=1}^{N} q^k= \prod_{1 \leq k<\ell \leq N} (\lambda_\ell-\lambda_k). \label{eq:qNconst}
\eeq
Therefore, there are only $N-1$ independent moduli parameters in $q^k~(k=1,\cdots,N)$.
The Higgs branch of the $T_N$ theory is spanned by $\lambda_k$, $q^k$,
and $U_X~(X=A,B,C)$.

\begin{figure}
\begin{center}
\includegraphics[scale=0.25]{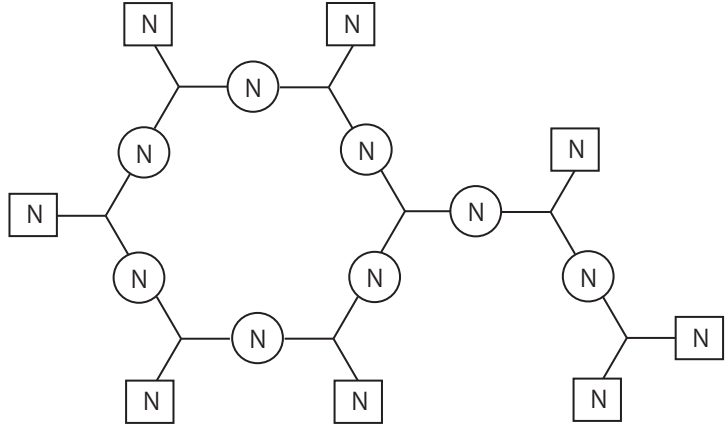}
\caption{Generalized quiver gauge theory. Trivalent vertices are copies of the $T_N$ theory,
circles are $\CN=2$ $\SU(N)$ vector multiplets, and boxes are flavor $\SU(N)$ symmetries.
In this example, there is $g=1$ loop, and there are $n=8$ flavor groups.
The $g$ and $n$ correspond to the genus of the Riemann surface and the number of punctures, respectively,
in the corresponding twisted 5d SYM.}\label{fig:quiver}
\end{center}
\end{figure}

\paragraph{Higgs branch of generalized quiver.}
Now we study Higgs branches of generalized quiver gauge theories as in figure~\ref{fig:quiver}.
We take copies of the $T_N$ theory glued by $\CN=2$ vector multiplets.
Each vector multiplet is coupled to two copies of the $T_N$ theory.

Each trivalent vertex represents a copy of the $T_N$ theory, $T_N^{(V)}$.
Each internal line with a circle inserted represents an $\CN=2$ vector multiplet, $(V_{(I)}, \phi_{(I)})$, with the gauge group $\SU(N)_{(I)}$.
Each external line connected to a box represents a flavor group $\SU(N)_{(E)}$.

The holomorphic moment maps of $T_N^{(V)}$ are denoted as $\mu_{(V,I)}$ or $\mu_{(V,E)}$ depending on the group
$\SU(N)_{(I)}$ or $\SU(N)_{(E)}$ of which they are adjoint representations.
Similarly, we denote $Q$ operators of the $T_N^{(V)}$ theory as e.g., $Q_{(V)}^{i_I i_{I'} i_E}$ if the $T_N^{(V)}$
is connected to $\SU(N)_I$, $\SU(N)_{I'}$ and $\SU(N)_E$.

Two copies of the $T_N$ theory, say $T_N^{(V)}$ and $T_N^{(V')}$, are glued as follows.
Take a flavor symmetry $\SU(N)_{(V,I)}$ of $T_N^{(V)}$ and $\SU(N)_{(V',I)}$ of $T_N^{(V')}$.
Then, we gauge the diagonal subgroup $\SU(N)_{(I)} \subset \SU(N)_{(V,I)} \times \SU(N)_{(V',I)}$ given as
\beq
\SU(N)_{(I)} \ni g \mapsto (g,  {}^tg^{-1}) \in\SU(N)_{(V,I)} \times \SU(N)_{(V',I)},
\eeq
where the superscript $t$ means transpose.
Corresponding to this gauging, the superpotential for the adjoint chiral multiplet $\phi_{(I)}$ is given as
\beq
W \supset \sqrt{2} \tr \phi_{(I)} \left( \mu_{(V,I)}-{}^t\mu_{(V',I)} \right).
\eeq
The minus sign and the transpose in the second term is the result of the above embedding of $\SU(N)_{(I)}$.

Let us study the Higgs branch of the theory. Equations of motion of $\phi_{(I)}$ give $\mu_{(V,I)}={}^t\mu_{(V',I)} $.
This equation says in particular that the eigenvalues of $\mu_{(V,I)}$ and ${}^t\mu_{(V',I)}$ are the same.
Combined with the result \eqref{eq:momenteigen} for a single copy of the $T_N$ theory, we get
\beq
U_{(V,I)} \mu_{(V,I)} U_{(V,I)}^{-1}=U_{(V,E)} \mu_{(V,E)} U_{(V,E)}^{-1}=\diag(\lambda_1, \cdots, \lambda_N) \equiv \lambda,
\eeq
for all $I$ and $E$. Then, as in \eqref{eq:Qeigen}, we get
\beq
\tilde{Q}_{(V)}^{kkk}=q_{(V)}^k,~~~~~(\tilde{Q}_{(V)})_{kkk}=(q_{(V)})_k.
\eeq
The $(q_{(V)})_k$ are determined as in \eqref{eq:qqrelation} and $q_{(V)}^k$ satisfy the relation \eqref{eq:qNconst}.

The vevs of $\mu$'s break each $\SU(N)_{(I)}$ gauge symmetry to the Cartan $\U(1)_{(I)}^{N-1}$ subgroup.
The vevs of $q_{(V)}^k$ further break these $\U(1)$ gauge symmetries. 
Suppose that the vertex $T_N^{(V)}$ is connected to $\SU(N)_{(I)}$, $\SU(N)_{(J)}$ and $\SU(N)_{(K)}$ gauge groups.
Then, the vev of $q_{(V)}$ imposes that massless vector multiplets satisfy
\beq
(-1)^{h(V,I)}V_{(I)}+(-1)^{h(V,J)}V_{(J)}+(-1)^{h(V,K)}V_{(K)}=0
\eeq
where $h(V,I)=0$ if $\SU(N)_{(I)}$ is embedded in $\SU(N)_{(V,I)}$ as $g \mapsto g$, and $h(V,I)=1$ if it is embedded as $g \mapsto {}^tg^{-1}$.
There are similar constraints if some of the $\SU(N)_{(I)}$, $\SU(N)_{(J)}$ and $\SU(N)_{(K)}$ are flavor groups.
From these constraints, one can see that there is unbroken $\U(1)^{N-1}$ gauge group for each loop of the generalized quiver diagram.
Denoting the number of loops as $g$, we get $(N-1)g$ massless $\U(1)$ vector fields.
Therefore, gauge symmetry cannot be completely Higgsed for $g>0$ and it is actually a mixed Higgs-Coulomb branch.

Each $q_{(V)}^k$ is not gauge invariant. We can construct gauge invariant operators as
\beq
q_{\rm tot}^k=\prod_{V} q_{(V)}^k,
\eeq
where the product is over all the vertices. We also need to divide the space by an appropriate Weyl group.

Let us summarize what we have found.

\begin{enumerate}
\item There is a set of eigenvalues $\lambda=\diag(\lambda_1, \cdots, \lambda_N)~~(\sum_k \lambda_k=0)$ which is a part of the moduli fields. 
The holomorphic moment map of the flavor symmetry $\SU(N)_{(V,E)}$ is in the orbid
\beq
\mu_{(V,E)} \in {\cal O}_\lambda^{(E)} \equiv \{ U^{-1}_{(V,E)} \lambda U_{(V,E)}  \}. \label{eq:fieldtheoryorbit}
\eeq
These orbits for flavor symmetries $\SU(N)_{(V,E)}$ also contribute to the moduli space.

\item There are $N$ gauge invariant operators $q_{\rm tot}^k~(k=1,\cdots,N)$, up to Weyl group actions. They satisfy one constraint,
\beq
\prod_{k=1}^{N}q_{\rm tot}^k=\left( \prod_{1 \leq k<\ell \leq N} (\lambda_\ell-\lambda_k) \right)^{N_V}.\label{eq:qtotprod}
\eeq
where $N_V$ is the number of vertices.
Therefore, $N-1$ of them are independent and contributes to the dimension of the moduli space.

\item There are $(N-1)g$ massless $\U(1)$ vector multiplets.

\end{enumerate}

\subsection{Twisted 5d SYM}\label{sec:N2twisted}
We study the same system using the twisted 5d SYM. We take a genus $g$ Riemann surface with maximal punctures labelled by $E$,
corresponding to the flavor symmetries $\SU(N)_{(E)}$ in the field theory.
At each puncture there is a copy of the $T[G]$ theory, which will be denoted as $T[G]_{(E)}$.

In the twisted 5d SYM, $\Phi_1$ is a section of ${\rm ad}(E)$ and $\Phi_2$ is a section of $K \otimes {\rm ad }(E)$.
In this case, the following set of equations hold;
\beq
0&= F_{z\bar{z}} - [\Phi_{2}, \bar{\Phi}_{2} ]      ,  \label{eq:N2HitchinD}\\
0&= D_{\bar z} {\Phi}_2, ,\label{eq:N2HitchinF}  \\
0&=D_{\bar{z}} \sigma=D_{\bar{z}} \Phi_1=D_{\bar{z}} \bar{\Phi}_1 , \label{eq:N2Hitchinhid1} \\
0&=[\sigma,\Phi_2]=[\Phi_1, \Phi_2]=[\bar{\Phi}_1, \Phi_2] , \label{eq:N2Hitchinhid2} \\
0&=[\sigma,\Phi_1]=[\Phi_1, \bar{\Phi}_1] , \label{eq:N2Hitchinhid3} 
\eeq
and complex conjugates of some of them. At the points where there are punctures, delta function source terms as in section~\ref{sec:3} need to be included,
but we do not write them explicitly.
These equations can be derived from
the equations \eqref{eq:HitchinD}-\eqref{eq:Hitchinhid2} as follows.
We consider the trace of the square of \eqref{eq:HitchinD},
\beq
0=&\sqrt{g} \Tr \left(g^{z\bar{z}}F_{z\bar{z}} - g^{z\bar{z}}[\Phi_{2}, \bar{\Phi}_{2} ]-[\Phi_{1}, \bar{\Phi}_{1} ]  \right)^2 \nonumber \\
=&\sqrt{g^{-1}} \Tr \left(F_{z\bar{z}} - [\Phi_{2}, \bar{\Phi}_{2} ] \right)^2+\sqrt{g} \Tr \left([\Phi_{1}, \bar{\Phi}_{1} ] \right)^2
-2\Tr \left(\left(F_{z\bar{z}} - [\Phi_{2}, \bar{\Phi}_{2} ] \right)[\Phi_1, \bar{\Phi}_1] \right).
\eeq
By a little computation, we get
\beq
&\Tr (F_{z\bar{z}}[\Phi_1, \bar{\Phi}_1])=\Tr ( \bar{\Phi}_1 ([D_{z},D_{\bar{z}}]\Phi_1) ),\\
&\Tr \left([\Phi_{2}, \bar{\Phi}_{2} ] [\Phi_1, \bar{\Phi}_1] \right)
=\Tr \left( [\Phi_1, \bar{\Phi}_2][{\Phi}_2,\bar{\Phi}_1] \right)+\Tr \left( [\Phi_1, {\Phi}_2][\bar{\Phi}_1, \bar{\Phi}_2] \right).
\eeq
Then, by using $D_{\bar{z}} \Phi_1=0 $ and $[\Phi_1, \Phi_2]=0$, we obtain
\beq
0=&\int |d^2z|\sqrt{g^{-1}} \Tr \left(F_{z\bar{z}} - [\Phi_{2}, \bar{\Phi}_{2} ] \right)^2+\int |d^2z| \sqrt{g} \Tr \left([\Phi_{1}, \bar{\Phi}_{1} ] \right)^2 \nonumber \\
&-2\int |d^2z| \Tr ( D_{\bar{z}}\bar{\Phi}_1 D_{{z}}\Phi_1 )+2\int |d^2z| \Tr \left( [\Phi_1, \bar{\Phi}_2][{\Phi}_2,\bar{\Phi}_1] \right), \label{eq:Dsquared}
\eeq
where the integral is performed excluding infinitesimally small regions around punctures.
In using integration by parts, we have used the fact that $\Phi_1$ is not singular at punctures.
Recalling that our definition is such that $\bar{\Phi}=-\Phi^\dagger$, we can see that each term in \eqref{eq:Dsquared} is non-negative.
Therefore, \eqref{eq:Dsquared} can be zero if and only if \eqref{eq:N2HitchinD}-\eqref{eq:N2Hitchinhid3} are satisfied.
This result is expected since there is an $\SU(2)$ $R$-symmetry which rotates $\sqrt{2}\Re \Phi_1$, $\sqrt{2}\Im \Phi_1$ and $\sigma$
as a triplet.

Let us study the moduli space of solutions of \eqref{eq:N2HitchinD}-\eqref{eq:N2Hitchinhid3}.
For the moment we restrict our attention to $G=\SU(N)$, although it is straightforward to extend the results to general gauge groups.
Because of the commutation relations \eqref{eq:N2Hitchinhid3}, we can simultaneously diagonalize $\Phi_1$ and $\sigma$ by a gauge transformation.
Furthermore, \eqref{eq:N2Hitchinhid1} says that the eigenvalues are constant. Let us set
\beq
\lambda=\diag(\lambda_1, \cdots,\lambda_N)=\frac{\sqrt{2}}{ 8 \pi^2 R } \Phi_1.
\eeq
We assume that $\lambda_k$ are generic.

At each puncture, there is an operator $\mu^{(4d)}_{(E)}=(\mu^{(3d)}_C)_{(E)}/2\pi R$ which is the holomorphic moment map of
the flavor $\SU(N)_{(E)}$ symmetry of $T[\SU(N)]_{(E)}$ as explained in section~\ref{sec:3}. 
As shown in \eqref{eq:precisecorres}, for generic $\lambda$ we have
\beq
\mu^{(4d)}_{(E)}=U^{-1}_{(E)}\lambda U_{(E)}, \label{eq:holomomentrelation}
\eeq
for $U_{(E)} \in \SL(N)_{(E)}$. This is exactly the structure \eqref{eq:fieldtheoryorbit} obtained in the field theory.

Next, let us consider $(A_{\bar{z}},\Phi_2)$. The equations \eqref{eq:N2Hitchinhid1} and \eqref{eq:N2Hitchinhid2}
impose that $(A_{\bar{z}},\Phi_2)$ be in the Cartan subalgebra when $ \Phi_1$ is generic.
The unbroken gauge group is $\U(1)^{N-1}$.
In this case, \eqref{eq:N2HitchinD} requires $F_{z\bar{z}}=0$, so $A_{\bar{z}}$ is a flat connection.
Equivalently, we may forget \eqref{eq:N2HitchinD} and divide the space of $A_{\bar{z}}$ by the complexified gauge group. 
This complexified group can be used to set $A_{\bar{z}}=0$, and hence the connections define
$N-1$ holomorphic line bundles.
Therefore the moduli space of $A_{\bar{z}}$ divided by the gauge group is given by the moduli space of these holomorphic line bundles.

For simplicity, let us pretend as if the gauge group is $\U(N)$ instead of $\SU(N)$, and it is broken to $\U(1)^N$.
The moduli space of each $\U(1)$ line bundle is given by the Jacobian variety of $C$, which we denote as $J(C)$.
The $J(C)$ is the space of Wilson loops $\exp(-\oint_{\gamma} A)$ for flat $\U(1)$ connections $A$
satisfying $F_{z\bar{z}}=0$, where $\gamma$ is one of the $2g$ cycles in $C$. 
Thus $J(C)$ is a torus with real dimension $2g$ or complex dimension $g$.\footnote{
The $J(C)$ for a Riemann surface $C$ as a complex manifold is 
explicitly given as follows. Let $\alpha_I$ and $\beta^I$ $(I=1,\cdots,g)$ be the usual real basis of closed one forms 
$H^1(C,{\mathbb Z})$ on $C$, satisfying e.g., 
$\int_C \alpha_I \wedge \beta^J=\delta_I^J$, $\int_C \alpha_I \wedge \alpha_J=\int_C \beta^I \wedge \beta^J=0$. 
Then a holomorphic basis $\lambda_I \in H^{1,0}(C)$, $\partial_{\bar{z}} \lambda_I=0$,
is given as $\lambda_I=\alpha_I+\tau_{IJ}\beta^J$ for some $\tau_{IJ}$ which is a symmetric matrix $\tau_{IJ}=\tau_{JI}$, with $\Im \tau_{IJ}$ positive definite.
A flat $\U(1)$ connection is parametrized as $A=2\pi i (a^I \alpha_I-b_I\beta^I)$ for parameters $a^I$ and $b_I$
with periodicity $a^I \cong a^I+1$, $b_I \cong b_I+1$ coming from large gauge transformations. The anti-holomorphic part of $A$ 
is given as $A_{\bar{z}} d\bar{z} =(b_I+\tau_{IJ}a^J) ( \pi (\Im \tau)^{-1,JK} \bar{\lambda}_K) $.
Therefore, we can take complex coordinates of $J(C)$ as $\varphi_I=b_I+\tau_{IJ}a^J$ which have the periodicity
$\varphi_I \cong \varphi_I +m_I+\tau_{IJ}n^J$ for $m_I, n^J \in {\mathbb Z}$.
}
Now, the moduli space of the $\U(1)^N$ connections $A_{\bar{z}}$ is given by $J(C)^{N}=J(C) \times \cdots \times J(C)$.
Let the curve $\Sigma=C+\cdots+C$ be $N$ copies of disconnected $C$'s. Then $J(\Sigma)=J(C)^N$,
i.e., the moduli space of $A_{\bar{z}}$ is given by the Jacobian variety of the curve $\Sigma$.
The traceless condition in the $\SU(N)$ requires that the actual moduli space is the $(N-1)g$ dimensional subspace of the $Ng$
dimensional space $J(\Sigma)$.
This space is identified with the moduli space of the $(N-1)g$ massless vector fields found in the field theory,
after dualizing them to complex scalars as explained in subsection~\ref{sec:fieldnonren}.

The $\Phi_2$ is expanded as $\Phi_2=\sum_{k=1}^{N-1} \Phi_{2,k}H_k$, where $H_k$ are generators of the Cartan subalgebra. 
Each $\Phi_{2,k}$ is invariant under $\U(1)^{N-1}$, and they are holomorphic sections of the canonical bundle $K$.
Since $\dim H^0(C,K)=g$, there are $(N-1)g$ moduli parameters in $\Phi_2$. 
These are the 4d $\CN=2$ superpartners of the massless vector fields discussed above.

The remaining field in the twisted 5d SYM is the vector multiplet $V$, which contains the $\sigma$ and 3d vector field $A_\mu~(\mu=0,1,2)$.
The zero modes of these fields do not depend on $z$, and we consider them as 3d fields. They are also diagonal,
\beq
\sigma=\diag(\sigma_1,\cdots,\sigma_N), ~~~~~
A_\mu=\diag((A_1)_\mu,\cdots, (A_N)_\mu).
\eeq
For simplicity, let us again pretend as if the gauge group is $\U(N)$ and there are no traceless conditions on $\sigma_k$ and $(A_k)_\mu$.
Taking the dual of $(A_k)_\mu$, we get real scalars $\rho_k$ with the period $\rho_k \cong \rho_k+1$.
They are combined with $\sigma_k$ to form chiral fields as
\beq
\varphi_k=\rho_k+\frac{2i{\cal A}}{g^2_5}  \sigma_k=\rho_k+\frac{i{\cal A}}{4\pi^2}  \sigma_k^{(6d)}
\eeq
where ${\cal A}=\int \sqrt{g} |d^2 z|$ is the area of the Riemann surface $C$, $g_5^2=8\pi^2R$ is the 5d gauge coupling,
and $\sigma^{(6d)}=R^{-1} \sigma$.
The traceless condition in $\SU(N)$ is imposed as $\sum_{k=1}^{N} \varphi_k=0$.

Because of the periodicity $\rho_k \cong \rho_k+1$, we may define 
\beq
\tilde{q}^k=\exp (2 \pi i \varphi_k).
\eeq
Then, the traceless condition is translated into
\beq
\prod_{k=1}^{N} \tilde{q}^k=1.
\eeq
This constraint looks similar to the constraint \eqref{eq:qtotprod} found in the field theory.
Therefore, we may roughly identify 
\beq
\tilde{q}^k \sim q_{\rm tot}^k.  
\eeq
Probably the relation between them may be something like 
$q_{\rm tot}^k=\tilde{q}^k \prod_{\ell \neq k}(\lambda_k -\lambda_\ell)^{\frac{N_V}{2}}$.
We leave it for future work to determine the precise relation.
At least the dimension of the moduli space matches between
the field theory and the twisted 5d SYM.
Our conclusion is that the vector multiplet $V \ni (\sigma,A_\mu)$ contains the information about the moduli fields
contained in the operators $Q^{i_1i_2i_3}$ of the $T_N$ theory.

\paragraph{General gauge groups.}
The moduli space of the twisted 5d SYM for a general group $G$ can be studied in the same way as is done above.
The result should have implications for the corresponding field theory, which have not yet been fully investigated in the literature. 
Here we simply note a simple consequence of our result. As discussed in subsection~\ref{sec:generalgroup}, 
the equation \eqref{eq:holomomentrelation} is valid for generic $\lambda$,
and hence we get the following chiral ring relation. Let $E$ and $E'$ be two punctures on the Riemann surface.
The holomorphic moment maps $\mu^{(4d)}_{(E)}$ and $\mu^{(4d)}_{(E')}$ of the flavor groups $G^\vee_{(E)}$ and $G^\vee_{(E')}$
at the punctures satisfy the chiral ring relation
\beq
p\left(\mu^{(4d)}_{(E)} \right)=p\left(\mu^{(4d)}_{(E')} \right),
\eeq
where $p$ is any invariant polynomial of the Lie algebra of $G^\vee$.

%%%%%%%%%%%%%%%%%%%%%%%%%%%%%%%%%%%%%%%%%%%%%%%%%%%%%%%%%%%
\section{Examples of ${\cal N}=1$ theories}\label{sec:6}
Having done enough preparations, we can finally discuss dynamics of $\CN=1$
field theories. We only discuss the case $G=\SU(N)$.
Most of the examples in this section has been discussed in \cite{Xie:2013rsa}.
Our new tools developed in this paper make it possible to compare field theories and generalized Hitchin systems in great detail.
Although spectral curves are determined in \cite{Xie:2013rsa},
we give self-contained derivations of the curves for completeness and also because
we want to explain a new method to determine spectral curves in this section and in appendix~\ref{app:B}.

\subsection{Preliminary}\label{sec:preliminary}
In this section, we use the normalization of fields discussed in subsection~\ref{sec:regsing}.
However, we write $\tilde{\Phi}_i$ of that section simply as $\Phi_i$ in this section.

\paragraph{Irregular singularity.}
Although we have been focusing on regular singularities at punctures,
we will also use irregular singularities which are locally the same as the $\CN=2$ singularities 
obtained from M-theory uplift of type IIA brane configurations.
%discussed e.g.,in \cite{Gaiotto:2009hg,Nanopoulos:2010zb,Cecotti:2011rv,Bonelli:2011aa,Xie:2012hs,Gaiotto:2012rg,Kanno:2013vi}.
In a gauge where $A_{\bar{z}}=0$ and $\Phi_i$ are diagonal, the singularities we use are given by
\beq
N_f=0~:&~ \Phi\to \frac{\zeta}{(z-z_p)^{1+1/N}}\diag(1,\omega_N,\cdots,\omega_{N}^{N-1}), \label{eq:sing1} \\
N_f<N~:&~ \Phi\to \frac{\zeta}{(z-z_p)^{1+1/(N-N_f)}}\diag(0,\cdots,0,1,\omega_{N-N_f},\cdots,\omega_{N-N_f}^{N-N_f-1}) \nonumber \\
&~~~~~~~~+\frac{1}{(z-z_p)}\diag(m_1,\cdots,m_{N_f},m,\cdots,m) - ({\rm trace~part}), \label{eq:sing2}   \\
N_f=N~:&~\Phi \to \frac{1}{(z-z_p)}\diag(m_1,\cdots,m_{N}),\label{eq:sing3}
\eeq
where $\sum_{k=1}^{N_f}m_k+(N-N_f)m=0$ and $\omega_k=\exp(2\pi i/k)$.
The $N_f$ corresponds roughly to the ``number of flavors'' at the puncture. See also appendix~\ref{app:A} for field theory interpretation. 
The $N_f=N$ case is just the regular singularity we have
discussed in subsection~\ref{sec:regsing}.
Another, probably more familiar, way of writing the singularities is
\beq
N_f=0~:&~\det (x-\Phi) \to x^{N}-\frac{\zeta^N}{(z-z_p)^{N+1}} +({\rm less~singular}) , \label{eq:csing1} \\
N_f<N-1~:&~\det (x-\Phi) \to x^{N}-\frac{\zeta^{N-N_f}}{(z-z_p)^{N-N_f+1}}\prod_{k=1}^{N_f}\left(x -\frac{m_k}{z-z_p} \right) 
+({\rm less~singular}) , \label{eq:csing2}\\
N_f=N~:&~\det (x-\Phi) \to \prod_{k=1}^N \left(x -\frac{m_k}{z-z_p} \right)+({\rm less~singular}). \label{eq:csing3}
\eeq
The case $N_f=N-1$ is obtained by shifting $x$ in \eqref{eq:csing2} so that the coefficient of the $x^{N-1}$ term vanishes.
These singularities can be obtained directly from the type IIA brane construction as in \cite{Witten:1997sc,Gaiotto:2009we,Gaiotto:2009hg}.

In the equations \eqref{eq:sing1} and \eqref{eq:sing2}, $\Phi$ is not single-valued. This is an artifact of taking $A_{\bar{z}}=0$ and
making $\Phi$ to be diagonal. They can be made single-valued by an appropriate change of basis of the bundle ${\rm ad}(E)$~\cite{Gaiotto:2009hg}.
After the change of the basis, some of the fields become non-diagonal and the gauge symmetry is broken down to a subgroup at the puncture. 
For $N_f<N$, the gauge symmetry is reduced to $\U(N_f)$ for massless case and $\U(1)^{N_f}$ for generic masses
at the puncture. As discussed in subsection~\ref{sec:sigma}, $\sigma$ must be in the Cartan subalgebra of unbroken gauge group.
Therefore, in the presence of irregular singularities, $\sigma$ is only allowed to be in the Cartan subalgebra of $\U(N_f)$.
For example, if there is an irregular singularity of the $N_f=0$ type, $\sigma $ must be set to zero.
This is only a local constraint at the punctures, and $\sigma$ is also constrained by the global structure of a solution for $\Phi$.

The behaviors at irregular singularities constrain not only $\sigma$, but also $\Phi_i$.
Suppose that $\Phi_2$ has the $N_f=0$ type irregular singularity. Then, the condition $[\Phi_1, \Phi_2]=0$
gives a constraint $\Phi_1 \to 0$;
\beq
{\rm If~} \Phi_2 \to \frac{\zeta}{(z-z_p)^{1+1/N}}\diag(1,\omega_N,\cdots,\omega_{N}^{N-1}),~{\rm then~}\Phi_1 \to 0~{\rm and~vice~versa.}
\label{eq:commcon}
\eeq
This fact can be seen e.g., by noting that $\Tr [\Phi_1^k \Phi_2]$ must be a single-valued function of $z$.
See appendix~\ref{app:B} for more details.

The phase of $\zeta$ is not single valued and only its power $\zeta^{N-N_f}$ is well-defined.
Therefore, if $\zeta$ appears without the power $N-N_f$ in a solution, that means there are discrete vacua labelled by
the phase of $\zeta$.

In the case of irregular singularities, we have not studied how to identify holomorphic moment maps or meson operators.
Suppose that $\Phi_1$ has a singularity of the type \eqref{eq:sing2}.  There is a $\U(N_f)$ flavor symmetry associated to the puncture and 
we have the corresponding
moment map $\mu$ which is an $N_f \times N_f$ matrix. If it is constructed from quarks $q_i$ and $\tilde{q}^i$, it is given as
$\mu^{i}_{~j}=M^{i}_{~j}=\tilde{q}^i q_j$, i.e., it is just a meson matrix. We propose, but do not prove, that this operator is identified as
\beq
\Phi_2 \approx \left(
\begin{array}{cc}
\mu-\frac{1}{N}  ( \tr \mu )  {\bf 1}_{N_f} & 0 \\
0 &      -\frac{1}{N}  ( \tr \mu ) {\bf 1}_{N-N_f} 
\end{array}
\right),\label{eq:irregularmeson}
\eeq
where $\approx$ means that the characteristic polynomials of both sides agree.
This proposal would be a straightforward generalization of the regular case \eqref{eq:nomalizedop2} if the trace of $\mu$ were zero.
The above form of the trace part which is proportional to $\tr \mu$ may be motivated by the fact
that the superpotential formula \eqref{eq:rewrittensuper} gives
\beq
\oint \frac{dz}{2 \pi i} \tr(\Phi_1 \Phi_2) = \tr ( m \mu )+\cdots ,
\eeq
where $m=\diag(m_1,\cdots,m_{N_f})$ is the mass matrix, and the ellipsis represents possible terms coming from
subleading terms in the singularities which depend on explicit solutions of generalized Hitchin's equations. 
The term $\tr ( m \mu )$ looks precisely as the tree level mass term.

\paragraph{Singularities and redefinition of $\Phi$.}
Let us perform a slight redefinition of $\Phi_i$ which is not essential but makes the analysis of spectral curves a little bit simpler.
First, the following mathematical fact is known. 
For a given set of points $\{z_b\}_{b \in B}$ on a Riemann surface, there exists a holomorphic line bundle 
which we denote as $L_B$ with the following property.
Up to normalization, $L_B$ has a unique holomorphic section $s_1$ which has simple zeros at $\{z_b\}_{b \in B}$ and
has no other zeros.\footnote{
If some of the points coincide, e.g., $z_1=z_2=\cdots=z_k$, the section $s$ has a zero of degree $k$ there.}
The degree of the bundle $L_B$ is the same as the number of points in $\{z_b\}_{b \in B}$, which we denote as $n_1$.

More explicitly, in the case of a Riemann sphere $C={\mathbb C} \cup \{\infty \}$, $s_B$ is given as
\beq
s_1=\prod_{b \in B}(z-z_b),
\eeq
where $z$ is the coordinate of ${\mathbb C}=C-\{\infty\}$. 
Since $s_1$ is a section of the line bundle $L_B$ with degree $\deg L_B=n_1$, the behavior near $z \to \infty$ is described as
\beq
(s_1)_\infty \equiv z^{-n_1} s_1 \to 1~~~(z \to \infty).
\eeq
Thus $s_B$ as a section goes to a nonzero finite constant at $z \to \infty$.
If one of the points in $\{z_b\}_{b \in B}$ is at $z_\infty=\infty$,
the product is taken as $s_1=\prod_{b \in B-\{\infty\}}(z-z_b)$. Then, near $z \to \infty$, the section behaves as $(s_1)_\infty \to 1/z$.

Now suppose that the singularities of $\Phi_1$ are at $\{z_b\}_{b \in B}$.
Then, we define
\beq
\Phi'_1=s_1 \Phi_1.\label{eq:primedPhi}
\eeq
The new adjoint field $\Phi'_1$ takes values in the line bundle
\beq
L'_1=L_1 \otimes L_B.
\eeq
The degree of the bundle $L_1'$ is 
\beq
\deg L_1'= \deg L_1+n_1=p, 
\eeq
where $p$ is defined in appendix~\ref{app:A} as the number of copies of the $T_N$ theory of ``$+$ type''~\cite{Bah:2012dg,Gadde:2013fma}. 
(In the notation of appendix~\ref{app:A}, we have $n_1=n_+$ and $n_2=n_-$.)
Furthermore, because $s_1$ has simple zeros at $\{z_b\}_{b \in B}$,
the behavior of $\Phi_1$ near the punctures is , e.g., for the $N_f$ type puncture,
\beq
\Phi_1'  \to & \frac{\zeta}{(z-z_b)^{1/(N-N_f)}}\diag(0,\cdots,0,1,\omega_{N-N_f},\cdots,\omega_{N-N_f}^{N-N_f-1}) \nonumber \\
&+\diag(m_1,\cdots,m_{N_f},m,\cdots,m) - ({\rm trace~part}). \label{eq;reducedsing}
\eeq

Similarly, let $\{z_a\}_{a \in A}$ be the positions of the singularities of $\Phi_2$, $L_A$ be the line bundle associated to $\{z_a\}_{a \in A}$,
 $s_2$ be the section of $L_A$ which has zeros at $\{z_a\}_{a \in A}$,
and $\Phi_2'=s_2\Phi_2$. Now $\Phi_2'$ takes values in the line bundle $L_2'=L_2 \otimes L_A$ which has
the degree $\det L_2'=\deg L_2+n_2=q$, where $q$ is defined in the appendix~\ref{app:A} as the number of 
copies of the $T_N$ theory of ``$-$ type''.
The $\Phi_2'$ has similar behavior as \eqref{eq;reducedsing} at the punctures.

\subsection{SQCD}

We will consider supersymmetric QCD (SQCD) realized as a Riemann sphere with two punctures at 
$z=0$ and $z=\infty$. The $\Phi_1$ will have a singularity at $z=\infty$, and $\Phi_2$ will have a singularity at $z=0$.
The degrees of the line bundles $L'_1=L_1 \otimes L_B$ and $L'_2=L_2 \otimes L_A$ are both zero, $\deg L'_1=\deg L'_2=0$, or
in other words $\deg L_1=\deg L_2=-1$~\cite{Xie:2013gma,Xie:2013rsa}.
This fact can be seen by considering RG flows from the theories discusses in appendix~\ref{app:A},
or from the type IIA brane construction \cite{Hori:1997ab,Witten:1997ep}.
See \cite{Intriligator:1995au} for a review of exact results of SQCD including precise numerical coefficients.

\subsubsection{Massive SQCD with $N_f<N$ flavors}\label{sec:NflessN}

Let us consider an $\SU(N)$ SQCD with $N_f$ flavors of quarks $q_i$ and $\tilde{q}^i$~($i=1,\cdots,N_f$)
in the fundamental and anti-fundamental representations of the gauge group, respectively.
We assume $N_f<N-1$ for simplicity, although the $N_f=N-1$ case is very similar.

The mass term is given as $\tr m M$, where $M^i_{~j} = \tilde{q}^i q_j$ are mesons and the trace is over flavor indices.
The effective superpotential including the Affleck-Dine-Seiberg superpotential~\cite{Affleck:1983mk} is given as
\beq
W=\tr mM +(N-N_f)\left(\frac{\Lambda^{3N-N_f}}{\det M}\right)^{\frac{1}{N-N_f}},
\eeq
where $\Lambda^{3N-N_f}$ is the dynamical scale (or more precisely the one instanton factor).
Integrating out the mesons, we get the vevs of $M$ and $W$ as
\beq
{M}&=\left( \Lambda^{3N-N_f} \det m \right)^{\frac{1}{N}}m^{-1} , \\
{W}&=N \left(  \Lambda^{3N-N_f} \det m \right)^{\frac{1}{N}}.
\eeq
Our purpose is to reproduce these results from the twisted 5d SYM.

The $\Phi_1$ has a singularity of the type \eqref{eq:csing2} at $z=\infty$, 
and $\Phi_2$ has a singularity of the type \eqref{eq:csing1} at $z=0$.
The singularities suggest that
\beq
\det (x'_1-\Phi'_1)&=x'^N_1-z \zeta_1^{N-N_f}\prod_{k=1}^{N_f}(x'_1-m_k)+\sum_{k=2}^{N} u_k x'^{N-k}_1,\\
\det (x'_2-\Phi'_2)&=x'^N_2-\frac{\zeta_2^N}{z} +\sum_{k=2}^{N} u'_k x'^{N-k}_1,
\eeq
where we have used the definition \eqref{eq:primedPhi}.
However, since $\Phi_2$ has the singularity \eqref{eq:csing1} and there is a constraint \eqref{eq:commcon},
the moduli parameters $u_k$ in $\Phi_1$ must be set to zero. Therefore, the spectral curve is
\beq
0=\det (x'_1-\Phi'_1)=x'^N_1-z \zeta_1^{N-N_f}\prod_{k=1}^{N_f}(x'_1-m_k).\label{eq:sqcd1x1}
\eeq

Next, let us determine $\Phi'_2$. A detailed derivation is given in appendix~\ref{app:B}, 
and here we only give a heuristic argument.
From \eqref{eq:sqcd1x1}, we can see that $\Phi'_1$ behaves at $z \to 0$ as
\beq
\Phi'_1 \to  \left( (-1)^{N_f} \zeta_1^{N-N_f}\prod_{k=1}^{N_f}m_k  \right)^{\frac{1}{N}} z^{\frac{1}{N}}  \diag(1,\omega_N,\cdots,\omega_N^{N-1}),
\eeq
up to complexified gauge transformations. Then, the singular behavior of $\Phi'_2$ at $z=0$ is reproduced if we set
\beq
\Phi'_2 \sim (\Lambda_{\rm eff}^{3N})^{1 \over N} (\Phi'_1)^{-1}, \label{eq:roughPhi2}
\eeq
where we have defined
\beq
\Lambda_{\rm eff}^{3N}=(-1)^{N_f} \zeta_1^{N-N_f} \zeta_2^{N}\prod_{k=1}^{N_f}m_k.
\eeq
An important point of the ansatz \eqref{eq:roughPhi2} is that the commuting condition $[\Phi'_1, \Phi'_2]=0$ is automatic. 
One can also check that $\Phi'_2$ given by \eqref{eq:roughPhi2} is finite other than at the puncture $z=0$.
However, we must impose the traceless condition $\Tr \Phi'_2=0$.
By using the equation $\det((x'_1)^{-1} - (\Phi'_1)^{-1}) =(-x'_1)^{-N}(\det \Phi'_1)^{-1} \det (x'_1-\Phi'_1)  $, one can see from \eqref{eq:sqcd1x1}
that  the traceless condition is achieved by
\beq
\Phi'_2 = \Lambda_{\rm eff}^{3} \left((\Phi'_1)^{-1}- \frac{{\bf 1}_N}{N} \sum_{k=1}^{N_f} \frac{1}{m_k}\right).
\eeq
In the spectral curve, the pair $(x'_1,x'_2)$ is given by pairs of eigenvalues of $(\Phi'_1, \Phi'_2)$, so we get the curve
\beq
x'_1x'_2=\Lambda_{\rm eff}^{3} \left(1- \frac{x'_1}{N} \sum_{k=1}^{N_f} m_k^{-1}\right).\label{eq:sqcd1x1x2}
\eeq
See appendix~\ref{app:B} for more rigorous derivation of \eqref{eq:sqcd1x1x2}.

Let us calculate the meson vev using the above results. We can see that $\Phi_2'$ at $z \to \infty$ is given as
\beq
\Phi'_2 \to \Lambda_{\rm eff}^{3} \left( \diag(m_1^{-1},\cdots,m_{N_f}^{-1},0,\cdots,0)-\frac{{\bf 1}_N}{N}  \sum_{k=1}^{N_f} m_k^{-1} \right).
\eeq
Comparing this result with the proposal \eqref{eq:irregularmeson}, we obtain
\beq
M \approx \Lambda_{\rm eff}^{3}  \diag(m_1^{-1},\cdots,m_{N_f}^{-1}).\label{eq:spectralmesonvev}
\eeq
This result agrees with the field theory result if we identify 
\beq
(-1)^{N_f} \zeta_1^{N-N_f} \zeta_2^{N}=\Lambda^{3N-N_f}. \label{eq:dynid}
\eeq
The vev \eqref{eq:spectralmesonvev} is determined up to conjugation, but we expect
the result is exactly given by $M=\Lambda_{\rm eff}^{3} m^{-1}$.

The superpotential vev can be calculated easily. First, we note that $\Phi_1'=\Phi_1$ and $\Phi'_2=z \Phi_2$.
Then we get $\Phi_1 \Phi_2 \to \Lambda_{\rm eff}^3 {\bf 1}_N/z$ at $z \to 0$.
The superpotential formula \eqref{eq:nomalizedsup} gives
\beq
W=\oint_{z \sim 0}\frac{dz}{ 2 \pi i} \Tr (\Phi_1 \Phi_2)=N\Lambda_{\rm eff}^{3}.
\eeq
This is in perfect agreement with the field theory result with the identification \eqref{eq:dynid}.

\subsubsection{Massless SQCD with $N_f=N=N_1+N_2$ flavors}
Here we consider theories with the total flavor number $N_f=N$. We assume that 
$\Phi_1$ has a singularity of the type \eqref{eq:sing2} with $N_f \to N_2$ at $z =\infty$ and $\Phi_2$
has a singularity of the type \eqref{eq:sing2} with $N_f \to N_1$ at $z=0$.
We assume $N_f=N=N_1+N_2$ and $N_1, N_2 \geq 1$.

Note that theories with different pairs $(N_1,N_2)$ are really different theories. 
There are quarks $q_i, \tilde{q}^i~(i=1,\cdots,N_1)$ which are in the (anti-)fundamental representations of
the flavor group $\U(N_1)$, and there are also quarks $p_\ell, \tilde{p}^\ell ~(\ell=1,\cdots,N_2)$ in the (anti-)fundamental representations of
the flavor group $\U(N_2)$. 
We define mesons as
\beq
M= \left(\begin{array}{c}
\tilde{q}^i \\
\tilde{p}^\ell
\end{array}\right)
(q_j, p_m)
=
\left(\begin{array}{cc}
(M_1)^i_j & L^i_m \\
\tilde{L}^\ell_j & (M_2)^\ell_m
\end{array}\right).
\eeq
As reviewed in appendix~\ref{app:A}, there is a quartic superpotential,
\beq
W &=c \left( q^\alpha_i \tilde{q}_\beta^i -\frac{\delta^\alpha_\beta}{N}q^\gamma_i \tilde{q}_\gamma^i \right)
\left( p^\beta_\ell \tilde{p}_\alpha^\ell -\frac{\delta_\alpha^\beta}{N}p^\gamma_\ell \tilde{p}_\gamma^\ell \right) \nonumber \\
&=c \left(\tr (L\tilde{L})-\frac{1}{N} (\tr M_1)(\tr M_2) \right).
\eeq
where $\alpha,\beta=1,\cdots,N$ are gauge indices.
This quartic superpotential explicitly depends on $N_1$ and $N_2$, and hence theories with different values of $(N_1,N_2)$
are different theories even if their sum $N_1+N_2=N$ is the same.

\paragraph{SU(2) theory.}
Consider the $\SU(2)$ theory with $N_1=N_2=1$. The superpotential, including the deformed moduli constraint~\cite{Seiberg:1994bz}, is given as
\beq
W=X(M_1M_2-L \tilde{L}-B\tilde{B} -\Lambda^4)+c\left( L\tilde{L}-\frac{1}{2}M_1M_2 \right),
\eeq
where $B$ and $\tilde{B}$ are baryons and anti-baryons, respectively. 
There are three branches.
\beq
(1)~:&~X=c/2,~~~M_1M_2=\Lambda^4,~~~L=\tilde{L}=B=\tilde{B}=0. \label{eq:su2fb1} \\
(2)~:&~X={c},~~~~~~L\tilde{L}=-\Lambda^4,~~~~~M_1=M_2=B=\tilde{B}=0. \label{eq:su2fb2} \\
(3)~:&~X=0,~~~~~B\tilde{B}=-\Lambda^4,~~~~~ M_1=M_2=L=\tilde{L}.\label{eq:su2fb3}
\eeq
We would like to recover these branches from the twisted 5d SYM.

Let us first determine the spectral curve. For $\SU(2)$, the curve \eqref{eq:U(2)generalcurve} is
\beq
x'^2_1=\frac{1}{2}\Tr \Phi'^2_1,~~~x'^2_2=\frac{1}{2}\Tr \Phi'^2_1,~~~x'_1x'_2=\frac{1}{2} \Tr \Phi'_1 \Phi'_2.
\eeq
The singularities of $\Phi'_1$ and $\Phi'_2$ give
\beq
x'^2_1=\frac{1}{4}\zeta_1^2z^2+u_1,~~~
x'^2_2=\frac{1}{4}\frac{\zeta_2^2}{z^2}+u_2, ~~~
x'_1x'_2=h(z).
\eeq
Then, we have to impose that $h(z)^2=x'^2_1 x'^2_2$ is a square of some holomorphic function $h(z)$. There are
three possibilities;
\beq
(1)~:~&u_1u_2=\frac{1}{16} \zeta_1^2\zeta_2^2,~~~x'_1x'_2=\frac{\zeta_1 \sqrt{u_2}}{2z}\left(z^2+\frac{\zeta_2^2}{4u_2} \right),\label{eq:su2branch1}\\
(2)~:~&u_1=u_2=0,~~~~~~~x'_1x'_2=+\frac{1}{4}\zeta_1\zeta_2,\label{eq:su2branch2}\\
(3)~:~&u_1=u_2=0,~~~~~~~x'_1x'_2=-\frac{1}{4}\zeta_1\zeta_2.\label{eq:su2branch3}
\eeq
In the first equation, possible $\pm$ signs in the equation for $x'_1x'_2$ are absorbed in the definition of the moduli field $\sqrt{u_2}$.
However, in the second and third equations, there are no moduli fields to absorb the $\pm$ signs and we have to distinguish the different signs.

The identification \eqref{eq:irregularmeson} suggests that we identify mesons as $M_1^2=4u_1$ and $M_2^2=4u_2$.
Then, \eqref{eq:su2branch1} gives $M_1^2M_2^2=\zeta_1^2\zeta_2^2$. The sign of $M_1M_2$ can be fixed by
the equation of $x_1'x_2'$. We should get $x_1'x_2' \to -M_1 \zeta_2/(4z)$  as $z \to 0$, and 
$x_1'x_2' \to -M_2 \zeta_1 z/4$ as $z \to \infty$. Then we identify $M_1=-\zeta_1\zeta_2/(2\sqrt{u_2})$
and $M_2=-2\sqrt{u_2}$. We finally get $M_1M_2=\zeta_1\zeta_2$.
This reproduces the field theory result \eqref{eq:su2fb1} by identifying $\zeta_1\zeta_2=\Lambda^4$

In the other two branches \eqref{eq:su2branch2} and \eqref{eq:su2branch3}, we have $M_1=-2\sqrt{u_1}=0$ and $M_2=-2\sqrt{u_2}=0$.
Therefore these branches should correspond to the branches \eqref{eq:su2fb2} and \eqref{eq:su2fb3}.
Where is the moduli field contained in $L,\tilde{L}$ or $B, \tilde{B}$?
In these cases, we can explicitly write down $\Phi'_1$ and $\Phi'_2$ as
\beq
(2)~:~&\Phi'_1=\frac{1}{2} \left( \begin{array}{cc}
-\zeta_1 z & 0 \\
0 & +\zeta_1 z
\end{array}\right),~~~~~
\Phi'_2=\frac{1}{2} \left( \begin{array}{cc}
-\zeta_2/z & 0 \\
0 & +\zeta_2/ z
\end{array}\right), \\
(3)~:~&\Phi'_1=\frac{1}{2} \left( \begin{array}{cc}
-\zeta_1 z & 0 \\
0 &+ \zeta_1 z
\end{array}\right),~~~~~
\Phi'_2=\frac{1}{2} \left( \begin{array}{cc}
+\zeta_2/z & 0 \\
0 & -\zeta_2/ z
\end{array}\right).
\eeq
As is clear from these solutions, there is an unbroken $\U(1)$ symmetry in each case. Then, we can turn on $\sigma$ as
\beq
\sigma=\left( \begin{array}{cc}
\sigma_0 & 0 \\
0 & -\sigma_0
\end{array}\right),
\eeq
for constant $\sigma_0$. This $\sigma_0$ is combined with the dual photon of unbroken $\U(1)$ to 
give a single chiral field as explained in section~\ref{sec:5}.
In the field theory result \eqref{eq:su2fb2} and \eqref{eq:su2fb3}, there is one modulus field in each case,
parametrized by $L/\tilde{L}$ or $B/\tilde{B}$ respectively.
This one modulus should be identified with the chiral field containing $\sigma_0$.
Indeed, $L,\tilde{L}$ and $B,\tilde{B}$ are operators which are not associated to punctures,
but are kind of ``baryon'' operators similar to $Q^{i_A i_B \alpha} Q_{i_C i_D \alpha}$,
where the $Q$'s are the $T_N$ theory operators discussed in section~\ref{sec:5}.
As discussed there, they are interpreted as coming from the $\sigma$.
In the present case, we have to replace $Q^{i_A i_B \alpha}  \to (q_i^\alpha,\tilde{q}^{i\alpha})$, 
$Q_{i_C i_D \alpha} \to (p_\alpha^\ell,\tilde{p}_{\ell \alpha})$
and $Q^{i_A i_B \alpha} Q_{i_C i_D \alpha} \to (B,\tilde{B},L,\tilde{L})$.
We conclude that, at least qualitatively, the field theory and the twisted 5d SYM match.

The $\sigma$ is forced to be zero in the case of the branch \eqref{eq:su2branch1}, and 
we can recover the deformed moduli space structure $M_1M_2=\Lambda^4$.
On the other hand, if $\sigma$ is nonzero, we have not yet succeeded in determining the precise relation $L\tilde{L}=-\Lambda^4$
or $B\tilde{B}=-\Lambda^4$ from the twisted 5d SYM.
We leave it for future work to study the precise relations between $L,\tilde{L},B,\tilde{B}$ and $\sigma$.
However, we stress that there is no mismatch between the field theory and the twisted 5d SYM.

\paragraph{SU(N) theory.}
Now let us see the $\SU(N)$ theory with $N_f=N=N_1+N_2$. For simplicity we assume $N_1,N_2>1$.

The low energy effective superpotential is given as
\beq
W=X(\det M-B\tilde{B}-\Lambda^{2N})+c \left(\tr (L\tilde{L})-\frac{1}{N} (\tr M_1)(\tr M_2) \right).
\eeq
We can see that there is a baryonic branch where $B\tilde{B} \neq 0$ and we focus on this branch.
In this branch, the equations of motion of baryons give $X=0$. Then, $L, \tilde{L}$, $\tr M_1$ and $\tr M_2$ 
are massive and set to zero by equations of motion. Therefore the remaining moduli fields are
$\mu_1=M_1-\frac{1}{N_1} \tr M_1$, $\mu_2=M_2-\frac{1}{N_2} \tr M_2$, $B$ and $\tilde{B}$ with the constraint
$(\det \mu_1)( \det \mu_2)-B\tilde{B}=\Lambda^{2N}$.

In the twisted 5d SYM, the structure of the singularities allows the following form of solutions,
\beq
\Phi'_1=\left(\begin{array}{cc}
\Phi'_{1,1} & 0 \\
0 & \Phi'_{1,2}
\end{array} \right),~~~
\Phi'_2=\left(\begin{array}{cc}
\Phi'_{2,1} & 0 \\
0 & \Phi'_{2,2}
\end{array} \right),~~~
\sigma=\left(\begin{array}{cc}
\sigma_0 {\bf 1}_{N_1}/N_1 & 0 \\
0 & -\sigma_0 {\bf 1}_{N_2}/N_2
\end{array} \right),~~~
\eeq
where $\Phi'_{1,1}$ and $\Phi'_{2,1}$ are $N_1 \times N_1$ matrices, 
$\Phi'_{1,2}$ and $\Phi'_{2,2}$ are $N_2 \times N_2$
matrices, and $\Tr \Phi'_{1,1}+\Tr \Phi'_{1,2}=\Tr \Phi'_{2,1}+\Tr \Phi'_{2,2}=0$.
This is the most general form consistent with nonzero $\sigma$.
The singularities of these fields are given as
\beq
\Phi'_{1,1} &\to \zeta_1z^{1/N_1} \diag(1,\omega_{N_1},\cdots, \omega_{N_1}^{N_1-1}) ,~~z\to \infty \\
\Phi'_{2,2} &\to \frac{\zeta_2}{z^{1/N_2}} \diag(1,\omega_{N_2},\cdots, \omega_{N_2}^{N_2-1}),~~~~z \to 0.
\eeq
The $\Phi'_{1,2}$ and $\Phi'_{2,1}$ are nonsingular.

The singularity structure requires that $\Phi'_{2,1} \to 0$ at $z \to 0$.
Furthermore, $\Phi'_{2,1} $ is nonsingular on the entire Riemann sphere. 
This means that $\Tr (\Phi'_{2,1})^k$ are holomorphic functions on the Riemann sphere
which go to zero at $z \to 0 $ and do not have poles. Then these functions must be zero, and 
we set $\Phi'_{2,1}=0$. In the same way we set $\Phi'_{1,2}=0$.

The spectral curves are given as
\beq
\det(x'_1 -\Phi'_1)&=x'^{N_2}_1 \left( x'^{N_1}_1+\sum_{k=2}^{N_1}u_{1,k}x'^{N_1-k}_1- z\zeta_1^{N_1} \right), \\
\det(x'_2 -\Phi'_2)&=x'^{N_1}_2 \left( x'^{N_2}_2+\sum_{k=2}^{N_2}u_{2,k}x'^{N_2-k}_2- z^{-1}\zeta_2^{N_2} \right).
\eeq
There are no constraints on the moduli fields $u_{1,k}$ and $u_{2,k}$. The identification \eqref{eq:irregularmeson}
suggests that the characteristic polynomials of the fields $\mu_1$ and $\mu_2$ in the field theory are given by
\beq
&\det (x'_1-\mu_1)=x'^{N_1}_1+\sum_{k=2}^{N_1}u_{1,k}x'^{N_1-k}_1, \\
&\det (x'_2-\mu_2)=x'^{N_2}_2+\sum_{k=2}^{N_2}u_{2,k}x'^{N_2-k}_2.
\eeq
A degree of freedom coming from the baryons $B/\tilde{B}$ is identified as $\sigma_0$ and the dual photon.

\subsection{$T_N$ theory}
Let us consider a Riemann sphere with three singularities. 
This theory corresponds to a copy of the $T_N$ theory coupled to some vector multiplets. 
The number of singularities of 
$\Phi_1$ is $n_1$, and that of $\Phi_2$ is $n_2$, with $n_1+n_2=3$. The degrees of the line bundles are
\beq
\deg L'_1=\deg L_1+n_1=0, ~~~~~
\deg L'_2=\deg L_2+n_2=1.
\eeq
As discussed in appendix~\ref{app:A}, an irregular singularity of type \eqref{eq:sing1} for $\Phi_2$ corresponds to
an $\CN=2$ vector multiplet coupled to the $T_N$ theory, and an irregular singularity of type \eqref{eq:sing1} for $\Phi_1$
corresponds to an $\CN=1$ vector multiplet coupled to the $T_N$ theory.

As reviewed in section~\ref{sec:5},
the $T_N$ theory has flavor symmetries $\SU(N)_A \times \SU(N)_B \times \SU(N)_C$.
In the subsection \ref{sec:TNN1} and \ref{sec:N1N2mixed}, 
we will only consider the cases in which at least one of the three $\SU(N)$'s is gauged by an $\CN=1$ vector multiplet.
Let us gauge $\SU(N)_C$ by an $\CN=1$ vector multiplet. 
Before the gauging of $\SU(N)_C$, there are chiral ring relations of the $T_N$ theory given in \eqref{eq:CR1},
\beq
\det (x-\mu_A)=\det(x-\mu_B)=\det(x-\mu_C),
\eeq
for arbitrary $x$,
where $\mu_{A,B,C}$ are the holomorphic moment maps associated to the flavor symmetry groups.
After gauging $\SU(N)_C$, the low energy theory is described by the gauge invariant fields $\mu_A$ and $\mu_B$
satisfying the deformed moduli constraint~\cite{Maruyoshi:2013hja}
\beq
\det(x -\mu_A)=\det(x -\mu_B)-\Lambda_C^{2N}, \label{eq:TNdeformed}
\eeq
for arbitrary $x$.

In the twisted 5d SYM, the gauging of $\SU(N)_C$ introduces an irregular singularity of $\Phi_1$ which we take at $z=\infty$,
\beq
\Phi'_1 \to \zeta_C z^{1/N} \diag(1,\omega_N,\cdots,\omega_N^{N-1}).\label{eq:TNsinginfty}
\eeq
The $\Phi_1$ and/or $\Phi_2$ may have other singularities depending on the theory we consider.

\subsubsection{$T_N$ theory coupled to $\CN=1$ vector multiplet(s)}\label{sec:TNN1}
Let us consider the case where only $\Phi_1$ has irregular singularities. 
All the singularities of $\Phi_2$ are regular with mass parameters taken to be zero.
Then the residues of poles of $\Phi_2$ take values in nilpotent cones~\cite{Chacaltana:2012zy}, which can be zero.
Therefore, in this case, it is consistent to set $\Phi_2=0$ on the entire Riemann surface.\footnote{We 
believe that it is not only consistent to set $\Phi_2=0$, but $\Phi_2$ is forced to be zero by generalized Hitchin's equations
for the theories studied in this subsection. However, we will not prove this claim.}
Furthermore, irregular singularities of $\Phi_1$ set $\sigma=0$ as discussed in subsection~\ref{sec:preliminary}.
Then we get a twisted Higgs bundle discussed in subsection~\ref{sec:spectral}.
Spectral curves are easily determined similar to the case of $\CN=2$ field theories.

\paragraph{One $\SU(N)$ gauge group: ${\bf (n_1,n_2)=(1,2)}$.} 
When the $T_N$ theory is coupled to one $\CN=1$ $\SU(N)$ vector multiplet,
the field theory is described by the deformed moduli space \eqref{eq:TNdeformed}.
Let us reproduce this deformed moduli constraint from the twisted 5d SYM.

From the singularity \eqref{eq:TNsinginfty}, we get
\beq
\det (x'_1-\Phi'_1)=x'^N_1+\sum_{k=2}^N u_k x'^{N-k}_1-\zeta_C^N z.
\eeq
Suppose that the other two regular punctures of $\Phi_2$ are at $z=0$ and $z=1$.
The identification \eqref{eq:nomalizedop1} gives us
\beq
\det(x'_1-\mu_A)&=\det (x'_1-\Phi'_1)|_{z=0}=x'^N_1+\sum_{k=2}^N u_k x'^{N-k}_1, \\
\det(x'_1-\mu_B)&=\det (x'_1-\Phi'_1)|_{z=1}=x'^N_1+\sum_{k=2}^N u_k x'^{N-k}_1-\zeta_C^N.
\eeq
From these equations, we obtain
\beq
\det(x'_1-\mu_A)=\det(x'_1-\mu_B)+\zeta_C^N.
\eeq
This is exactly the relation \eqref{eq:TNdeformed} with the identification $\zeta_C^N=-\Lambda_C^{2N}$.
Thus the twisted 5d SYM perfectly reproduces the deformed moduli constraint of the field theory.

\paragraph{Two $\SU(N)$ gauge groups: ${\bf (n_1,n_2)=(2,1)}$.}
When $\SU(N)_C$ and $\SU(N)_B$ are gauged by $\CN=1$ vector multiplets,
we obtain the following low energy theory.
First, by taking $\Lambda_C$ to be large, the theory is described by $\mu_A$ and $\mu_B$ with the constraint \eqref{eq:TNdeformed}.
The $\mu_A$ is gauge invariant, but $\mu_B$ is now an adjoint chiral field of the gauge group $\SU(N)_B$.
A generic vev of $\mu_B$ breaks $\SU(N)_B$ to $\U(1)^{N-1}$. 
Gauge invariant polynomials of $\mu_B$ are fixed by $\mu_A$ due to the relation~\eqref{eq:TNdeformed}.
Thus the low energy theory is described by $\mu_A$ and $N-1$ massless vector multiplets.
See \cite{Intriligator:1994sm} for the $\SU(2)$ case.

The spectral curve of this theory is given by
\beq
0=\det (x'_1-\Phi'_1)=x'^N_1+\sum_{k=2}^N u_k x'^{N-k}_1-\frac{\zeta_B^N z}{z-1}-\zeta_C^N z,\label{eq:twogaugecurve}
\eeq
where we have chosen the origin of $u_N$ such that the characteristic polynomial of $\mu_A$ is given by
\beq
\det (x'_1-\mu_A)=\det (x'_1-\Phi'_1)|_{z=0}=x'^N_1+\sum_{k=2}^N u_k x'^{N-k}_1.
\eeq
The curve \eqref{eq:twogaugecurve} is identified as the Seiberg-Witten curve describing the holomorphic coupling matrix $\tau_{IJ}$
of the low energy massless $\U(1)^{N-1}$ fields. 

For example, let us consider the $\SU(2)$ case. The curve can be rewritten as
\beq
y^2=w^3+(\zeta_C^2+\zeta_B^2-\det \mu_A)w^2+\zeta_C^2\zeta_B^2 w,
\eeq
where $w=\zeta_C^2(z-1)$ and $y=\zeta_C^2(z-1)x'_1$. This is exactly the same as the curve found by Intriligator and Seiberg ~\cite{Intriligator:1994sm} 
for this $\SU(2)$ theory,
with the identification $\Lambda_C^{2N}=\zeta_C^N$ and $\Lambda_B^{2N}=\zeta_B^N$.
Notice that the moduli space of the theory is spanned by $\mu_A$ and it has dimension $N^2-1=3$.
However, only the flavor singlet operator $\det \mu_A$ appears in the curve.

\paragraph{Three $\SU(N)$ gauge groups: ${\bf (n_1,n_2)=(3,0)}$.}
If all the three $\SU(N)$ groups are gauged by $\CN=1$ vector multiplets,
the low energy theory is described by $\U(1)^{2(N-1)}$ massless vector multiplets and $N-1$ gauge invariant fields 
$\tr \mu_A^k \sim \tr \mu_B^k \sim \tr \mu_C^k~(k=2,\cdots,N)$. The spectral curve is given as
\beq
0=\det (x'_1-\Phi'_1)=x'^N_1+\sum_{k=2}^N u_k x'^{N-k}_1-\frac{\zeta_A^N}{z}-\frac{\zeta_B^N z}{z-1}-\zeta_C^N z.
\eeq
This is exactly the same as the Seiberg-Witten curve derived in \cite{Tachikawa:2011ea,Maruyoshi:2013hja} 
for $\zeta^N_{A,B,C}=\Lambda^{2N}_{A,B,C}$.

In this theory, the moduli fields $u_k$ are not composites of more fundamental gauge invariant operators.
They are just independent fields parametrizing the moduli space similar to the case of Coulomb moduli of $\CN=2$ theories.
There are no chiral ring relations among them, and the moduli space is just ${\mathbb C}^{N-1}$.

\subsubsection{$T_N$ theory coupled to $\CN=1$ and $\CN=2$ vector multiplets}\label{sec:N1N2mixed}

Here we are going to study the theory defined by a Riemann sphere with two irregular singularities for $\Phi_1$
and one irregular singularity for $\Phi_2$. All of the irregular singularities are of the type \eqref{eq:sing1}.
 In this case, both $\Phi_1$ and $\Phi_2$ are nonzero.

\paragraph{Field theory.}
The field theory dynamics of this theory is interesting. There will be no massless moduli fields, and gaugino condensation will occur which
leads to discrete vacua.
However, there will be massless $\U(1)$ vector multiplets. So the moduli space is a set of discrete points with massless vector fields at each point.
The Seiberg-Witten curve will only depend on dynamical scales of the theory.

Let us gauge $\SU(N)_C$ and $\SU(N)_A$ by ${\cal N}=1$ vector multiplets and $\SU(N)_B$ by an ${\cal N}=2$ vector multiplet.
We assume that the dynamical scale of $\SU(N)_C$, $\Lambda_C$, is very large.
The effective superpotential after confinement of $\SU(N)_C$ is given as
\beq
W=\sqrt{2}\tr \phi_B \mu_B+\sum_{k=2}^{N} X_k (\tr \mu_A^k-\tr \mu_B^k-N\Lambda_C^{2N} \delta^{k,N}).
\eeq
where $\phi_B$ is the adjoint scalar in the $\CN=2$ vector multiplet of $\SU(N)_B$, and $X_k$ are Lagrange multipliers imposing
the deformed moduli constraint \eqref{eq:TNdeformed}.
In the effective superpotential, $\mu_B$ and $\phi_B$ have a mass term $\tr \phi_B \mu_B$ and can be integrated out.
The $F$-term equation of $\phi_B$ set $\mu_B=0$, and then the deformed moduli constraint requires that
\beq
\mu_A=\Lambda_C^2 \diag(1,\omega_N,\cdots, \omega_N^{N-1}),~~~\mu_B=0,  \label{eq:isolatedvevAB}
\eeq
up to $\SU(N)_A$ rotations.

As a result of integrating out $\phi_B$ and $\mu_B$, the $\SU(N)_B$ becomes pure $\CN=1$ SYM at low energies. 
Then the $\SU(N)_B$ develops gaugino condensation, and gives $N$ isolated vacua.
The vev \eqref{eq:isolatedvevAB} for $\mu_A$ breaks the $\SU(N)_A$ gauge group
to $\U(1)^{N-1}$ and there are $N-1$ massless vector fields. 
We need a Seiberg-Witten curve which determines the low energy coupling constant matrix of  these massless $\U(1)^{N-1}$ fields.

The curve will be determined by using the spectral curve later, but we can also see how the curve looks like
by field theory consideration. 
The $\SU(N)_A$ vector multiplet combined with the adjoint chiral field $\mu_A$ is very similar to an $\CN=2$ pure SYM.
The Seiberg-Witten curve for this pure ${\cal N}=2$ SYM may be given as~\cite{Klemm:1994qs,Argyres:1994xh}
\beq
\det \left( x-\frac{\mu_A}{\Lambda_C} \right)-\frac{\Lambda_A^{2N}}{z}+z \sim 0,
\eeq
where we have divided $\mu_A$ by $\Lambda_C$ so that $\mu_A/\Lambda_C$ has mass dimension one, which is the correct mass dimension 
for a chiral field
with a canonical kinetic term. 
After using \eqref{eq:isolatedvevAB} and rescaling the variables $x \to x/\Lambda_C$ and $z \to z \Lambda_C^N$, we get
\beq
x^N-\Lambda_C^{2N}-\frac{\Lambda_A^{2N}}{z}+\Lambda_C^{2N}z \sim 0.
\eeq
This curve is derived in the limit $\Lambda_C \gg \Lambda_A$, so we can have a small correction like 
$\Lambda_C^{2N} \to \Lambda_C^{2N}+c\Lambda_A^{2N}$ in some terms.
The original field theory has a symmetry under the exchange $A \leftrightarrow C$.
Requiring this symmetry which is accompanied with $z \leftrightarrow z^{-1}$ and $x^N \leftrightarrow -x^N$, 
we expect that the curve is given as
\beq
x^N-\frac{\Lambda_A^{2N}}{z}+\Lambda_C^{2N}z-(\Lambda_C^{2N}-\Lambda_A^{2N}) = 0.\label{eq:21curve}
\eeq
This derivation is only heuristic. We will derive it using the spectral curve.

The gaugino condensation of $\SU(N)_B$ induces a constant superpotential vev.
The dynamical scale of the low energy $\SU(N)_B$ is given by
\beq
\Lambda_{B,{\rm low}}^{3N} \sim \Lambda_B^N \Lambda_C^{2N}.
\eeq
This $\Lambda_{B,{\rm low}}^{3N}$ is determined as follows. The high energy one instanton factor of $\SU(N)_B$ is given as $\Lambda_B^N$,
and hence $\Lambda_{B,{\rm low}}^{3N}$ should be proportional to it. Assuming that the mass of the fields $\phi_B$ and $\mu_B$ is of order $\Lambda_C$,
we get the factor $\Lambda_C^{2N}$ in $\Lambda_{B,{\rm low}}^{3N}$
when the adjoint fields $\phi_B$ and $\mu_B$ are integrated out. 

The $\Lambda_{B,{\rm low}}^{3N}$ is derived for $\Lambda_C \gg \Lambda_{A,B}$, but 
the symmetry under the exchange $A \leftrightarrow C$ may require that the exact form is 
\beq
\Lambda_{B,{\rm low}}^{3N} = \Lambda_B^N (\Lambda_C^{2N}+\Lambda_A^{2N})
\eeq
up to an overall numerical coefficient. The gaugino condensation induces the superpotential
\beq
W_{\rm condense}=N \Lambda_{B,{\rm low}}^{3}=N \left[\Lambda_B^N (\Lambda_C^{2N}+\Lambda_A^{2N}) \right]^{\frac{1}{N}}. 
\label{eq:condense21}
\eeq
The presence of the $N$-th root suggests the existence of $N$-vacua as usual.

\paragraph{Twisted 5d SYM.}

The field $\Phi_1$ has irregular singularities of the type \eqref{eq:sing1} at $z=0$ and $z=\infty$.
Then we have
\beq
\det (x_1'-\Phi_1')=x_1'^N+\sum_{k=2}^{N} u_k x'^{N-k}_1 -\frac{\zeta_A^{N}}{z}+\zeta_C^{N}z,  \label{eq:21curve'}
\eeq
for some moduli $u_k$.
However, because $\Phi_2$ has a singularity at $z=1$ of the type \eqref{eq:sing1},
$\Phi'_1$ must be zero at $z=1$ by the constraint \eqref{eq:commcon}.
Then the spectral curve for $x'_1$ is completely fixed as
\beq
0=\det (x_1'-\Phi_1')=x_1'^N -\frac{\zeta_A^{N}(1-z)}{z}-\zeta_C^{N}(1-z).\label{eq:21spectralcurve}
\eeq
One can see that this curve is precisely the Seiberg-Witten curve \eqref{eq:21curve} 
in the field theory
if the parameters are identified as $\zeta_A^N = \Lambda_A^{2N}$ and $\zeta_C^N =  \Lambda_C^{2N}$.

Next let us determine $\Phi'_2$. A detailed derivation is given in appendix~\ref{app:B}, and here we only give a heuristic derivation.
If we only look at the behavior at the punctures, we can see that $(\Phi_1')^{-1}$ almost reproduces the behavior of $\Phi_2'$ at $z=0,1,\infty$;
at the punctures $z_p =0,\infty$, the fields $\Phi'_2$ and $(\Phi_1')^{-1}$ both behave as $(z-z_p)^{1/N} \diag(1,\omega_N, \cdots, \omega_N^{N-1})$
because of the constraint \eqref{eq:commcon},
and at the puncture $z_p=1$, they both behave as $(z-z_p)^{-1/N} \diag(1,\omega_N, \cdots, \omega_N^{N-1})$.
Furthermore, $(\Phi_1')^{-1}$ manifestly commutes with $\Phi_1'$, as required by the generalized Hitchin's equations. 
So we might hope that we can find a solution for $\Phi_2'$ by setting
\beq
\Phi_2' \sim (\Phi_1')^{-1}.\label{eq:roughPhi12relation}
\eeq

There are two problems in this proposal. First, the line bundle $L_1'$ has degree $\deg L_1'=0$, but the $L_2'$ has degree $\deg L_2'=1$.
Thus \eqref{eq:roughPhi12relation} does not make sense.
Second, the curve \eqref{eq:21spectralcurve} indicates that $\Phi_1'$ behaves near $z=-(\zeta_A^N/\zeta_C^N)$ as
\beq
\Phi_1' \to {\rm const.} (z+\zeta_A^N/\zeta_C^N)^{1/N} \diag( 1, \omega_N,\cdots, \omega_N^{N-1}).
\eeq
Then, $(\Phi_1')^{-1}$ is singular at this point. These two problems can be solved simultaneously.
We take a holomorphic section of the degree 1 line bundle $L_2'$ as,
\beq
\lambda=\frac{\zeta_B}{(\zeta_A^N+\zeta_C^N)^{1-1/N}}(\zeta_C^N z+\zeta_A^N),
\eeq
and set
\beq
\Phi_2'=\lambda (\Phi_1')^{-1}.\label{eq:21solphi2}
\eeq
Now both the left and right hand side are sections of the same bundle. Due to the zero of $\lambda$ at $z=-(\zeta_A^N/\zeta_C^N)$,
the singular behavior is avoided. The overall factor of $\lambda$ is chosen so that $\Phi_2'$ behaves at $z=1$ as
$\Phi_2' \sim \zeta_B (1-z)^{-1/N}\diag(1,\omega_N,\cdots,\omega_N^{N-1})$.
In this way, $\Phi_2'$ has all the desired property. 

We summarize the above solution as the spectral curve,
\beq
x_1'^N&=\frac{(\zeta_C^N z+\zeta_A^{N})(1-z)}{z}, \\
x_1'x_2'&=\frac{\zeta_B}{(\zeta_C^N+\zeta_A^N)^{1-1/N}}(\zeta_C^N z+\zeta_A^N), 
\eeq
where $(x_1', x_2')$ are the eigenvalues of $(\Phi_1', \Phi_2')$.
See appendix~\ref{app:B} for a rigorous derivation of this curve.

Let us compute the superpotential vev.
We use the general formula \eqref{eq:nomalizedsup}.
First we have to go back to the original fields $\Phi_1=s_1^{-1}\Phi'_1$ and $\Phi_2=s_2^{-1}\Phi'_2$.
The sections $s_1$ and $s_2$ introduced in subsection~\ref{sec:preliminary} are given as
\beq
s_1=z,~~~s_2=(z-1),
\eeq
where $s_1$ is a section of the degree two line bundle and $s_2$ is a section of the degree one line bundle.
Then we get
\beq
x_1x_2=\frac{\zeta_B}{(\zeta_C^N+\zeta_A^N)^{1-1/N}} \left(\frac{\zeta_C^N z+\zeta_A^N}{z(z-1)} \right) .\label{eq:x1x2prod}
\eeq
The formula \eqref{eq:nomalizedsup} gives
\beq
W &=  \oint _{z \sim 1} \frac{dz}{ 2 \pi i} x_1x_2N \nonumber \\
&=N\zeta_B (\zeta_C^N+\zeta_A^N)^{1/N},
\eeq
where the factor $N$ comes from the fact that $\Phi_1\Phi_2 \propto {\bf 1}_N$ and $\Tr {\bf 1}_N=N$.
This result agrees with the field theory estimate \eqref{eq:condense21} 
if the parameters are identified as $\zeta_A^N = \Lambda_A^{2N}$, $\zeta_C^N =  \Lambda_C^{2N}$, and $\zeta_B^N = \Lambda_B^N$.
Therefore, the spectral curve contains both the information of the Seiberg-Witten curve and the dynamical superpotential vev.

\subsubsection{$T_N$ theory coupled to singlets}
Here we do not gauge any flavor groups of the $T_N$ theory.
Instead, we introduce singlets $M_A$ in the adjoint representation of the flavor group $\SU(N)_A$.
We take a superpotential,
\beq
W=\tr (M_A \mu_A).\label{eq:singletTNcoup}
\eeq

The moduli spaces studied in \ref{sec:TNN1} and \ref{sec:N1N2mixed} are ``Higgs branch'' in the sense that
the Higgs branch operators of the original $T_N$ theory have nonzero vevs.
In this subsection, we consider ``Coulomb branch'' in the sense that the vevs of Coulomb branch operators of the $T_N$ theory are turned on.

Let us give a nonzero generic vev to $M_A$. From the point of view of the Coulomb branch of the $T_N$ theory,
the superpotential \eqref{eq:singletTNcoup} gives a mass term associated to the flavor symmetry $\SU(N)_A$,
with the mass matrix given by the vev of $M_A$. Thus the Seiberg-Witten curve is the same as that of the $T_N$
theory with the mass $M_A$. Our purpose is to reproduce this result from the twisted 5d SYM.

As explained in appendix~\ref{app:A}, the above theory is realized by a Riemann sphere with three regular punctures.
At two of the punctures, $z=1$ and $z=\infty$, the $\Phi_2$ has singularities. These punctures are associated to
$\SU(N)_B$ and $\SU(N)_C$. The $\Phi_1$ has a singularity at one puncture $z=0$ which is associated to $\SU(N)_A$.
The degrees of the line bundles are $\deg L'_1=0$ and $\deg L'_2=1$, or equivalently $\deg L_1=-1$ and $\deg L_2=-1$.
From the result \eqref{eq:nomalizedop2}, we have
\beq
\Phi_2(z=0) \approx M_A.
\eeq
Notice we have taken $M_A$ instead of $\mu_A$. Actually, in the present theory we get $\mu_A=0$ by the equation of motion of $M_A$,
and $\Phi_2(z=0)$ should be identified as $M_A$. See appendix~\ref{app:A}.

Now, let us take a section $s$ of the degree $-1$ line bundle given as $s=1/z$, and define
\beq
\Phi''_2=s \Phi_2=\frac{\Phi_2}{z}.
\eeq
Now $\Phi''_2$ takes values in the degree $-2$ line bundle, which is the canonical bundle $K$ on the Riemann sphere.
If we set $\Phi_1=0$, the spectral curve is given as
\beq
0=\det(x''-\Phi''_2).
\eeq
Because $\Phi''_2$ takes values in the canonical bundle, the spectral curve is the same as that of the original Coulomb branch of the $T_N$ theory.
The only change from the $\CN=2$ case is that the singularity at $z=0$ is given as
\beq
\Phi''_2 \to \frac{M_A}{z}.\label{eq:artificialpole}
\eeq
This singularity exactly matches with the fact that $M_A$ gives the mass of the $T_N$ theory as discussed above from the field theory point of view.

In the actual $\CN=2$ case with the singularity $\Phi_2 \to m/z$, the mass $m$ is a non-normalizable deformation.
The Kahler potential \eqref{eq:infkahler} gives an infinite kinetic term for $m$, and hence $m$ is frozen and it is not a moduli field.
However, in the theory considered in this subsection, the pole \eqref{eq:artificialpole} has been introduced in an artificial way.
The singlets $M_A$ have finite kinetic terms and they are moduli field.

%\section{Discusstions}

\section*{Acknowledgments}
It is a pleasure to thank D.~Xie for many fruitful discussions which are crucial for this work. 
The author would also like to thank K. Maruyoshi and Y. Tachikawa for helpful discussions.
The work of K.Y. is supported in part by NSF grant PHY- 0969448.

%%%%%%%%%%%%%%%%%%%%%%%%%%%%%%%%%%%%%%%%%%%%%%%%%%%%%%%%%%%
\appendix
\section{$\CN=1$ theories of class~$\CS$ and dualities}\label{app:A}
In this appendix, we review
$\CN=1$ class $\CS$ theories and their dualities, mainly based on \cite{Bah:2012dg,Gadde:2013fma,Xie:2013gma,Xie:2013rsa}. 
An important ingredient is the $T_N$ theory.\footnote{This is different from the $T[\SU(N)]$ theory discussed in section~\ref{sec:3}.}
The $T_N$ theory has flavor symmetries 
$\SU(N)_A \times \SU(N)_B \times \SU(N)_C$, and there are chiral multiplets
$\mu_A$, $\mu_B$ and $\mu_C$ in the adjoint representations of the corresponding flavor groups.
See section~\ref{sec:5} for more detailed review.

\paragraph{Field theory.}
Take two copies of the $T_N$ theory. One of them has global symmetries
$\SU(N)_A \times \SU(N)_B \times \SU(N)_1$ and the other one has $\SU(N)_C \times \SU(N)_C \times \SU(N)_2$.
We will specify the copies of the $T_N$ theory by their flavor symmetries as $T_N{(A,B,1)}$ and $T_N{(C,D,2)}$.
First let us briefly recall $\CN=2$ dualities.
We gauge the diagonal subgroup $\SU(N)_g \subset \SU(N)_1 \times \SU(N)_2$ which is embedded as
$\SU(N)_g \ni g \mapsto (g,{}^t g^{-1}) \in \SU(N)_1 \times \SU(N)_2$. The superpotential is
\beq
\tr \left(\phi (\mu_1-{}^t\mu_2)\right),\label{eq:N=2gluing}
\eeq
where $\phi$ is the adjoint chiral field of the $\CN=2$ vector multiplet, and 
we have omitted the usual $\sqrt{2}$ factor for simplicity.
We may say that the two copies of the $T_N$ theory are glued by the $\CN=2$ vector multiplet.

This theory is dual to the following theory. We take $T_N(A,C,1)$ with flavor symmetry $\SU(N)_A \times \SU(N)_C \times \SU(N)_1$
and $T_N(B,D,2)$ with flavor symmetry $\SU(N)_B \times \SU(N)_D \times \SU(N)_2$. The diagonal subgroup of
$\SU(N)_1 \times \SU(N)_2$ is gauged. This theory is dual to the above theory.
There are precise correspondences of the flavor symmetries $\SU(N)_{A,B,C,D}$ in the original and dual theories,
but $\SU(N)_1$ and $\SU(N)_2$ are gauged and hence there are no gauge invariant relations of these groups between the original and dual 
theories. In this duality, the UV coupling constant $\tau$ is mapped as $\tau \leftrightarrow -\tau^{-1}$.

The $\CN=1$ dualities discussed in \cite{Gadde:2013fma} are similar to the $\CN=2$ dualities, but we need more labels
to specify the theories. A sign $\pm$ is assigned to each copy of the $T_N$ theory. Each flavor symmetry also has a sign $\pm$.
Thus each $T_N$ theory is labeled like $T_N^{(\pm)}(A^\pm,B^\pm,1)$.

For example, let us glue $T_N^{(+)}(A^+,B^+,1)$ and $T_N^{(-)}(C^-,D^-,2)$.
They are glued by an $\CN=1$ vector multiplet, and
the superpotential is taken as
\beq
W=c\tr (\mu_1 {}^t\mu_2),
\eeq
where $c$ is an exactly marginal coupling at the IR fixed point.
This theory is dual to a theory in which $T_N^{(+)}(A^+,C^-,1)$ and $T_N^{(-)}(B^+,D^-,2)$ are
glued by an $\CN=1$ vector multiplet with a superpotential 
\beq
W=c' \tr (\mu_1 {}^t\mu_2)+\tr (\mu_B M_B)+\tr (\mu_C M_C),   \label{eq:dual1}
\eeq
where we have introduced new singlets $M_B$ and $M_C$ which are in the adjoint representations of $\SU(N)_B$ and $\SU(N)_C$, respectively.
The $\mu_{B,C}$ of the original theory are dual to $M_{B,C}$, similar to the case in Seiberg duality~\cite{Seiberg:1994pq}.
Furthermore, the theory is dual to a theory in which $T_N^{(+)}(C^-,D^-,1)$ and $T_N^{(-)}(A^+,B^+,2)$ are
glued by an $\CN=1$ vector multiplet with a superpotential 
\beq
W=c'' \tr (\mu_1 {}^t\mu_2)+\tr (\mu_A M_A)+\tr (\mu_B M_B)+\tr (\mu_C M_C)+\tr (\mu_D M_D).
\eeq

The general rule is the following. 
\begin{enumerate}
\item When two copies of the $T_N$ theory are glued,
the vector multiplet used in the gluing is an $\CN=2$ vector multiplet with the coupling \eqref{eq:N=2gluing}
if the two $T_N$ have the same sign, i.e., if the combinations are $T_N^{(+)}$ and $T_N^{(+)}$, or $T_N^{(-)}$ and $T_N^{(-)}$.
If they have different signs, that is, if $T_N^{(+)}$ and $T_N^{(-)}$ are glued, we use an $\CN=1$ vector multiplet
with a superpotential 
\beq
c\tr (\mu_1 {}^t\mu_2).\label{eq:N1gluesuperpot}
\eeq
\item If the sign of $T_N$ and one of its flavor symmetries, say $A$,
are different, such as $T_N^{(+)}(A^-,*,*)$ and $T_N^{(-)}(A^+,*,*)$, then we introduce singlets $M_A$
in the adjoint representation of the flavor group $\SU(N)_A$. The $M_A$ are mesons. We also take a superpotential
\beq
\tr (\mu_A M_A).\label{eq:N1mesoncoup}
\eeq
If the sign of $T_N$ and a flavor symmetry is the same as $T_N^{(+)}(A^+,*,*)$ and $T_N^{(-)}(A^-,*,*)$,
there is no new ingredient.
\end{enumerate}

In a generalized quiver, there are following data:
\begin{enumerate}
\item The number of $T_N^{(+)}$, denoted as $p$, and the number of $T_N^{(-)}$, denoted as $q$.
\item The number of flavor symmetries with $+$ sign, denoted as $n_+$, and the number of flavor symmetries with $-$ sign,
denoted as $n_-$.
\end{enumerate}
For example, in the example discussed above, all the dual theories have $(p,q,n_+,n_-)=(1,1,2,2)$.
The claim is that theories with the same set of numbers $(p,q,n_+,n_-)$ are dual to each other.
(More precisely, dual theories are specified by $(p,q)$ and the set of flavor symmetries $\{A^\pm,B^\pm,\cdots \}$.)
The class of theories are constructed by following the general rule described above.

We will soon discuss that the set $\{A^\pm,B^\pm,\cdots \}$ corresponds to punctures on a Riemann surface.
In the above discussion, we have only considered the case that all the punctures are maximal.
However, it is possible to use more general punctures. 
Let us consider the case in which $A^+$ and $D^-$ are simple punctures and $B^+$ and $C^-$ are maximal ones.
Then, for example, $T_N^{(+)}(A^+,B^+,1)$ is no longer the $T_N$ theory, but it is a bifundamental $q_i^\alpha, \tilde{q}^i_\alpha$,
where $i=1,\cdots,N$ is a flavor index for $\SU(N)_B$ and $\alpha$ is a gauge index.
Similarly, $T_N^{(-)}(C^-,D^-,2)$ is now a bifundamental $p_\ell^\alpha, \tilde{p}^\ell_\alpha$ 
where $\ell=1,\cdots,N$ is a flavor index for $\SU(N)_C$. 
The theory constructed by gluing these two bifundamentals is
an $N_f=2N$ SQCD with a superpotential \eqref{eq:N1gluesuperpot}, where 
\beq
(\mu_1)^\alpha_\beta&=q^\alpha_i \tilde{q}_\beta^i -\frac{\delta^\alpha_\beta}{N}q^\gamma_i \tilde{q}_\gamma^i ,\\
({}^t\mu_2)^\beta_\alpha &=p^\beta_\ell \tilde{p}_\alpha^\ell -\frac{\delta_\alpha^\beta}{N}p^\gamma_\ell \tilde{p}_\gamma^\ell.
\eeq
The dual theory with $T_N^{(+)}(A^+,C^-,1)$ and $T_N^{(-)}(B^+,D^-,2)$ as bifundamentals are constructed 
similarly using $\eqref{eq:dual1}$. This is essentially the same as Seiberg duality.
The dual theory using $T_N^{(+)}(C^-,D^-,1)$ and $T_N^{(-)}(A^+,B^+,1)$
is more nontrivial. See \cite{Gadde:2013fma} for details.

All of the above theories are conformal and all the punctures are regular. However, we can also consider non-conformal cases.
Let us start from the theory constructed by $T_N^{(+)}(A^+,B^+,1)$ and $T_N^{(-)}(C^-,D^-,2)$.
Then, for example, we introduce an $\CN=2$ vector multiplet coupled to the group $\SU(N)_B$.
We also introduce $N_f $ flavors of quarks $q_i, \tilde{q}^i~(i=1,\cdots,N_f)$ coupled to the $\SU(N)_B$ gauge group.
The superpotential is
\beq
W \supset  \tr \phi_B (\mu_B+q_i \tilde{q}^i).
\eeq
For $N_f<N$, the puncture $B^+$ corresponds to a irregular puncture.
(It corresponds to \eqref{eq:sing2} of section~\ref{sec:6}).

The above puncture $B^+$ is ``locally''  an $\CN=2$ irregular puncture, but we can get an $\CN=1$ dual of this puncture.
Going to the dual theory which is constructed by $T_N^{(+)}(A^+,C^-,1)$ and $T_N^{(-)}(B^+,D^-,2)$,
we get
\beq
W \supset  \tr \phi_B (M_B+q_i \tilde{q}^i)+\tr (M_B \mu_B),
\eeq
where we have used the fact that $\mu_B$ of the original theory is dual to $M_B$. 
The adjoint fields $\phi_B$ and $M_B$ become massive and can be integrated out.
Then we get
\beq
W \supset -\tr (\mu_B q_i \tilde{q}^i).\label{eq:rotatedirregular}
\eeq
Therefore, the $\CN=1$ dual of the $\CN=2$ irregular puncture is given by the $\CN=1$ vector multiplet
and quarks with the superpotential \eqref{eq:rotatedirregular}.
When $N_f=N$, one can see that \eqref{eq:rotatedirregular} is just the same as \eqref{eq:N1gluesuperpot}.
In this $N_f=N$ case, we get one simple and one maximal regular punctures instead of one irregular puncture.

\paragraph{(2,0) theory interpretation.}

\begin{figure}
\begin{center}
\includegraphics[scale=0.26]{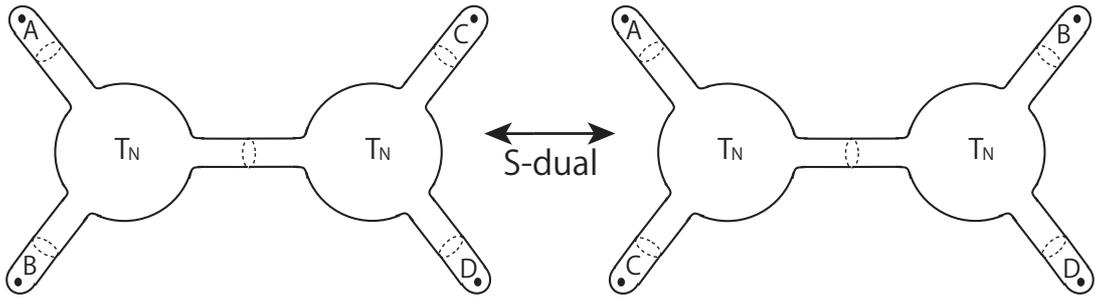}
\caption{$S$-duality in $\CN=2$ theory. Both the left and right figures are just a Riemann sphere with four punctures $A,B,C$ and
$D$, but different degeneration limits give different field theory realizations.}\label{fig:n2dual}
\end{center}
\end{figure}

The dualities discussed above have a nice interpretation in terms of the $\CN=(2,0)$ theories compactified on a Riemann surface.
First, let us recall the $\CN=2$ dualities. The theory constructed by gluing $T_N(A,B,1)$ and $T_N(C,D,2)$
is realized as a Riemann sphere with four punctures $A,B,C$ and $D$ as in figure~\ref{fig:n2dual}.
The theory is manifestly  dual to the theory constructed by gluing $T_N(A,C,1)$ and $T_N(B,D,2)$,
and so on. Different degeneration limits give different field theory realizations as in the figure.

\begin{figure}
\begin{center}
\includegraphics[scale=0.26]{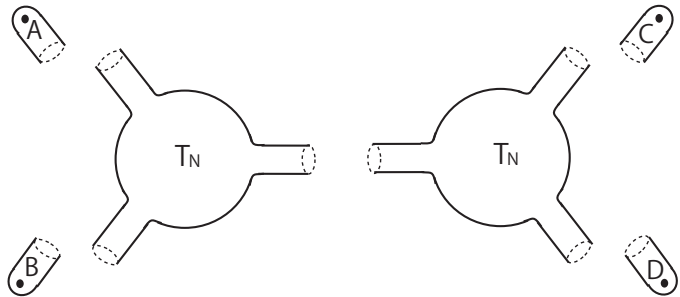}
\caption{Decomposition of the Riemann surface into pieces. By gluing them, we can get a 4d field theory.
Although each piece does not have a direct field theory interpretation, it is convenient to consider this decomposition
to understand the $\CN=1$ dualities.
}\label{fig:bara}
\end{center}
\end{figure}

We decompose the Riemann surface into several pieces as in figure~\ref{fig:bara}.
In the figure, a Riemann sphere with three holes corresponds to the ``body'' of a copy of $T_N$.
There are also cap-like pieces with a puncture on it.
These caps are glued to a hole of the Riemann sphere.
Gauging the diagonal subgroup of $\SU(N)_1 \times \SU(N)_2$ of two copies of
$T_N$ is realized by gluing two holes of the Riemann spheres.
Near the boundaries of these pieces, the metric is flat and the Riemann surface locally looks like $S^1 \times {\mathbb R}$.

On each piece of the Riemann surface, we take two line bundles. These two line bundles 
are the ones in $F=L_1 \oplus L_2$ which are used in the twisting of the $(2,0)$ theory (or 5d SYM) as in section~\ref{sec:2}.
We take one line bundle to be the canonical bundle, denoted as $K$,
and the other is the trivial bundle, denoted as $\CO$. 
In the $\CN=2$ case, canonical bundles are glued together and trivial bundles are glued together,
defining the canonical bundle and trivial bundle on the entire Riemann surface.
We get $L_1=K$ and $L_2=\CO$ in this case.

Now let us discuss the $\CN=1$ case. We use different gluing of the line bundles from the $\CN=2$ case.
The general rule is the following:
\begin{enumerate}
\item For $T_N^{(+)}$, the line bundles of the corresponding Riemann sphere with three holes
are taken such that $ L_1 \to K$ and $ L_2 \to \CO$. If it is $T_N^{(-)}$, we take 
$L_1 \to \CO$ and $L_2 \to K$.
\item For $A^+$, the line bundles of the corresponding cap are taken such that $L_1 \to K$ and $ L_2 \to \CO$.
The field $\Phi_1$ has a singularity at the puncture.
If it is $A^{-}$, we take $L_1 \to \CO$ and $L_2 \to K$. In this case, the $\Phi_2$ has a singularity.
\end{enumerate}
As we mentioned above, the metric near boundaries are flat, so the canonical bundle is trivial near the boundaries and hence
the gluing of the line bundles are straightforward. The gluing requires complex parameters which determine the complex moduli of the line bundles.
These parameters, combined with the complex moduli of the Riemann surface, correspond to exactly marginal couplings of the field theory.

The above rule matches with the field theory rule very well.
Let us glue two copies of $T_N$. If both of them has the same sign, 
the above gluing of pieces of the Riemann surface is locally very similar to the case of $\CN=2$, aside from the possible complex
parameters in the gluing.
Therefore, it is natural that we get an $\CN=2$ vector multiplet.
Complex parameters in the gluing correspond to changing \eqref{eq:N=2gluing} as $\tr \left(\phi (c\mu_1- c^{-1}{}^t\mu_2)\right)$ for a parameter $c$.
On the other hand, if two $T_N$'s of different signs are glued, the gluing process breaks half
of the supersymmetry. Our interpretation is that this gluing gives an $\CN=1$ vector multiplet with
the superpotential \eqref{eq:N1gluesuperpot}. A similar thing can happen
in the gluing of $T_N$ and a cap labeled by $A$. If they have the same sign,
the gluing is locally $\CN=2$. On the other hand, if they have different signs,
the gluing breaks half of the supersymmetry. We interpret that this process introduces the meson $M_A$
and the coupling \eqref{eq:N1mesoncoup}.

Actually, the above picture can be checked in type IIA brane construction~\cite{Xie:2013gma}.
We prepare several branes as in table~\ref{tab:1}.
These branes preserve 4d $\CN=1$ supersymmetry and can be used to construct gauge theories~\cite{Hori:1997ab,Witten:1997ep}.
Let us denote a simple puncture as $S$ and a maximal puncture as $M$. 
Then, an NS5 brane corresponds to $T^{(+)}(*,*,S^{+})$ and an NS5' brane corresponds to $T^{(-)}(*,*,S^{-})$.
$N$ D6 branes correspond to $M^{+}$  and $N$ D6' branes corresponds to $M^{-}$, etc. 
Gluing pieces of the Riemann surface corresponds to suspending $N$ D4 branes between these NS5, NS5', D6 and D6' branes.
From this interpretation, we can see the properties described above.

\begin{table}
\begin{center}
\begin{tabular}{|c|c|c|c|c|c|c|c|}
\hline
& 0,1,2,3&4&5&6&7&8&9   \\
\hline
D4 & $\circ$&&&$\circ$&&& \\
\hline
NS5 &$\circ$&$\circ$&$\circ$&&&& \\
\hline 
D6 & $\circ$&&&&$\circ$&$\circ$&$\circ$ \\
\hline
NS5' &$\circ$&&&&$\circ$&$\circ$& \\ 
\hline
D6' &$\circ$&$\circ$&$\circ$&&&&$\circ$ \\
\hline
\end{tabular}
\caption{Branes which preserve 4d $\CN=1$ supersymmetry, and their extended directions.}
\label{tab:1}
\end{center}
\end{table}

The degrees of the line bundles $L_1$ and $L_2$ are determined as follows.
A Riemann sphere with three holes has the Euler number $2-3=-1$ and hence the canonical bundle $K$
has degree $\deg K=+1$.
A cap region surrounding a puncture has the Euler number $2-1=+1$ and hence $K$ has degree $\deg K=-1$.
Here the degree is defined as the integral of the first Chern class $\int c(K)$, which is well defined under the condition that
the metric is flat near the boundaries.
Therefore, using the above general rule, we get
\beq
\deg L_1=p-n_+,~~~~~\deg L_2=q-n_-.
\eeq
This is an important equation to construct the twisted 5d SYM from field theory data.

The above gluing rule is not the most general one consistent with the $\CN=1$ supersymmetry.
Take two copies of $T_N$, denoted as $T_N^{(1)}$ and $T_N^{(2)}$. Then they have rank two bundles
$F^{(1)}=\CO^{(1)} \oplus K^{(1)}$ and $F^{(2)}=\CO^{(2)} \oplus K^{(2)}$.
In general, we can glue the two bundles $F^{(1)}$ and $F^{(2)} $ in more complicated ways
using a $\U(2)$ matrix. The same is true for a gluing of $T_N $ and $A$.
See \cite{Benini:2009mz} where such gluing has essentially appeared as Wilson lines of $\SU(2) \subset \U(2)$.
It would be interesting to study this case more systematically.

%%%%%%%%%%%%%%%%%%%%%%%%%%%%%%%%%%%%%%%%%%%%%%%%%%%%%%%%%%%
\section{Solving generalized Hitchin's equations}\label{app:B}
Here we develop a method to determine spectral curves based on the set of equations \eqref{eq:U(2)generalcurve},
\beq
0=P_{i_1 \cdots i_N}(x_1,x_2) \equiv \frac{1}{N!}(x_{i_1} -\Phi_{i_1})^{\alpha_1}_{~\beta_1} \cdots (x_{i_N} -\Phi_{i_N})^{\alpha_N}_{~\beta_N}
\epsilon_{\alpha_1\cdots \alpha_N} \epsilon^{\beta_1 \cdots \beta_N}. \label{eq:U(2)generalcurveagain}
\eeq
We only consider the case $F=L_1 \otimes L_2$, i.e., $\Phi_1$ and $\Phi_2$ are sections of $L_1 \otimes {\rm ad}(E)$ and $L_2 \otimes {\rm ad}(E)$,
respectively. In this case, it is convenient to define
\beq
\phi_{k,\ell}(z)=(-1)^{k+\ell}\frac{     (\Phi_{1})^{\alpha_1}_{~\beta_1} \cdots (\Phi_{1})^{\alpha_k}_{~\beta_k}
 (\Phi_{2})^{\alpha_{k+1}}_{~\beta_{k+1}} \cdots (\Phi_{2})^{\alpha_{k+\ell}}_{~\beta_{k+\ell}}
 \delta_{\alpha_1\cdots \alpha_k \alpha_{k+1}\cdots \alpha_{k+\ell} \gamma_{k+\ell+1} \cdots \gamma_{N}}  ^{\beta_1 \cdots \beta_k \beta_{k+1} \cdots \beta_{k+\ell}  \gamma_{k+\ell+1} \cdots \gamma_{N} }
}{k!\ell ! (N-k-\ell)!} ,\label{eq:defsections}
\eeq
where
$
\delta_{\alpha_1 \cdots \alpha_N}^{\beta_1 \cdots \beta_N}=\epsilon_{\alpha_1 \cdots \alpha_N} \epsilon^{\beta_1 \cdots \beta_N}.
$
The $\phi_{k,\ell}$ is a section of line bundle $L_1^k \otimes L_2^\ell$.
Then, $P_{i_1 \cdots i_N}$ is given as
\beq
P_{1\cdots 1 2\cdots 2}= x_1^{N-m} x_2^{m}+\sum_{k, \ell} \frac{(N-m)!m! (N-k-\ell)!}{N!(N-m-k)! (m-\ell)!}  \phi_{k, \ell}(z) x_1^{N-m-k} x_2^{m-\ell} ,
\label{eq:overdetermed}
\eeq
where we have taken $i_1=\cdots=i_{N-m}=1$ and $i_{N-m+1}=\cdots=i_N=2$ in $P_{i_1 \cdots i_N}$.

Particularly important equations in the following discussions are
\beq
0&=P_1 \equiv x_1^N+\sum_{k=2}^N \phi_{k,0}(z) x_1^{N-k}, \label{eq:vNw0}\\
0&=\frac{\partial P_1(x_1,z)}{\partial x_1} x_2+\sum_{k=1}^{N-1} \phi_{k,1}(z) x_1^{N-1-k}.\label{eq:vN-1w1} 
\eeq
The second equation \eqref{eq:vN-1w1} can be explicitly checked using \eqref{eq:defsections} and \eqref{eq:overdetermed}.
These equations defines an $N$-covering $\Sigma$ of the Riemann surface $C$.
An equation similar to \eqref{eq:vN-1w1} was also discussed in \cite{Bonelli:2013pva}, but our equation
is more explicitly given in terms of the Higgs fields $\Phi_{1,2}$.

We claim (without complete proof) that these two equations are enough to determine the curve $\Sigma=\{P_{i_1 \cdots i_N}=0\}$
in the case where the eigenvalues of $\Phi_1$ are distinct on generic points of the Riemann surface.
The motivation of this claim is as follows. As explained in \eqref{eq:generalcurve2}, the curve $\Sigma$
is given by $(x_1,x_2)=(\lambda_{1,k},\lambda_{2,k})~(k=1,\cdots,N)$, where $(\lambda_{1,k},\lambda_{2,k})$
are pairs of eigenvalues of $(\Phi_1,\Phi_2)$. The equation $P_{11\cdots 1}=0$, \eqref{eq:vNw0}, sets $x_1=\lambda_{1,k}$ for some $k$.
Then, the equation $P_{21\cdots 1}=0$, \eqref{eq:vN-1w1}, gives
\beq
(x_2-\lambda_{2,k})\prod_{\ell \neq k} (\lambda_{1,k}-\lambda_{1,\ell})=0,
\eeq
where we have used $x_1=\lambda_{1,k}$. Therefore, when $\lambda_{1,k} \neq \lambda_{1,\ell}$ for $k \neq \ell$,
we get $(x_1,x_2)=(\lambda_{1,k},\lambda_{2,k})$ as desired.

Note that we have only assumed that the eigenvalues of $\Phi_1$ do not degenerate at generic points.
At some discrete points on the Riemann surface, degeneration of the eigenvalues can occur. In fact, such
points are important to determine the curve as we will see below.

The above result is also represented as follows. A generalization of the Cayley-Hamilton theorem is that
the commuting matrices $\Phi_1$ and $\Phi_2$ satisfy $P_{i_1 \cdots i_N} (\Phi_1, \Phi_2)=0$ where 
we have substituted the matrices $(\Phi_1, \Phi_2)$ for $(x_1,x_2)$ in $P_{i_1 \cdots i_N}(x_1,x_2)$ defined in \eqref{eq:U(2)generalcurveagain}.
When eigenvalues are not degenerate, this formula is proved by simultaneously diagonalizing $\Phi_1$ and $\Phi_2$.
The case where some of the eigenvalues are degenerate can be reached as a limit of the non-degenerate case,
and hence the theorem is proved. If we are given a solution for $\Phi_1$, \eqref{eq:vN-1w1} tells us
\beq
\Phi_2=- \left [\frac{\partial P_1}{\partial x_1}(x_1=\Phi_1) \right]^{-1}\sum_{k=1}^{N-1} \phi_{k,1} \Phi_1^{N-1-k}.\label{eq:Phi2determine}
\eeq
This is the motivation for our claim that \eqref{eq:vNw0} and \eqref{eq:vN-1w1} are enough in generic case;
the curve \eqref{eq:vNw0} and the points of its Jacobian variety may determine $(A_{\bar{z}},\Phi_1)$ as in the original Hitchin systems' case,
and $\Phi_2$ is uniquely determined by \eqref{eq:Phi2determine}.
We do not try to give a complete mathematical proof here.

\subsection{Constraint from commuting condition} \label{sec:commuting}
The crucial difference between generalized Hitchin systems and original Hitchin systems is that there is the commuting condition $[\Phi_1,\Phi_2]=0$.
Here we explain why this commuting condition strongly constrains solutions of generalized Hitchin's equations~\cite{Xie:2013rsa}. 

If eigenvalues of $\Phi_1$ and $\Phi_2$ are generic, the commuting condition just implies that the two matrices $\Phi_1$ and $\Phi_2$
are simultaneously diagonalizable. However, when some of the eigenvalues of $\Phi_1$ or $\Phi_2$ become degenerate,
the commuting condition even constrain the eigenvalues of them.

Let us first see a simple $\SU(2)$ example. Suppose that near $z \sim 0$, $\Phi_1$ behaves as $\det (x_1-\Phi_1)=x_1^2-z$.
The two eigenvalues of $\Phi_1$ are degenerate at $z=0$. In a diagonal form, $\Phi_1$ is given as $\Phi_1=\diag(z^{1/2},-z^{1/2})$.
However, this is not single valued. In a more appropriate basis, $\Phi_1$ may be given as
\beq
\Phi_1=\left( \begin{array}{cc}
0 & 1 \\
z & 0
\end{array}\right).
\eeq
Then, imposing $[\Phi_1,\Phi_2]=0$, one can easily see that $\Phi_2$ at $z=0$ must be of the form
\beq
\Phi_2(z=0)=\left( \begin{array}{cc}
0 & c \\
0 & 0
\end{array}\right).
\eeq
Therefore, the eigenvalues of $\Phi_2$ must also degenerate at $z=0$.

Another way of seeing this condition is the following.
For the $\SU(2)$ case, \eqref{eq:U(2)generalcurveagain} gives
\beq
x_1^2=f(z),~~~x_2^2=g(z),~~~x_1x_2=h(z), \label{eq:SU2threeeq}
\eeq
where $f(z)=\Tr (\Phi_1^2)/2$, $g(z)=\Tr (\Phi_2^2)/2$ and $h(z)=\Tr (\Phi_1 \Phi_2)/2$. For these equations to be consistent,
we need to have $f(z)g(z)=h(z)^2$. Then, if $f(z)$ has a simple zero at $z=0$, $g(z)$ must also have a zero of odd degree at the same point
so that $h(z)$ is holomorphic. In this way we get the same conclusion as above about the degeneracy of the matrices $\Phi_1$ and $\Phi_2$.
The argument based on \eqref{eq:SU2threeeq} might look quite different from the argument based on the commuting condition.
However, recall that the commuting condition was
the essential reason that the over-determined equations \eqref{eq:SU2threeeq} or \eqref{eq:U(2)generalcurveagain} define a consistent curve,
as explained in subsection~\ref{sec:generalizedH}.

Now let us consider a general constraint for $\SU(N)$ case.
The commuting condition $[\Phi_1, \Phi_2]=0$ tells us that the matrices $\Phi_1$ and $\Phi_2$ are generically simultaneously diagonalizable.
Suppose that the eigenvalues of $\Phi_1$ are generic enough so that all the eigenvalues are distinct at generic points of the Riemann surface $C$.
Then, we can expand $\Phi_2$ as
\beq
\Phi_2=\sum_{k=0}^{N-1} f_k \Phi_1^k. \label{eq:matrixexpansion}
\eeq
The coefficients $f_k$ may be determined by solving e.g.,
\beq
\sum_{k=0}^{N-1} \tr (\Phi_1^{m+k})f_k= \tr(\Phi_1^m \Phi_2),~~~(m=0,\cdots, N-1).
\eeq
In the following discussion, we do not need the explicit form of $f_k$. The important point is that
$f_k$ are given by gauge invariant polynomials of $\Phi_1$ and $\Phi_2$, i.e., $\tr \Phi_1^m \Phi_2^k$.
Therefore, they are single-valued on the Riemann surface $C$.

We have assumed above that the eigenvalues of $\Phi_1$ are distinct at generic points of the Riemann surface.
However, at discrete set of points, their eigenvalues degenerate. These points are the branching points of the cover $\det(x_1-\Phi_1)=0$.
Suppose that $z=0$ is one of these points and $\det(x_1 -\Phi_1)$ behaves as
\beq
\det(x_1 -\Phi_1) \sim \prod_\ell [(x_1-a_\ell)^{n_\ell}-b_\ell^{n_\ell} z^{m_\ell}], \label{eq:branch1}
\eeq
where $a_\ell$ and $b_\ell$ are constants, $n_\ell$ and $ m_\ell$ are relatively prime integers such that
$\sum_\ell n_\ell=N$. The integers $m_\ell$ can be negative so that we can also treat behaviors at punctures.
The constants $a_\ell$ may or may not be zero, but we assume $b_\ell \neq 0$. 
Then the eigenvalues of $\Phi_1$ behave as
\beq
\Phi_1 \sim \diag\left[ \bigoplus_\ell \left( a_\ell+b_\ell z^{m_\ell / n_\ell},\cdots, a_\ell+(\omega_{n_\ell})^{n_\ell -1}b_\ell z^{m_\ell/n_\ell} \right) \right],
\label{eq:Phi1eigen}
\eeq
where $\omega_{n_\ell}=\exp(2\pi i/ n_\ell)$. 

Now we use \eqref{eq:matrixexpansion} to determine the behavior of $\Phi_2$.
Because $f_k$ are single-valued on the Riemann surface and cannot have fractional powers of $z$ around $z=0$,
we obtain
\beq
\Phi_2 \sim \diag \left[ \bigoplus_\ell \left( c_\ell z^{r_\ell} +d_\ell z^{p_\ell+q_\ell m_\ell / n_\ell},\cdots, c_\ell z^{r_\ell}+
(\omega_{n_\ell}^{q_\ell})^{n_\ell -1}d_\ell z^{p_\ell+q_\ell m_\ell / n_\ell} \right) \right],\label{eq:Phi2eigen}
\eeq
where $p_\ell$, $q_\ell$ and $r_k$ are integers such that $q_\ell$ is not a multiple of $n_\ell$.

As a special case of \eqref{eq:Phi1eigen} and \eqref{eq:Phi2eigen}, 
the following simple observation will be useful.
When $n_\ell=N$, the traceless condition $\Tr \Phi_1=\Tr \Phi_2=0$ suggests that $a_\ell=c_\ell=0$.
Therefore, if $\Phi_1$ behaves as $\Phi_1 \sim z^{m_\ell/N} \diag(1,\omega_N, \cdots, \omega_N^{N-1})$
with $m_\ell$ and $N$ relatively prime,
$\Phi_2$ must behave as $\Phi_2 \sim z^{p_\ell+q_\ell m_\ell /N} \diag(1,\omega_N^{q_\ell}, \cdots, (\omega_N^{q_\ell})^{N-1})$
for $q_\ell$ which is not a multiple of $N$. In particular, if $\Phi_2$ does not diverge at $z=0$, it must actually vanish, $\Phi_2 \to 0$;
\beq
{\rm If~}\Phi_1 \to z^{m_\ell/N} \diag(1,\omega_N, \cdots, \omega_N^{N-1}) ,{\rm ~then~}\Phi_2 \to 0 {\rm ~and~vice~versa.}\label{eq:commconapp}
\eeq

The above condition about degeneracy of eigenvalues of $\Phi_1$ and $\Phi_2$ indicates that the degeneration points
give constraints on the solutions of generalized Hitchin's equations.
This observation suggests the following strategy to determine \eqref{eq:vNw0} and \eqref{eq:vN-1w1}. 
First, note that some of the eigenvalues degenerate when 
we can find a solution to $P_1=\partial P_1/\partial x_1=0$.
Then, from \eqref{eq:vN-1w1}, we can see that $\sum_{k=1}^{N-1} \phi_{k,1}(z) x_1^{N-1-k}$ must be zero at these points.
This condition fixes parameters inside $\phi_{k,1}$.

%%%%%%%%%%%%%%%%%%%%%%%%%%%%%%%%%%%%%%%%%%%%%%%%%%%%%%%%%%%%%%%%%%%%%%%%%%%

\subsection{Solutions}

We give a derivation of the spectral curves of subsection~\ref{sec:NflessN} and \ref{sec:N1N2mixed} based on the strategy discussed above.
Here we use $\Phi'_1=s_1\Phi_1$ and $\Phi'_2=s_2\Phi_2$ as defined in subsection~\ref{sec:preliminary}.
We use $\phi'_{k,\ell}$ which denote sections defined by using $\Phi'_1$ and $\Phi'_2$ in \eqref{eq:defsections}.

\paragraph{Massive SQCD with $N_f<N$ flavors.}
Here we determine the curve of subsection~\ref{sec:NflessN}.
The behavior at $z \to 0$ is given as
\beq
&\Phi'_1 \to c_1  z^{1/N} \diag(1,\omega_N, \cdots, \omega_N^{N-1}) \\
&\Phi'_2\to \frac{\zeta_2}{z^{1/N}} \diag(1,\omega_N^{-1},\cdots,\omega_N^{-N+1}) 
\eeq
where we have used the result of subsection~\ref{sec:commuting} for the behavior of $\Phi_1$.
The constant $c_1$ is to be determined by solving the generalized Hitchin's equations.
Similarly, at $z=\infty$, we have
\beq
\Phi'_1 \to& \zeta_1 z^{1/(N-N_f)} \diag(0,\cdots, 0,1,\omega_{N-N_f}, \cdots, \omega_{N-N_f}^{N-N_f-1})  \nonumber \\
&+\diag ( m_1,\cdots,m_{N_f},m',\cdots,m'), \label{eq:zinfsing}\\
\Phi'_2  \to & \diag(c'_1, \cdots, c'_{N_f},c',\cdots, c') . 
\eeq
where $\sum_{i=1}^{N_f} m_i+(N-N_f)m=0$, $\sum_{i=1}^{N_f} c'_i+(N-N_f)c'=0$
and we have again used the result of subsection~\ref{sec:commuting}.

The singular behaviors above suggest the following:
\begin{enumerate}
\item At $z \to 0$, $\phi'_{k,\ell}$ behaves as
\beq
\phi'_{k,\ell} \to 
\left\{
\begin{array}{ll}
O(z^{1}) & k>\ell \\
O(z^{0}) & \ell \neq N \\
O(z^{-1}) & (k,\ell)=(0,N)
\end{array}
\right.
\eeq

\item At $z \to \infty$, $\phi'_{k,\ell}$ behaves as
\beq
\phi'_{k,\ell} \to 
\left\{
\begin{array}{ll}
O(z^{0}) & k  < N-N_f \\
O(z^{1}) & k  \geq  N-N_f
\end{array}
\right.
\eeq
\end{enumerate}
In determining the above behaviors, it is important to note that $\phi'_{k,\ell}$ are single valued functions of $z$,
and hence, for example, if $\phi'_{k,\ell} \to O(z^{1/N})$ at $z \to 0$, we must have $\phi'_{k,\ell} \to O(z^{1})$ etc.

The curve $\det(x'_1-\Phi'_1)=0$ is uniquely fixed by the above singular behaviors. 
For example, $\phi'_{k,0}$ for $2 \leq k < N-N_f$ must be zero to be consistent with the above behavior. For $k \geq N-N_f$,
$\phi'_{k,0}$ must be proportional to $z$, and their coefficients are fixed by the singularity \eqref{eq:zinfsing}.
We get
\beq
0=P_1= x'^N_1-\zeta_1^{N-N_f} z Q_1(x'_1), \label{eq:SQCDuniquew}
\eeq
where we have defined
\beq
Q_1(x'_1)=\prod_{i=1}^{N_f} (x'_1-m_i).
\eeq
Our remaining task is to determine the curve \eqref{eq:vN-1w1}.

From the singular behavior described above, we get
\beq
\sum_{k=1}^{N-1}\phi'_{k,1} x'^{N-1-k}_1 =  a x'^{N-2}_1 + z\sum_{k=N-N_f}^{N-1}b_{N-1-k} x'^{N-1-k}_1,
\eeq
where $a$ and $b_{N-1-k}$ are constants. Then \eqref{eq:vN-1w1} becomes
\beq
0=\left(Nx'^{N-1}_1x'_2-  z\zeta_1^{N-N_f} \frac{\partial Q_1}{\partial x'_1}x'_2   \right) + \left( a x'^{N-2}_1+ z\sum_{k=0}^{N_f-1}b_k x'^{k}_1  \right).
\eeq
By multiplying $x'^2_1$ and using \eqref{eq:SQCDuniquew}, we get
\beq
0=\left(NQ_1-  x'_1\frac{\partial Q_1}{\partial x'_1} \right)x'_1x'_2 +\left(aQ_1+  \sum_{k=0}^{N_f-1}b'_k x'^{k+2}_1  \right),
\eeq
where $b'_k=b_k/\zeta^{N-N_f}$.
This equation suggests that the polynomial $aQ_1+  \sum_{k=0}^{N_f-1}b'_k x'^{k+2}_1$ must vanish 
at the zeros of the polynomial $NQ_1- x'_1(\partial Q_1/\partial x'_1)$. 
Because they are polynomials of $x'_1$
of degree $N_f+1$ and $N_f$ respectively, the constants $b'_k$ must be such that
\beq
aQ_1+  \sum_{k=0}^{N_f-1}b'_k x'^{k+2}_1=c'(1+cx'_1)\left(Q_1-\frac{ x'_1}{N}\frac{\partial Q_1}{\partial x'_1} \right),
\eeq
where $c$ and $c'$ are constants. Comparing the $x'^0_1$ and $x'^1_1$ terms, we get
\beq
c'=a,~~~~~c=-\frac{1}{N} \sum_{i=1}^{N_f} \frac{1}{m_i}.
\eeq
Note that the zeros of $NQ_1- x'_1(\partial Q_1/\partial x'_1)$ occur at the points where $P_1=\partial P_1/\partial x'_1=0$.
As explained in subsection~\ref{sec:commuting}, these points constrain the moduli parameters.

The curve is now
\beq
x'_1x'_2+\frac{a}{N} \left(1-\frac{x'_1}{N}\sum_{i=1}^{N_f} \frac{1}{m_i} \right)=0.
\eeq
The constant $a$ is determined from the behavior of $x'_1$ and $x'_2$ at $z \to 0$. 
Since $x'^N_2 \to \zeta_2^N z^{-1}$ and $x'^N_1 \to (-1)^{N_f}\zeta_1^{N-N_f} z\prod_{i=1}^{N_f} m_i$,
we get $(x'_1x'_2)^{N} \to (-1)^{N_f} \zeta_1^{N-N_f}  \zeta_2^N \prod_{i=1}^{N_f} m_i$.
Therefore, the final result is
\beq
0&=x'_1x'_2- \Lambda^3_{\rm eff} \left(1-\frac{x'_1}{N}\sum_{i=1}^{N_f} \frac{1}{m_i} \right), \label{eq:SQCDlinkcurve} \\
\Lambda_{\rm eff}^{3}&= \left[ (-1)^{N_f}\zeta_1^{N-N_f} \zeta_2^{N} \prod_{i=1}^{N_f} m_i \right]^{\frac{1}{N}}. 
\eeq
This is the curve discussed in subsection~\ref{sec:NflessN}.

\paragraph{$T_N$ theory coupled to $\CN=1$ and $\CN=2$ vector multiplets.}

Here we derive the curve of the theory discussed in subsection~\ref{sec:N1N2mixed}.
The line bundle $L'_1=L_1 \otimes L_B$ and $L'_2=L_2 \otimes L_A$ have degrees $\deg L'_1=0$ and $\deg L'_2=1$, respectively.

The singular behaviors of the Higgs fields are the following. At $z \to 0$, we have
\beq
&\Phi'_1 \to \frac{\zeta_A}{z^{1/N}} \diag(1,\omega_N,\cdots,\omega_N^{N-1}), \\
&\Phi'_2 \to c  z^{1/N} \diag(1,\omega_N^{-1}, \cdots, \omega_N^{-N+1}),
\eeq
where we have used the result of subsection~\ref{sec:commuting} to determine the behavior of $\Phi'_2$.
Similarly, at $z \to \infty$, we require
\beq
&\Phi'_1 \to \zeta_C z^{1/N} \diag(1,\omega_N,\cdots,\omega_N^{N-1}), \\
&\Phi'_2 \to z  \frac{c }{ z^{1/N}} \diag(1,\omega_N^{-1}, \cdots, \omega_N^{-N+1}),
\eeq
where the factor $z$ in $\Phi'_2$ comes from the fact that $\deg L'_2=1$.
Finally, at $z \to 1$ we require
\beq
&\Phi'_1 \to c (z-1)^{1/N}\diag(1,\omega_N,\cdots,\omega_N^{N-1}) \\
&\Phi'_2 \to \frac{\zeta_B}{ (z-1)^{1/N} } \diag(1,\omega_N^{-1}, \cdots, \omega_N^{-N+1}).
\eeq
These behaviors suggest the following behaviors of $\phi'_{k,\ell}$;
\begin{enumerate}
\item At $z=0$, $\phi'_{k,\ell}$ behaves as
\beq
\phi'_{k,\ell} \to 
\left\{
\begin{array}{ll}
O(z^{1}) & \ell>k \\
O(z^{0}) & k \neq N \\
O(z^{-1}) & (k,\ell)=(N,0)
\end{array}
\right.
\eeq

\item At $z=\infty$, $\phi'_{k,\ell}$ behaves as
\beq
\phi'_{k,\ell} \to 
\left\{
\begin{array}{ll}
O(z^{\ell-1}) & \ell>k \\
O(z^{\ell}) & k \neq N \\
O(z^{1}) & (k,\ell)=(N,0)
\end{array}
\right.
\eeq

\item At $z=1$, $\phi'_{k,\ell}$ behaves as
\beq
\phi'_{k,\ell} \to 
\left\{
\begin{array}{ll}
O((z-1)^{1}) & k>\ell \\
O((z-1)^{0}) &  \ell \neq N \\
O((z-1)^{-1}) & (k,\ell)=(0,N)
\end{array}
\right.
\eeq
\end{enumerate}

The only possible solution for $P_1 \equiv \det (x'_1-\Phi'_1)$ is given as
\beq
P_1=x'^N_1 -\left(\frac{\zeta_A^{N}}{z}+\zeta_C^{N} \right)(1-z).\label{eq:21curve'}
\eeq
Next, let us determine \eqref{eq:vN-1w1}. Using the above singular behaviors, we get,
\beq
0= x'^{N-1}_1x'_2+  (a_1z +a_2)x'^{N-2}_1 + (1-z)  \sum_{k=2}^{N-1} b_k  x'^{N-1-k}_1,
\eeq
for some constants $a_{1,2}$ and $b_k$. Multiplying $x'^2_1$ and using \eqref{eq:21curve'}, we get
\beq
0=(\zeta_C^N z+\zeta_A^N)( x'_1x'_2+a_1 z+a_2)+ z\sum_{k=2}^{N-1}b_k x'^{N+1-k}_1. \label{eq:link21curve}
\eeq
Let us see the behavior at $(\zeta_C^N z+\zeta_A^N) \to 0$.
In this limit, we have $x'_1 \sim (\zeta_C^N z+\zeta_1^N)^{1/N}$ as one can see from \eqref{eq:21curve'}.
Since the first term of \eqref{eq:link21curve} vanishes linearly as $(\zeta_C^N z+\zeta_A^N) \to 0$,
we must set $b_k=0$ for \eqref{eq:link21curve} to be consistent. Then we get $x'_1x'_2+a_1z+a_2=0$.
By considering the limit $(\zeta_C^N z+\zeta_A^N) \to 0$ again, $a_1$ and $a_2$ must be such that
$x'_1x'_2=a(\zeta_C^N z+\zeta_A^N)$. Note that the point $(\zeta_C^N z+\zeta_A^N)=0$
is exactly the point $P_1=\partial P_1/\partial x'_1=0$ where solutions of $x'_1$ degenerate.
This point is important to constrain the curve, as explained in subsection~\ref{sec:commuting}.

The constant $a$ is determined by considering the limit $z \to 1$.
The final result for the curve is 
\beq
x'_1x'_2=\left(\frac{\zeta_B^N}{(\zeta_C^N +\zeta_A^N)^{N-1}}\right)^{1/N} (\zeta_C^N z+\zeta_A^N).
\eeq
This is the curve discussed in subsection~\ref{sec:N1N2mixed}.

%%%%%%%%%%%%%%%%%%%%%%%%%%%%%%%%%%%%%%%%%%%%%%%%%%%%%%%%%%%
\bibliographystyle{JHEP}
\bibliography{ref}

\end{document}